\documentclass[preprint,10pt,authoryear,nopreprintline]{elsarticle}
\usepackage[margin=2.5 cm]{geometry} 
\usepackage{amsmath}
\usepackage{mathtools}
\usepackage{rotating}
\usepackage{multirow}
\usepackage{multicol}

\usepackage{graphicx}
\usepackage{dcolumn}
\usepackage{natbib}
\usepackage{siunitx}
\usepackage{array,overpic}
\usepackage{booktabs}
\usepackage[colorlinks=true,linkcolor=black, citecolor=cyan, urlcolor=cyan]{hyperref}
\usepackage{bm}

\usepackage[dvipsnames]{xcolor}

\usepackage{lmodern}

\newcommand{\ds}{\displaystyle}
\newcommand{\be}{\begin{eqnarray}}

\newcommand{\ee}{\end{eqnarray}}

\newcommand{\Op}{\Omega} 

\newcommand{\gav}{\bar{g}}

\renewcommand{\vec}[1]{{{\mathbf #1}}}
\newcommand{\pdhfrac}[2]{\mathchoice{\frac{#1}{#2}}{#1/#2}{#1/#2}{#1/#2}}
\newcommand{\p}[2]{\pdhfrac{{\partial}#1}{{\partial}#2}}

\def\third{{\textstyle \frac{1}{3}}}

\renewcommand{\d}{\mathrm{d}}

\newcommand{\linVk}{\third\Vk}
\newcommand{\linVVk}{\third\VVk}
\newcommand{\Sigkoneone}{\Sigma^{k}_{11}}
\newcommand{\Sigkg}{\Sigma^{k,g}}
\newcommand{\Kk}{K^{k}}
\newcommand{\Jk}{J^{k}}
\newcommand{\Ddiffk}{D^{k}}
\newcommand{\Deffk}{D^{\mathrm{e,eff},k}}
\newcommand{\keffek}{\kappa^{\mathrm{e,eff},k}}
\newcommand{\keffk}{\kappa^{\mathrm{eff},k}}
\newcommand{\ckR}{c^{k,R}}

\newcommand{\dd}{\partial}

\newcommand{\cref}{c^{\mathrm{ref}}}

\newcommand{\DD}{{\mathcal D}}

\definecolor{pink}{rgb}{1,0.5,0.5}
\definecolor{darkred}{rgb}{0.8, 0.0, 0.0}
\definecolor{darkgreen}{rgb}{0.0, 0.6, 0.0}
\definecolor{darkblue}{rgb}{0.0, 0.0, 0.8}
\definecolor{grey}{rgb}{0.7 0.7 0.7}

\newcommand{\sigmaeff}{\sigma^{\text{eff}}}

\newcommand{\ueff}{u^{\text{eff}}}
\newcommand{\epsiloneff}{\epsilon^{\text{eff}}}

\newcommand{\epsb}{\epsilon^{\mathrm{na}}_{ij}}
\newcommand{\epsbvol}{\epsilon^{\mathrm{na}}_{ll}}

\newcommand{\ub}{U^{b}_{i}}

\newcommand{\Obek}{\Omega^{\mathrm{na},k}}

\newcommand{\Ok}{\Omega^{k}}
\newcommand{\Om}{\Omega}

\newcommand{\Ose}{\Omega^{\mathrm{se}}}

\newcommand{\dOcck}{\partial \Omega^{\mathrm{cc},k}}

\newcommand{\dOk}{\partial \Omega^{k}}
\newcommand{\dO}{\partial \Omega}
\newcommand{\Obe}{\Omega^{\mathrm{na}}}
\newcommand{\dObe}{\partial \Omega^{\mathrm{na}}}

\newcommand{\Ubek}{U^{\mathrm{na},k}}
\newcommand{\ebek}{\epsilon^{\mathrm{na},k}}
\newcommand{\sbek}{\sigma^{\mathrm{na},k}}
\newcommand{\sbe}{\sigma^{\mathrm{na}}}

\newcommand{\Ubekz}{U^{\mathrm{na},(0)}}
\newcommand{\ebekz}{\epsilon^{\mathrm{na},(0)}}
\newcommand{\sbekz}{\sigma^{\mathrm{na},(0)}}

\newcommand{\Ubekf}{U^{\mathrm{na},(1)}}
\newcommand{\ebekf}{\epsilon^{\mathrm{na},(1)}}
\newcommand{\sbekf}{\sigma^{\mathrm{na},(1)}}

\newcommand{\Uk}{U^{k}}
\newcommand{\Ukr}{{U_r^{k}}}
\newcommand{\ek}{\epsilon^{k}}
\newcommand{\sk}{\sigma^{k}}

\newcommand{\Ukz}{U^{(0)}}
\newcommand{\ekz}{\epsilon^{(0)}}
\newcommand{\skz}{\sigma^{(0)}}

\newcommand{\Urz}{{U_r^{(0)}}}

\newcommand{\Ukf}{U^{(1)}}
\newcommand{\ekf}{\epsilon^{(1)}}

\newcommand{\Gbek}{G^{\mathrm{na},k}}
\newcommand{\Gk}{G^{k}}

\newcommand{\lbek}{\lambda^{\mathrm{na},k}}
\newcommand{\lk}{\lambda^{k}}

\newcommand{\alk}{\alpha^{k}}
\newcommand{\gk}{g^{k}}
\newcommand{\Vk}{V_c^{k}}
\newcommand{\Vkp}{V^{\mathrm{p}}_k}
\newcommand{\Vkn}{V^{\mathrm{n}}_k}
\newcommand{\VVk}{\mathcal{V}_c^{k}}
\newcommand{\V}{\mathcal{V}_{c}}
\newcommand{\Vp}{V_c^{\mathrm{p}}}
\newcommand{\Vn}{V_c^{\mathrm{n}}}
\newcommand{\linVkp}{\third\Vp}
\newcommand{\linVkn}{\third\Vn}
\newcommand{\ck}{c^{k}}
\newcommand{\ckref}{c^{k,\mathrm{ref}}}
\newcommand{\Ckref}{\mathcal{C}^{k,\mathrm{ref}}}
\newcommand{\ckmax}{c^{k,\mathrm{max}}}
\newcommand{\cpmax}{c^{\mathrm{p,max}}}

\newcommand{\ckz}{c^{k,(0)}}
\newcommand{\ckf}{c^{k,(1)}}
\newcommand{\Cref}{\mathcal{C}^{\mathrm{ref}}}

\newcommand{\muk}{\mu^{k}}
\newcommand{\etak}{\eta^{k}}
\newcommand{\Nk}{N^{k}}
\newcommand{\phie}{\phi^{\mathrm{e}}}
\newcommand{\phik}{\phi^{k}}
\newcommand{\phip}{\phi^{\mathrm{p}}}
\newcommand{\phin}{\phi^{\mathrm{n}}}

\newcommand{\etakz}{\eta^{k,(0)}}
\newcommand{\Nkz}{N^{k,(0)}}
\newcommand{\mukz}{\mu^{k,(0)}}

\newcommand{\phiez}{\phi^{\mathrm{e},(0)}}
\newcommand{\phief}{\phi^{\mathrm{e},(1)}}

\newcommand{\phikf}{\phi^{k,(1)}}

\newcommand{\etakf}{\eta^{k,(1)}}

\newcommand{\mukf}{\mu^{k,(1)}}

\newcommand{\ke}{\kappa^{\mathrm{e}}}
\newcommand{\K}{\mathcal{K}}

\newcommand{\Bk}{\mathcal{B}^{k}}
\newcommand{\Bkhat}{\hat{\mathcal{B}}^{k}}
\newcommand{\Bs}{\mathcal{B}^{\mathrm{s}}}
\newcommand{\Dk}{D_0^{k}}

\newcommand{\ce}{c^{\mathrm{e}}}
\newcommand{\cezero}{c^{\mathrm{e},0}}

\newcommand{\cez}{c^{\mathrm{e},(0)}}
\newcommand{\cef}{c^{\mathrm{e},(1)}}

\newcommand{\Nke}{N^{\mathrm{e},k}}
\newcommand{\ike}{i^{\mathrm{e},k}}
\newcommand{\Nse}{N^{\mathrm{e},\mathrm{s}}}
\newcommand{\ise}{i^{\mathrm{e},\mathrm{s}}}

\newcommand{\Nkez}{N^{\mathrm{e},k,(0)}}
\newcommand{\ikez}{i^{\mathrm{e},k,(0)}}

\newcommand{\B}{\mathcal{B}}

\newcommand{\ik}{i^{k}}
\newcommand{\ikz}{i^{k,(0)}}

\newcommand{\che}{\chi^{\mathrm{e}}}
\newcommand{\chk}{\chi^{k}}

\newcommand{\tplus}{t^+}

\newcommand{\Ueqk}{\mathrm{U}^{\mathrm{eq},k}}

\usepackage{amssymb}

\journal{}

\begin{document}

\begin{frontmatter}


\title{
Incorporating multiscale mechanics in lithium-ion battery models}

\author[inst1,inst5]{Andrea Giudici}
\author[inst4]{Andres F. Galvis}
\author[inst2,inst5]{Smita Sahu}
\author[inst3]{Robert Timms}
\author[inst1,inst5]{Colin Please}
\author[inst1,inst5]{Jon Chapman}
\author[inst2,inst5]{Jamie M. Foster \corref{cor1}}
\ead{jamie.michael.foster@gmail.com}
\cortext[cor1]{Corresponding author}

\affiliation[inst1]{organization={Mathematical Institute, University of Oxford},
            addressline={Andrew Wiles Building, Woodstock Road}, 
            city={Oxford},
            postcode={OX2 6GG}, 
            country={United Kingdom}}

            \affiliation[inst5]{
            organization={The Faraday Institution},
            addressline={Quad One, Becquerel Avenue, Harwell Campus},
            city={Didcot},
            postcode={OX11 0RA},
            country={United Kingdom}}

\affiliation[inst4]{organization={School of Electrical and Mechanical Engineering, University of Portsmouth},
            addressline={Anglesea Building, Anglesea Road}, 
            city={Portsmouth},
            postcode={PO1 3DJ},
            country={United Kingdom}}

\affiliation[inst3]{
            organization={Ionworks Technologies Inc},
           addressline={5831 Forward Ave \#1276 },
            city={Pittsburgh, PA},
            postcode={15217},
            country={USA}}

\affiliation[inst2]{organization={School of Mathematics and Physics, University of Portsmouth},
            addressline={Lion Gate Building, Lion Terrace}, 
            city={Portsmouth},
            postcode={PO1 3HF },
            country={United Kingdom}}

\begin{abstract}
Lithiation-induced swelling in lithium-ion batteries generates stresses not only within active particles, but also across the surrounding non-active matrix, electrodes, and cell stack. These stresses can modify the chemical potential of lithium and therefore influence transport, reaction kinetics, and terminal voltage. We derive a reduced-order electro-chemo-mechanical model that captures this multiscale coupling while retaining a complexity comparable to standard Doyle--Fuller--Newman models. The electrode is modelled as a periodic array of spherical active particles embedded in a homogenised elastic non-active matrix. Exploiting the small stiffness of the non-active matrix relative to the active material, together with scale separation between particles, electrodes, and the full cell, we obtain an effective mechanical correction to the active-particle chemical potential and overpotential. This correction depends on particle swelling, electrode-scale strain, and macroscopic boundary conditions such as clamping or applied pressure. The resulting formulation can be incorporated directly into DFN, SPMe, and SPM frameworks, providing a computationally efficient route to include battery-scale mechanical effects in electrochemical simulations.
\end{abstract}

\end{frontmatter}

\section{\label{intro}Introduction}

During a charge and discharge cycle, the active materials within a lithium-ion battery expand and shrink. At the microscale (that of individual electrode particles), this expansion builds internal stresses, potentially leading to mechanical degradation of the active particles via fracture \citep{Ai20}, accelerated solid-electrolyte interphase (SEI) formation, and capacity loss \citep{XuChao}. At the meso- and macroscale, that of an individual electrode and pouch/roll cell respectively, particle expansion pushes (or pulls) on the surrounding non-active material, made of binder and additives. Due to the complex and heterogeneous composition of a battery, this expansion generates stresses at the electrode and cell scale {\citep{Foster2025}}, resulting in global deformation \citep{giudici2024mechanical,timms2023mechanical}. Such deformation can cause further material degradation and delamination \citep{foster2017mathematical,foster2017causes}, buckling of layers \citep{pfrang2018long}, and drive the flow of electrolyte \citep{solchenbach2024electrolyte,giudicielectrolyte}.

Mechanical stresses influence not only the mechanical integrity and longevity of a battery but also its electrochemical behaviour. This coupling arises through stress-assisted diffusion, whereby the stress field alters the chemical potential of lithium and thus the rate and direction of transport. Two distinct mechanisms underpin this coupling. First, concentration gradients within an active particle generate swelling and internal stresses, which feed back on lithium diffusion. This mechano-chemical coupling, first described by \citet{zhang2007numerical}, alters the intercalation dynamics and thus the behaviour of the battery. However, it is confined to the particle and independent of its surroundings. For this reason, it is unaffected by the meso- and macroscale stress state of the battery. Second, since the active material is embedded in a composite matrix of binder and conductive additives that transmit forces, the stress state at the electrode scale affects the stress state of the particle, altering its chemical potential energy. For this reason, the stress-assisted diffusion process in each particle depends not only on its own lithiation state, but also on the lithiation and expansion of nearby particles and on the mechanical state of the entire electrode and battery, thus linking mechanics and electrochemistry across scales. For example, even mild external pressures—of the order of an atmosphere—can alter the capacity and resistance of a cell \citep{mohtat2021reversible}.

\begin{figure}
\centering
    \includegraphics[width=\textwidth]{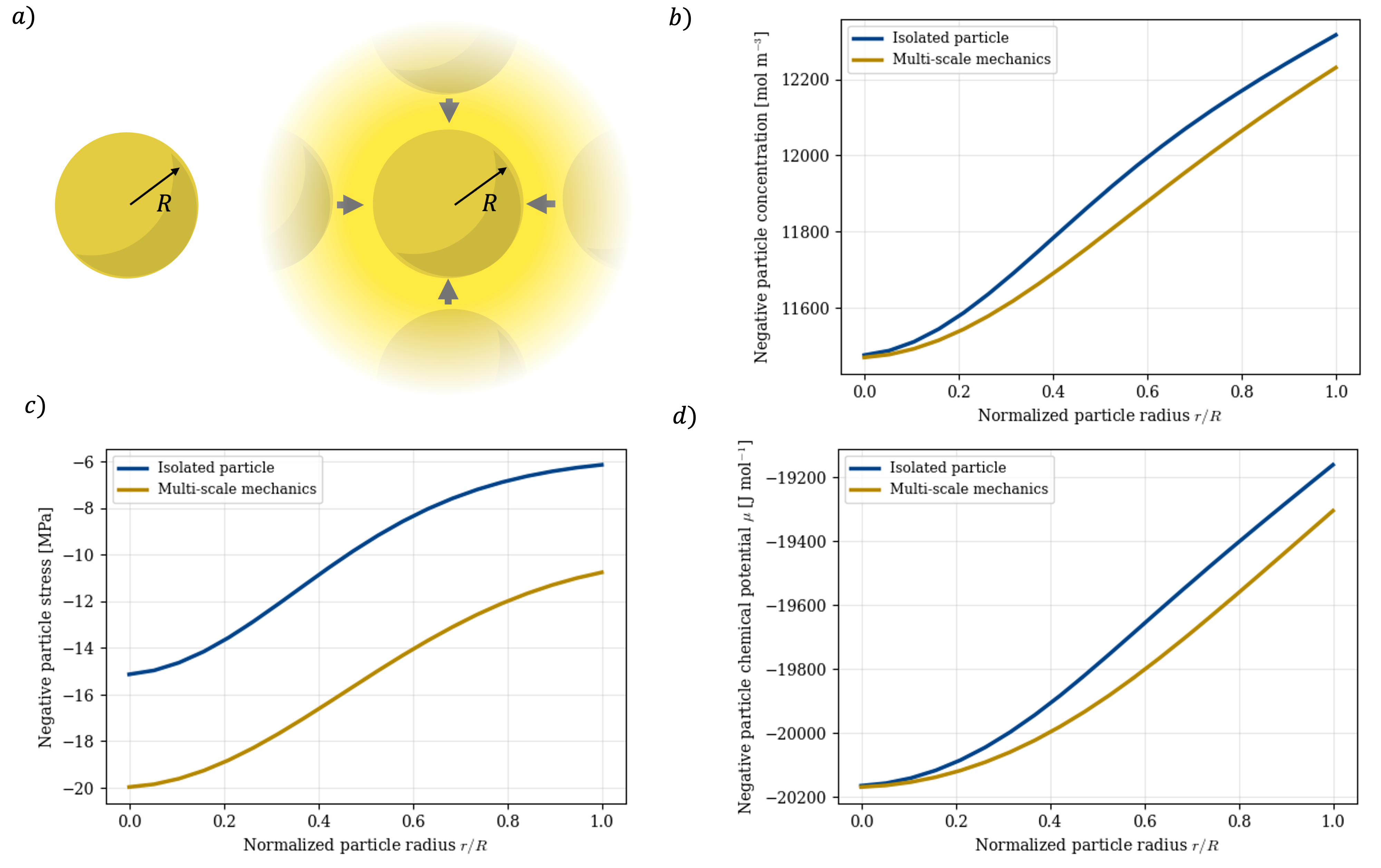}
    \caption{Simulation of battery charging at 0.5C in PyBaMM showing how the non-active matrix and global mechanics modify the behaviour inside an active particle. (a) Schematic comparison between an isolated active particle, for which the mechanical contribution is only via stress-assisted diffusion, and a particle embedded in a deforming non-active matrix, which also transmits stress. (b)--(d) Negative-particle concentration, radial stress, and electro-chemo-mechanical potential $\mu$ across the particle radius. For the multiscale mechanics case, the plotted quantities are averaged over the angular coordinates on each sphere, so that only their radial dependence is shown.}
\label{fig:cartoon}
\end{figure}

Previous work has modelled the coupling between electrochemistry and mechanics (e.g. \citealt{castelli2021efficient}). However, most studies focus solely on the local effect—stresses internal to a particle, neglecting the stresses transmitted through the non-active matrix \citep{zhang2007numerical}. Figure~\ref{fig:cartoon} illustrates how the surrounding non-active matrix affects the electrochemistry in the active particle using PyBaMM simulations. In the isolated-particle calculation, mechanics enters only through stress-assisted diffusion generated by the particle's own concentration gradients. In the multiscale mechanics calculation, the same stress-assisted diffusion is coupled to the global mechanical response and to the external pressure transmitted through the non-active matrix; this changes the stress and therefore the chemical potential inside the particle, leading to a different concentration profile.

In this work, we develop a reduced-order model that captures the multiscale coupling resulting from the mechanical work done by the non-active material, linking stresses at the particle, electrode, and cell scales with electrochemical transport. To include mechanics, we model the electrode as a periodic array of identical cubic unit cells (Figure~\ref{fig:cartoon}), each containing a spherical active particle embedded in a homogenised region of non-active material composed of a binder--electrolyte matrix. Both materials are assumed linearly elastic and the mechanical influence of the SEI is either (i) assumed negligible or (ii) assumed to be already included as part of the binder-electrolyte effective medium. Lithiation-driven swelling of the particle generates local stresses, which modify the chemical potential and in turn the intra-particle transport and reaction kinetics, thereby coupling mechanics and electrochemistry at the microscale. The particle size is typically an order of magnitude smaller than the electrode thickness, enabling a clear separation of scales and motivating an asymptotic homogenisation approach that links the micro-, meso-, and macroscales, while the thin aspect ratio of electrodes is used to link the meso- and macroscale \citet{giudici2024mechanical}. The problem is closed by homogenised equilibrium equations and global boundary conditions provided by the cell housing.
We obtain an analytic result estimating the correction to the particle potential, and thus to the voltage and current in the battery. Our formulation has comparable complexity to the Doyle-Fuller-Newman (DFN) model, incorporates mechanical effects, and can readily be used with SPM, SPMe and DFN \citep{brosa2022continuum} on simulation software such as PyBaMM showing an alteration of cell potential of several millivolts at full charge.

In what follows, we build this multiscale coupling systematically. First, Section~\ref{sec:2} introduces the full DFN model and its nondimensionalisation.
Sections~\ref{sec:3}-\ref{sec:4} then focus on the mechanics. In Section~\ref{sec:3}, we note that the homogenised non-active binder-electrolyte medium is much softer than the particle, allowing an expansion in the stiffness ratio. The leading-order solution corresponds to the unconstrained particle of \citet{zhang2007numerical}, while the first-order correction captures the binder’s mechanical feedback.
Section~\ref{sec:4} links the micro- and mesoscale stresses through multiple-scale homogenisation, deriving the effective cell functions that relate binder stress on the particle surface to mesoscale strain and particle swelling.
Section~\ref{sec:5} revisits the macroscopic mechanical framework of \citet{giudici2024mechanical} and considers the limit of thin aspect ratio and infinitely stiff current collectors (i.e. layers are wider than they are thick and current collectors are non-deformable) showing that deformation occurs only in the through-cell direction and that macroscopic constraints, such as clamped or stress-free ends, influence the mesoscale strain field. With these results, Section~\ref{sec:6} returns to the electrochemical problem and, using the computed stresses in the non-active material matrix, derives an exact first-order correction to the lithium concentration exploiting a surface average.
Finally, Section~\ref{sec:7} illustrates how this coupling can be accounted for by modifying the OCP entering the intercalation equation, affecting the voltage-current relationship in SPM, SPMe, and DFN-type models. 

Readers primarily interested in the model formulation and its physical consequences, rather than the mathematical derivations, are encouraged to begin with Section~\ref{sec:2} and then proceed directly to Section~\ref{sec:7}, where the key results and experimental implications are summarized.

\section{The microscale model}
\label{sec:2}

We begin by writing down the equations that govern lithium transport and mechanical deformation in a battery. We write the complete model in the same form as that presented in \cite{brosa2022continuum}, but extend this to include the effects of mechanical stresses in all solid parts of the battery.

The non-active material in a battery is typically a porous material with pores filled by liquid electrolyte. Moreover, there is typically a thin SEI layer encasing individual active particles. We do not separate these materials but rather treat them together as a composite poroelastic material, in a manner similar to, for example, {\cite{Foster2025}}\footnote{Note \cite{Foster2025} take the binder to be viscoelastic. However, here for simplicity we take it to be elastic.}. We shall refer to this region as the non-active (homogenised) material. We adopt a similar approach in the separator.

There are three spatial regions to consider, namely the positive and negative electrodes and the separators. In the electrodes we need to consider the governing equations in two subregions: (i) the solid active particles, and (ii) the composite poroelastic non-active matrix.
In the separator there is only the composite poroelastic separator/electrolyte material.
We begin by detailing the equations in the electrodes. Where necessary we distinguish the positive and negative electrodes by the superscript $k \in \{\mathrm{p}, \mathrm{n}\}$.

\subsection{Mechanics}

Although the liquid phase can affect the stress in the material due to hydrostatic pressures, this is expected to occur only during very fast charging. For sufficiently slow charge rates, the fluid has time to flow out of the pores without a large buildup of pressure, so that the liquid does not substantially affect the stress state of the non-active matrix. Thus, we assume hydrostatic pressure of the electrolyte can be ignored. Note (i) that the fluid pressure may not be negligible in extreme circumstances e.g. during cell impact, see \citep{Foster2025}, and (ii) that fluid flow can also alter the bulk concentration of electrolyte salt during repeated cycling, as
discussed in \cite{solchenbach2024electrolyte,giudicielectrolyte}, and thus affect the electrochemistry of the battery. Here, we choose to ignore this accumulation effect over several cycles and only focus on the
performance change due to mechanics over one single cycle. Electrolyte
flow and mechanical effects will be fully coupled in later work.  

A detailed treatment of the non-active matrix as a porous material arguably requires capturing the nonlinear behaviour typical of foams \citep{rahimidehgolan2023compressive}, which show a stiff initial response accompanied by a softening due to the buckling of the pores, followed by a second stiffening due to self-contact. Including such nonlinear mechanics in our multiple scale model requires several parameters difficult to estimate and significantly complicates the
problem. Here, since we are interested only in a first approximation to the additional stress caused by compression of the non-active matrix, we treat the non-active matrix as a linear elastic solid. We treat the separator as a homogenised elastic material as well. 

We allow the non-active matrix material in the positive and negative electrodes to differ. We identify them with the superscript
`$\mathrm{na},k$' where `$\mathrm{na}$' stands for non-active
and $k\in \{\mathrm{p,n}\}$ identifies the electrode, with $\Obek$
being the non-active matrix domain in the respective electrode. When
deformed, the non-active matrix undergoes a displacement $\Ubek_i$ in
the $X_i$ direction, for each $i\in\{1,2,3\}$. The resulting strain and stress are
\be 
\label{eq:mech_bek_dim}
    \ebek_{ij} =\frac{1}{2}\left( \frac{\dd \Ubek_j}{\dd X_i} + \frac{\dd \Ubek_i}{\dd X_j} \right), \qquad \sbek_{ij}= 2 \Gbek \ebek_{ij} + \delta_{ij} \lbek \ebek_{ll} \quad \textrm{in} \quad \Obek,
\ee
respectively, where $\Gbek$ is the effective shear modulus and $\lbek$
the effective second Lam\'e coefficient for the non-active matrix
medium in electrode $k$.\footnote{We use the shear notation, $G$, and
  not the first Lam\'e coefficient $\mu$ to avoid confusion with the
  chemical potential later on.} Note that here and elsewhere we are
using the Einstein summation convention, summing over repeated
indices, and $\delta_{ij}$ is the Kronecker delta.

The active particle in electrode $k\in \{\mathrm{p,n}\}$ occupies the domain $\Ok$. The particle is stiff, undergoes small strains and is thus modelled as a linear elastic material with  Lam\'e coefficients $\Gk$ and
$\lk$, and an isotropic swelling strain $\gk$ due to changes in the state of
lithiation. We write the displacement field in the particle as $\Uk_i$, with $i\in\{1,2,3\}$, so that the strain and stress are
\be  
\label{eq:mech_k_dim}
    \ek_{ij} = \frac{1}{2} \left( \frac{\dd \Uk_i}{\dd X_j} + \frac{\dd \Uk_j}{\dd X_i} \right), \qquad {\sk_{ij} = 2 \Gk \left( \ek_{ij} -\gk \delta_{ij}  \right) + \delta_{ij} \lk \left( \ek_{ll} -  \gk \delta_{ll} \right)} \quad \textrm{in} \quad \Ok,
\ee
where $\epsilon^k_{ij} - \gk \delta_{ij} $ is the elastic component of
the strain, that is, the total strain in the active particle minus the strain due to the lithiation-induced expansion.
The relationship between the linear swelling strain and lithiation is given by
\be
\label{eq:mech_g_dim}
    \gk = \linVk\,(\ck-\ckref) \quad \textrm{in} \quad \Ok,
\ee
where $\ck$ is the molar concentration of Li and $\ckref$ is a reference Li
concentration at which the expansion strain is zero, equal to the Li
concentration in the material immediately after synthesis.  Thus,
typically, $c^{\mathrm{n,ref}}=0$ and
$c^{\mathrm{p,ref}}=\cpmax$ corresponding to electrode manufacture in which the negative electrode is delithiated whereas the positive one is lithiated, see \cite{Menye}. Note that concentrations have been
labelled with the electrode identifier as a superscript, so that tensor indices remain unambiguously as subscripts. The constant $\Vk$ is the partial molar volume of Li, so that $\third\Vk$ is the corresponding linear expansion coefficient, and depends on the active-particle material. 

Elastic equilibrium requires
\be
\label{eq:mech_eq_dim}
    \frac{\dd \sbek_{ij}}{\dd X_j}=0\quad \textrm{in} \quad \Obek,\qquad
    \frac{\dd \sk_{ij}}{\dd X_j} = 0 \quad \textrm{in} \quad \Ok.
\ee
We assume that the non-active matrix and particle surfaces remain adhered to one another so that at these interfaces, $\dOk$, both stresses and displacements are continuous:
\be 
\label{eq:mech_bc_dim}
    \sk_{ij} n_j = \sbek_{ij} n_j, \qquad \Uk_i = \Ubek_i \quad \text{on} \quad \dOk ,
\ee
where $n_j$ is the normal vector (oriented outwards from the
particle).

\subsection{Transport in the active particles}

In the solid active particles the intercalated lithium is conserved and is taken to be transported down gradients in the electro-chemo-mechanical potential, $\muk$ (units J/mol), so that
\begin{equation} 
\label{eq:diff_k_dim}
    \frac{\partial \ck}{\partial t}+\frac{\partial \Nk_i}{\partial X_i}=0,\qquad
    \Nk_i=-\frac{\Dk \ck}{RT}\left(1-\frac{\ck}{\ckmax}\right)
    \frac{\partial \muk}{\partial X_i} \quad \textrm{in} \quad \Ok, 
\end{equation}
where  $\Dk$ is a constant and $\ckmax$ is the maximum possible lithium concentration. Since the trace $\sk_{ll}$ is three times the hydrostatic stress,
\begin{equation}
\label{eq:mu_dim}
    \muk = \mu^0- F \Ueqk (\ck) - \linVk\, \sk_{ll},
  \end{equation}
 where  the quantity $\Ueqk$ is the so-called equilibrium
 over-potential, which, as we shall see shortly, is the potential
 difference at which the net surface reaction is zero in a stress-free state.
Note that we have included the steric effects that decrease the
mobility of intercalated lithium at high lattice occupancy,  captured
by the presence of the factor of $1-c/\ckmax$. These effects are
appropriate to include here because we are interested in batteries
operating over their full range of states of charge, so that the
occupancy of insertion material will vary from close to zero, to close
to $\ckmax$. Although this is not a complete justification for the form of the right-hand side of (\ref{eq:diff_k_dim}b) we do note that it has
the following appealing properties: (i) the flux is isotropic, and therefore pertains to the insertion material on lengthscales greater than that of individual crystals of anisotropic materials such as LFP
and NMC, (ii) it predicts no flux when there are no Li particles to move, or when there are no vacant lattice sites for them to move to, and (iii) it is invariant under the transformation $\ck \to 1-\ck/\ckmax$. This latter property is inline with the intuition that the hopping rates of Li at high-occupancy should be equal to that of vacancies at low-occupancy. In any case, the form of (\ref{eq:diff_k_dim}b) that we use is in agreement with, for example \cite{ZengBazant2014}, but departs from, for example \cite{ZelicKatrasnik2019} and \cite{Chakraborty2015} where steric effects are neglected. 

The deintercalation rate at the surface of the solid particle into the non-active matrix medium (more on this below) is taken to be given by the symmetric Butler-Volmer relation (with transfer coefficient equal to 1/2), i.e.
\begin{equation} 
\label{eq:diff_k_bc_dim}
   \Nk_i n_i= \Kk \Jk\sinh\left(\frac{F\etak}{2RT} \right) \quad \textrm{on} \quad \dOk,
\end{equation}
where 
\begin{equation}
   \Jk = \sqrt{\frac{\ce}{\cezero}\frac{\ck}{\ckmax}\left(1-
    \frac{\ck}{\ckmax} \right)}
\end{equation}
where $n_i$ is the unit normal pointing into the non-active matrix medium from the active particle, $\ce$ is the electrolyte concentration, $\cezero$ is the initial electrolyte concentration, and $\Kk$ is a reaction rate constant. The overpotential $\etak$ is given by (see for example \cite{Ai,XuRong,Zhang})
\begin{equation} 
\label{eq:eta_dim}
    \etak = \phik - \phie - \Ueqk (\ck) - \frac{\linVk}{F}\,\sk_{ll}  \quad \textrm{on} \quad \dOk,
\end{equation}
where $\phik$ and $\phie$ are the electric potential in the active particle and electrolyte respectively both \textit{measured with respect to a Li reference electrode}. In order that it is possible to lithiate a fully depleted electrode and similarly to delithiate a fully lithiated electrode, the function $\Ueqk$  must have logarithmic singularities at the extremes; see {\cite{richardson2020chargetransportmodellinglithium}}. In a stress-free state, the overpotential collapses to the familiar form. Moreover, it aligns with the intuition that deintercalation is encouraged by increases (decreases) in $\phik$ ($\phie$) as well as  compressive hydrostatic stresses (intercalated lithium is ``squeezed out'' of the insertion material). In an effort to connect this with deeper thermodynamic principles, the overpotential $\etak$, can be decomposed into the electro-chemo-mechanical potentials of the product and the reactants. If we consider the reaction to be $\mathrm{Li}^{+,\mathrm{k}} \leftrightharpoons \mathrm{Li}^{+,\mathrm{e}}$ (where the superscript labels denote the domain: k for active material in an electrode and e for the electrolyte, i.e. the liquid subphase within the non-active homogenised medium)\footnote{An analogous result can be obtained if one instead considers the reaction to be $\mathrm{Li}^{\mathrm{a}} \leftrightharpoons \mathrm{Li}^{+,\mathrm{e}}+e^{-,\mathrm{a}}$. } and recall that
\begin{align} 
\label{eq:mu1}
    \mu^{\textrm{Li}^+,\mathrm{k}} &=  \mu^0 - F \Ueqk(\ck) + F \phik -  \linVk\,\sk_{ll} & \quad& \textrm{in} \quad \Ok,\\
\label{eq:mu2}
    \mu^{\textrm{Li}^+,\mathrm{e}} &= \mu^0 + F \phie &\quad& \textrm{in} \quad \Obek,
\end{align}
where $\mu^0$ is a constant reference potential, then it is clear that $\eta = (\mu^{\textrm{Li}^+,\mathrm{a}} - \mu^{\textrm{Li}^+,\mathrm{e}})/F$. Thus, \eqref{eq:diff_k_bc_dim} obeys the foundational principle that the reaction proceeds in the direction that decreases the overall energy (i.e. according to the thermodynamic driving force/reaction affinity \citep{bazant2013theory, zhang2007numerical}). 

A common source of confusion is whether, and if so how,  concentration dependence should appear in \eqref{eq:mu2}. Clarification on this point is given in and around equation (2.60) in \cite{richardson2020chargetransportmodellinglithium}, where it is shown that the presence/lack of the term $RT \log a^{\mathrm{e}}$ in \eqref{eq:mu2}, where $a^{\mathrm{e}}$ is the activity of lithium ions in the liquid electrolyte, depends upon whether potentials are ``true'' potentials (i.e. measured with respect to some genuinely fixed reference, e.g. vacuum-level) or whether they are measured with respect to a Li-metal reference electrode (as is almost always the case in practical experimental work). The basis of the argument presented in \cite{richardson2020chargetransportmodellinglithium} is that the electrochemical potentials of the reactants and products in the stripping/plating must, to a good approximation, be equal because stripping/plating is ``easy'' compared to intercalation; in the sense that it requires very little overpotential. This consideration yields a relationship between the true potential and that defined with respect to a Li reference electrode which, it turns out, differ by $RT \log a^{\mathrm{e}}$. The upshot is that if one chooses to use ``true'' potentials then the logarithm of the activity should appear, whereas if potentials are measured with respect to a Li-metal reference it should not. Here we opt to follow the ubiquitous convention in the battery community and work with potentials measured with respect to a Li reference electrode. 

It is also pertinent to justify why only the volumetric part of the stress appears in \eqref{eq:mu_dim} and, in turn, in \eqref{eq:eta_dim}. Although many insertion materials expand anisotropically at the scale of a single crystal, e.g. graphite which expands primarily along its c-axis (\cite{Boles}), the majority of commercial devices contain so-called secondary particles, made up of many very small, and randomly oriented, single-crystal particles. Due to this disparity in scales and uncorrelated orientations, the secondary particle undergoes \textit{effective} isotropic expansion on Li insertion. The quantity defined in \eqref{eq:mu1} is the energy required to take a mol of Li and remove it from the system to some reference level (in this case to a Li electrode). It follows that the energy required to remove Li is also associated with purely a volumetric stress, and is insensitive to deviatoric stress components which only alter the shape of the host material \citep{lu2016voltage,song2016stress}.

\subsection{The non-active matrix region}

In the non-active medium we assume no fluid motion and use the conventional dilute approximation model for an ideal salt solution where 
\begin{equation} 
\label{eq:e_diff_dim}
    \frac{\partial (\alk \ce )}{\partial t}+\frac{\partial \Nke_i}{\partial X_i}=0, \qquad \Nke_i=- \Bk D^{\mathrm{e}} \frac{\partial \ce}{\partial X_i} +\frac{\tplus}{F}{  \ike_i}  \quad \textrm{in} \quad \Obek
\end{equation}
\begin{equation} 
\label{eq:e_current_dim}
   \frac{\partial \ike_i}{\partial X_i} = 0, \qquad
    \ike_i= -\Bk \ke \left( \frac{\partial \phie}{\partial X_i} -\frac{2RT}{F}(1-\tplus)\frac{1}{\ce}\frac{\partial \ce}{\partial X_i}\right)  \quad \textrm{in} \quad \Obek
\end{equation}
and $\alk$ is the porosity of the non-active material in electrode $k$ and $\ke$ is the ionic conductivity of the electrolyte, $\ce$ is the lithium ion concentration in the liquid electrolyte, $\ike$ is the current density, $\Nke_i$ is the lithium ion flux in the non-active matrix (which we remind, includes also the electrolyte), $\tplus$ is the transference number and $\Bk$ is the Bruggeman tortuosity factor. The boundary conditions that supplement \eqref{eq:e_diff_dim} and \eqref{eq:e_current_dim} are
\begin{equation} 
\label{eq:e_diff_bc_dim}
    \Nke_i n_i= - \Kk \Jk \sinh\left(\frac{F\etak}{2RT} \right) \quad \textrm{on} \quad \dOk,
\end{equation}
\begin{equation} 
\label{eq:e_current_bc_dim}
  \ike_i n_i= - F \Kk
   \Jk \sinh\left(\frac{F\etak}{2RT} \right) \quad \textrm{on} \quad \dOk,
\end{equation}
where $\etak$ is defined in \eqref{eq:eta_dim}.

For the electron conduction through the homogenised non-active matrix we have a homogenised version of Ohm's law that reads
\begin{equation} 
\label{eq:current_dim}
    \frac{\partial \ik_i}{\partial X_i} = 0, \qquad \ik_i = - \Bkhat \kappa^{k} \frac{\partial \phik}{\partial X_i} \quad \text{in} \quad  \Obek,
\end{equation}
where $\kappa^{k}$ is the conductivity in the non-active  region of electrode $k$. It is worth emphasizing that in \eqref{eq:e_current_dim} and \eqref{eq:current_dim} $\Bk \neq \Bkhat$, as the former is the transport efficiency of the electrolyte network in the non-active matrix whereas the latter is that of the binder. Equations \eqref{eq:current_dim} are supplemented by the boundary condition
\begin{equation}
    \ik_i n_i= -F\Kk
   \Jk \sinh\left(\frac{F\etak}{2RT} \right) \quad \textrm{on} \quad \dOk.
\end{equation}

\subsection{The separator region}

In the separator, only ionic conduction and mechanical deformation take place; there are no insertion particles, no reactions and therefore the equations are a simplified version of those used for the non-active matrix. Considering the homogenised separator/electrolyte material in the domain $\Ose$. The ionic conduction needs to satisfy
\begin{equation} 
\label{eq:s_diff_dim}
    \frac{\partial (\alpha^{\mathrm{s}} \ce)}{\partial t}+\frac{\partial \Nse_i}{\partial X_i}=0, \qquad \Nse_i=- \Bs D^{\mathrm{e}} \frac{\partial \ce}{\partial X_i} +\frac{\tplus}{F}\ise_i  \quad \textrm{in} \quad \Ose,
\end{equation}
\begin{equation} 
\label{eq:s_current_dim}
    \frac{\partial \ise_i}{\partial X_i} = 0, \qquad
    \ise_i= -\Bs \ke \left( \frac{\partial \phie}{\partial X_i} -\frac{2RT}{F}(1-\tplus)\frac{1}{\ce}\frac{\partial \ce}{\partial X_i}\right)  \quad \textrm{in} \quad \Ose.
\end{equation}
The mechanical equations in the separator/electrolyte are 
\be 
\label{eq:mech_s_dim}
    \frac{\dd \sigma^{\mathrm{se}}_{ij}}{\dd X_j}=0, \qquad \epsilon^{\mathrm{se}}_{ij} =\frac{1}{2}\left( \frac{\dd U^{\mathrm{se}}_j}{\dd X_i} + \frac{\dd U^{\mathrm{se}}_i}{\dd X_j} \right), \qquad \sigma^{\mathrm{se}}_{ij} = 2 G^{\mathrm{se}} \epsilon^{\mathrm{se}}_{ij}+ \delta_{ij} \lambda^{\mathrm{se}} \epsilon^{\mathrm{se}}_{ll} \quad \textrm{in} \quad \Ose.
    \ee

\subsection{Boundary and interface conditions}
We now supply boundary/interface conditions on the mesoscale boundaries. In order to achieve complete problem closure we also need boundary conditions at the macroscale boundaries, i.e. the edges of the pouch cell/roll cell. These shall be specified in Section~\ref{sec:5}, where we discuss the link between electrode (meso-) and battery (macro-) scale mechanics. We note in passing that the problem requires appropriate initial conditions on all concentrations which, for convenience are usually taken to be those that correspond to a specified state of charge  at equilibrium throughout.

\paragraph{Current collectors/electrode interfaces} At the electrode-current collector interfaces (denoted $\partial\Omega^{cc,\mathrm{p}}$ and
$\partial\Omega^{cc,\mathrm{n}}$), we impose no ionic transport into the current collectors
\be
  \Nke_i n_i = 0, 
  \qquad 
  \ike_i n_i = 0,
\ee
where $\mathbf{n}^{k}$ is the outward unit normal from the electrode $k$ domain.

Each current collector is taken to be an equipotential. We remove the arbitrary degree of freedom in the electronic potentials in the current collectors by setting the negative (anode) current collector potential to zero (i.e. $\Phi^{\mathrm{n}} = 0$) and define the cell voltage, $V$, as the positive (cathode) solid potential (i.e. $\Phi^{\mathrm{p}} = V$) so that
\be
  V \equiv \Phi^{\mathrm{p}} - \Phi^{\mathrm{n}} = \Phi^{\mathrm{p}}.
\ee
To enforce the applied current (with the convention that discharge corresponds to $I>0$), we constrain the total electronic current fluxes at the current-collector faces such that
\begin{equation}
\label{eq:totI_dim}
  \iint_{\partial\Omega^{cc,\mathrm{p}}} i^{\mathrm p}_i n_i\, \mathrm{d}S \;=\; \frac{I}{m},
  \qquad
  \iint_{\partial\Omega^{cc,\mathrm{n}}} i^{\mathrm n}_i n_i\, \mathrm{d}S \;=\; -\,\frac{I}{m},
\end{equation}
where $m$ is the number of anode/cathode electrode pairs, shown in figure \ref{fig:cell}. 

\paragraph{Electrode/separator interfaces}
Here, we require continuity of flux
and current:
\begin{equation}
    \left[ \Nke_i n_i\right] = \left[ \ike_i n_i \right] = 0,
\end{equation}
where ${\bf n}$ is normal to the interface and $\left[\cdot\right]$
denotes the jump in the enclosed quantity across the interface.
We also impose no current 
\begin{equation}
    \ik_i n_i = 0, 
\end{equation}
and continuity of displacement and traction, i.e.
\begin{equation}
    \left[ U_i \right]= \left[ \sigma_{ij} n_j \right] = 0, 
\end{equation}
there, where $n_j$ are components of $\bf{n}$.

\subsection{Nondimensionalisation}
To facilitate the asymptotic expansions which follow, we now  introduce dimensionless variables.
We let $I^*$ be the typical size of the current, $L^*$ the typical length of a microscale representative element (comparable to a typical particle diameter), $H^*$ the typical thickness of the electrodes (normal to the current collectors) while $W^*$ their typical width, $\lambda^*$ is the typical size of the Lam\'e constant in the active particles (those in the anode and cathode are taken to have comparable stiffnesses), $D^{*,k}$ a typical value of the diffusivity in the active particles of electrode $k$, and $D^{*,\mathrm{e}}$ a typical value of the diffusion coefficient in the electrolyte. 
Since the non-active matrix is much softer than the particles, we rescale the Lam\'e coefficients in this homogenised region via the typical size $\lambda^{*,\mathrm{na}}$, where $\lambda^{*,\mathrm{na}}=\Lambda \lambda^*$ and $\Lambda \ll 1$ characterises the typical ratio of stiffness between non-active region materials and particles. We identify the characteristic linear particle swelling with $g^*= \third\Vp\cpmax$, and the magnitude of potentials with $\phi^* =\frac{RT}{F}$. The size of parameters in typical batteries is shown in table {\ref{tab:params_lmo}--\ref{tab:params_graphite}}, while the natural timescales arising in the problem are listed in table \ref{tab:timescales}.

\begin{table}[ht]
\centering
\begin{tabular}{llll}
\hline
\textbf{Parameter} & \textbf{LMO} & \textbf{NMC} & \textbf{Graphite} \\ \hline
\multicolumn{4}{l}{Dimensionless parameters} \\ \hline
$\bar \gk = \dfrac{\gk}{g^*}$ & 1 & 1 & 1 \\[8pt]
$\VVk = \dfrac{\Vk\, c^{k,\mathrm{max}}}{\Vp\, \cpmax}$ & 1 & 1 & 1 \\[10pt]
$\Ckref = \dfrac{\ckref}{c^{k,\mathrm{max}}}$ & 0.17 & 0.2661 & 0.0279 \\[10pt]
$\bar{\Gk} = \dfrac{\Gk}{\lambda^{*}}$ & 0.6667 & 0.9231 & 0.6667 \\[8pt]
$\bar{\lk} = \dfrac{\lk}{\lambda^{*}}$ & 1 & 1 & 1 \\[8pt]
$\bar D^{\mathrm{e}} = \dfrac{D^{\mathrm{e}}}{D^{*,\mathrm{e}}}$ & 1 & 1 & 1 \\[8pt]
$\bar \Kk = \dfrac{\Kk}{N^*\,\chk}$ & $5.53\times10^{4}$ & $2.17\times10^{3}$ & $2.60\times10^{6}$ \\[8pt]
$\mathcal{K}^{\|} = \dfrac{t^{*}}{t^{*,c,k}}$ & $6.50\times10^{4}$ & $1.19\times10^{6}$ & $3.91\times10^{2}$ \\[8pt]
$\mathcal K^{\mathrm{e}} = \dfrac{t^{*}}{t^{*,c,\mathrm{e}}}$ & $1.08\times10^{4}$ & $3.40\times10^{3}$ & $2.67\times10^{3}$ \\[8pt]
$\mathcal D^{k} = \dfrac{t^{*}}{t^{*,k}}$ & 1.6185 & 7.1762 & 1.1119 \\[8pt]
$\mathcal D^{\mathrm{e}} = \dfrac{t^{*}(H^{*})^{2}}{t^{*,\mathrm{e}}(L^{*})^{2}}$ & $1.21\times10^{3}$ & $1.27\times10^{4}$ & $5.96\times10^{3}$ \\[8pt]
$\gamma = \dfrac{(\Vp)^{2}\,\cpmax\,\lambda^{*}}{9RT}$ & 0.6723 & 0.1389 & 0.1316 \\[8pt]
\hline
\end{tabular}
\caption{{Dimensionless parameters for the LMO, NMC811 and graphite electrodes, computed from the dimensional values in the appendix (Tables~\ref{tab:params_lmo}, \ref{tab:params_NMC} and \ref{tab:params_graphite}) and the PVDF binder properties given in Appendix~\ref{sec:params}.} Each material uses its own reference scales, so $\bar g^k$, $\VVk$, $\bar\lambda^k$ and $\bar D^{\mathrm{e}}$ equal to 1 by construction.}
\label{tab:dimensionless_all}
\end{table}
\renewcommand{\arraystretch}{1.4} 
\begin{table}[ht]
\centering
\begin{tabular}{llll}
\hline
\textbf{Timescale} & \textbf{Symbol} & \textbf{Expression} & \textbf{Value [s]} \\ \hline
Discharge timescale & $t^{*}$ &
$\displaystyle \frac{F c^{\mathrm{p,max}} m H^{*}  (W^*)^2}{I^*}$ &
$10^{4}/C$ \\[6pt]
Diffusion in the particle & $t^{*,k}$ &
$\displaystyle \frac{(L^*)^2}{\Dk}$ &
$10^{3}$ \\[6pt]
Diffusion in the electrolyte & $t^{*,\mathrm{e}}$ &
$\displaystyle \frac{(H^{*})^{2}}{D^{*,\mathrm{e}}}$ &
$10$ \\[6pt]
Conduction in the electrolyte & $t^{*,c,\mathrm{e}}$ &
$\displaystyle \frac{\kappa^{\mathrm{e}} R T }{(FL^*)^2 \cezero}$ &
$10$
\end{tabular}
\caption{Characteristic time scales for discharge and diffusive processes, with $k \in \{ \mathrm{p,n} \}$. }
\label{tab:timescales}
\end{table}
\newpage

To summarise, the rescalings we use are
\begin{align}
    X_i &= L^* \bar{X}_i, \quad I = I^* \bar{I}, \quad t = t^*\,\bar{t}, \quad  \ce = \cezero \bar{c}^{\mathrm{e}}, \quad \phi^{\mathrm{e}} = \phi^*\,\bar{\phi}^{\mathrm{e}}, \quad V = \phi^*\,\bar{V}-\frac{\mu^{0}}{F}.
  \end{align}
In the positive and negative electrode, $k \in \{\mathrm{p,n}\}$ we have:
\begin{align}
    \ck &= \ckmax \bar{c}^{k},\quad \quad \Nk_i = N^* \chk \bar{N}^{k}_i,\quad \Nke_i = N^* \che \bar{N}^{\mathrm e,k}_i, \quad
    \ik_i = i^* \bar{i}^{k}_i,\quad\quad \ike_i = i^* \che \bar{i}^{\mathrm{e},k}_i, \quad  \\[6pt]\etak  &= \phi^*\,\bar{\eta}^{k}, \quad \phik = \phi^*\,\bar{\phi}^{k}, \quad \muk = \mu^{0} + RT\,\bar{\mu}^{k},\quad \Ueqk = \phi^* \bar{\mathrm{U}}^{\mathrm{eq},k},
     \quad 
    \Ubek_i = g^* L^*\,\bar{U}^{\mathrm{na},k}_i, \\[6pt] \Uk_i &= g^* L^*\,\bar{U}^{k}_i, \quad 
    \ebek_{ij} = g^*\,\bar{\epsilon}^{\mathrm{na},k}_{ij}, \quad 
    \epsilon^k_{ij} = g^*\,\bar{\epsilon}^k_{ij}, \quad
    \sbek_{ij} = \Lambda \, g^*\,\lambda^*\,\bar{\sigma}^{\mathrm{na},k}_{ij}, \quad
    \sk_{ij}= g^*\,\lambda^*\,\bar{\sigma}^k_{ij}.
\end{align}
where
\[
 i^*=\frac{I^* L^*}{m (W^*)^2 H^*}, \qquad N^* =  \frac{I^* L^*}{m F (W^*)^2 H^*}, \qquad \chk=\frac{\ckmax}{\cpmax},\qquad 
\che=\frac{\cezero}{\cpmax}.
\]
These factors account for the fact that the natural concentration scale in an active particle in electrode \(k\) is \(\ckmax\), whereas the natural concentration scale in the electrolyte is \(c^{\mathrm{e},0}\). Since the characteristic diffusive flux scales as concentration times \(L^*/t^*\), the solid and electrolyte fluxes differ in size when measured relative to the common reference flux \(N^*=L^*\cpmax/t^*\). This distinction is important at particle/electrolyte interfaces, where the same dimensional reaction flux appears in both the solid and electrolyte boundary conditions but is nondimensionalised using different concentration scales.

The scaling in the separator is:
\begin{align}
    \Nse_i &= N^* \bar{N}^{\mathrm{e},\mathrm{s}}_i, \quad \ise_i = i^* \bar{i}^{\mathrm{e},\mathrm{s}}_i, \quad 
    U^{\mathrm{se}}_i = g^* L^*\,\bar{U}^{\mathrm{se}}_i, \quad 
    \epsilon^{\mathrm{se}}_{ij} = g^*\,\bar{\epsilon}^{\mathrm{se}}_{ij}, \quad 
    \sigma^{\mathrm{se}}_{ij} = g^*\,\lambda^*\,\bar{\sigma}^{\mathrm{se}}_{ij}.
\end{align}
Scaling in this manner introduces the following dimensionless parameters:
\begin{align}
\label{dlessparam1}
\bar \gk &= \frac{\gk}{g^*},
&\VVk &= \frac{\Vk\,\ckmax}{\Vp\,\cpmax},
&\Ckref &= \frac{\ckref}{\ckmax},
&\bar{\Gk} &= \frac{\Gk}{\lambda^*},\\[4pt]
\bar{G}^{\mathrm{na},k} &= \frac{\Gbek}{\lambda^{*,\mathrm{na}}},
&\bar{\lk} &= \frac{\lk}{\lambda^*},
&\bar{\lambda}^{\mathrm{na},k} &= \frac{\lbek}{\lambda^{*,\mathrm{na}}},
&\bar \Kk &= \frac{\Kk}{N^*\,\chk} ,\\[6pt]
\bar D^{\mathrm{e}} &= \frac{D^{\mathrm{e}}}{D^{*,\mathrm{e}}},
&\bar G^{\mathrm{se}} &= \frac{G^{\mathrm{se}}}{\lambda^*},
&\bar \lambda^{\mathrm{se}} &= \frac{\lambda^{\mathrm{se}}}{\lambda^*},
&\mathcal \Kk &= \frac{t^*}{t^{*,c,k}},\\[4pt]
\mathcal K^{\mathrm{e}} &= \frac{t^*}{t^{*,c,\mathrm{e}}},
&\mathcal D^{k} &= \frac{t^*}{t^{*,k}},
&\mathcal D^{\mathrm{e}} &= \frac{t^*(H^*)^2}{t^{*,\mathrm{e}}(L^*)^2 },
&\gamma &= \frac{(\Vp)^2\,\cpmax\,\lambda^*}{9RT}.
\label{dlessparam2}
\end{align}
The only parameter whose meaning is not self-evident from its definition is $\gamma$, which is the ratio of the energy density due to intercalation stress to that of the entropy of mixing. For the purposes of simplifying the notation we shall drop the overbars from hereon and work only with dimensionless quantities.

\subsubsection{Dimensionless mechanics equations}
The dimensionless mechanical equations read
\begin{align}
\label{eq:mech_bek}
    \ebek_{ij} &=\frac{1}{2}\left( \frac{\dd \Ubek_j}{\dd X_i} + \frac{\dd \Ubek_i}{\dd X_j} \right),& \qquad \sbek_{ij}&= 2 \Gbek \ebek_{ij} +  \delta_{ij} \lbek \ebek_{ll} &\quad &\textrm{in} \quad \Obek,\\ 
\label{eq:srtress-strain}
    \ek_{ij} &= \frac{1}{2} \left( \frac{\dd \Uk_i}{\dd X_j} + \frac{\dd \Uk_j}{\dd X_i} \right), &\qquad \sk_{ij} &= 2   \Gk \left( \ek_{ij} - \delta_{ij} \gk \right) + \delta_{ij} \lk \left( \ek_{ll} -  \delta_{ll} \gk \right) &\quad& \textrm{in} \quad \Ok,
\end{align}
the degree of swelling being given by
\be
    \gk = \linVVk\,(\ck-\Ckref) \quad \textrm{in} \quad \Ok,
\ee
and equilibrium requiring
\be
\label{eq:mech_k}
    \frac{\dd \sbek_{ij}}{\dd X_j}=0\quad \textrm{in} \quad \Obek,\qquad
    \frac{\dd \sk_{ij}}{\dd X_j} = 0 \quad \textrm{in} \quad \Ok.
\ee
The boundary conditions read
\be 
\label{eq:mech_bc}
    \sk_{ij} n_j = \Lambda \sbek_{ij} n_j, \qquad \Uk_i = \Ubek_i \quad \text{on} \quad \dOk.
\ee
Note that the stiffness ratio $\Lambda$ appears in the boundary condition, with the particle feeling only a weak push from the non-active matrix of order $\Lambda \ll 1$.

\subsubsection{Dimensionless transport in the particles}
In the solid active particles we have
\begin{equation} 
\label{eq:diff_k}
    \frac{\partial \ck}{\partial t}+\frac{\partial \Nk_i}{\partial X_i}=0,\qquad \Nk_i=-      \DD^{k} \ck \left(1-\ck \right) \frac{\partial \muk}{\partial X_i} \quad \textrm{in} \quad \Ok,
\end{equation}
where
\begin{equation}
\label{eq:mu}
    \muk = - \Ueqk (\ck) - \gamma \linVVk\,\sk_{ll}.  
  \end{equation}
The Butler-Volmer deintercalation rate is
\begin{equation} 
\label{eq:diff_bc}
    \Nk_i n_i= \Jk \sinh\left(\frac{\etak}{2} \right) \quad \textrm{on} \quad \dOk,
\end{equation}
where
\begin{equation}
    \Jk=\Kk
    \sqrt{\ce\ck \left(1 - \ck \right)},
\end{equation}

and the overpotential $\etak$ is given by
\begin{equation} \label{eq:eta}
    \etak = \phik - \phie - \Ueqk(\ck) - \gamma \linVVk\,\sk_{ll}  \quad \textrm{on} \quad \dOk.
\end{equation}
Note that, since we chose to use the degree of expansion in the positive electrode as our scale for swelling, we have ${\cal V}_c^{p} = 1$.

\subsubsection{Dimensionless equations in the non-active regions}

In the non-active matrix the equations become
\begin{equation} 
\label{eq:e_diff}
    \frac{\partial (\alk \ce )}{\partial t}+ \frac{\partial \Nke_i}{\partial X_i}=0, \qquad \Nke_i=-  \Bk \DD^{\mathrm{e}} D^{\mathrm{e}} \frac{\partial \ce}{\partial X_i} +{\tplus}{  \ike_i}  \quad \textrm{in} \quad \Obek,
\end{equation}
\begin{equation} 
\label{eq:e_current}
   \frac{\partial \ike_i}{\partial X_i} = 0, \qquad
    \ike_i= - \K^{\mathrm{e}} \Bk \left( \frac{\partial \phie}{\partial X_i} -2(1-\tplus)\frac{1}{\ce}\frac{\partial \ce}{\partial X_i}\right)  \quad \textrm{in} \quad \Obek,
\end{equation}
where $D^{\mathrm{e}}$ depends on $\ce$.
The boundary conditions are

\begin{equation} 
\label{eq:e_diff_bc}
    \Nke_i n_i= - \frac{\Jk\chk}{\che}  \sinh\left(\frac{\etak}{2} \right) \quad \textrm{on} \quad \dOk,
\end{equation}
\begin{equation} 
\label{eq:e_current_bc}
    \ike_i n_i= -   \frac{\Jk\chk}{\che} \sinh\left(\frac{\etak}{2}\right) \quad \textrm{on} \quad \dOk,
\end{equation}
where $\etak$ is defined in \eqref{eq:eta}. For the electron conduction through the homogenised non-active medium we have 
\begin{equation}
    \frac{\partial \ik_i}{\partial X_i} = 0, \qquad \ik_i = - \mathcal{K}^{k} \Bkhat \frac{\partial \phik}{\partial X_i} \quad \text{in} \quad  \Obek,
\end{equation}
supplemented by the boundary condition
\begin{equation}
    \ik_i n_i= -\Jk \chk \sinh\left(\frac{\etak}{2} \right) \quad \textrm{on} \quad \dOk.
\end{equation}

\subsubsection{Dimensionless equations in the separators}

The ionic conduction satisfies
\begin{equation} 
\label{eq:s_diff_dim_sep}
    \frac{\partial (\alpha^{\mathrm{s}} \ce)}{\partial t}+ \frac{\partial \Nse_i}{\partial X_i}=0, \qquad \Nse_i=- \Bs \DD^{\mathrm{e}}\frac{\partial \ce}{\partial X_i} +\tplus\ise_i  \quad \textrm{in} \quad \Ose,
\end{equation}
\begin{equation} 
\label{eq:s_current_dim_sep}
    \frac{\partial \ise_i}{\partial X_i} = 0, \qquad
    \ise_i= -\K^{\mathrm{e}} \Bs \left( \frac{\partial \phie}{\partial X_i} -2(1-\tplus)\frac{1}{\ce}\frac{\partial \ce}{\partial X_i}\right)  \quad \textrm{in} \quad \Ose.
\end{equation}
The mechanical equations in the separator are
\be 
\label{eq:mech_s_dim_sep}
    \frac{\dd \sigma^{\mathrm{se}}_{ij}}{\dd X_j}=0, \qquad \epsilon^{\mathrm{se}} =\frac{1}{2}\left( \frac{\dd U^{\mathrm{se}}_j}{\dd X_i} + \frac{\dd U^{\mathrm{se}}_i}{\dd X_j} \right), \qquad \sigma^{\mathrm{se}} = 2 G^{\mathrm{se}} \epsilon^{\mathrm{se}}_{ij} + \delta_{ij} \lambda^{\mathrm{se}} \epsilon_{ll}^{\mathrm{se}} \quad \textrm{in} \quad \Ose.
\ee

\subsubsection{Dimensionless boundary and interface conditions}

At the interface between the electrode and the current collectors we have
\begin{equation}
    \Nke_i n^{k}_i = \ike_i n^{k}_i = 0.
\end{equation}
At the interface between the electrodes and separator,
\begin{equation}
    \left[ \Nke_i n_i \right] = \left[ \ike_i n_i \right] = 0,
\end{equation}
\begin{equation}
    \ik_i n_i = 0.
\end{equation}
Finally, we consider specified current operation and therefore impose the constraint
\begin{equation}
\label{eq:totI}
    \iint_{{\dOcck}} \ik_i n_i \, \d S = I/(m H^k W_2 W_3),
\end{equation}
where $W_2$ and $W_3$ and the in-plane width and length of the electrode. The cell potential given by
\be
\label{eq:V}
V = \Phi^{\mathrm{p}} - \Phi^{\mathrm{n}} = \Phi^{\mathrm{p}}.
\ee
Once again, we defer writing the mechanical boundary conditions on the whole cell to Section~\ref{sec:5}. Problem closure is achieved by specifying suitable initial conditions on the concentrations.

\subsection{Discussion of asymptotic limits}

We are interested in understanding the solutions to the model in
scenarios pertaining to modern lithium-ion batteries. For example, we note that polymer binders that give the non-active matrix its stiffness are significantly more pliable than active materials and thus $\Lambda \ll 1$. Therefore, at leading order in $\Lambda$, the particle can expand/shrink without resistance from the surrounding non-active matrix. This is a key simplification that we exploit in the next section. Conversely, the values of the rescaled (dimensionless) shear moduli $\Gk$ and $\lk$ are neither large nor small and hence will be taken to be $O(1)$. The value of $\gamma$ depends upon the electrode material and can be small, as for example in graphite and nickel-manganese-cobalt oxide (NMC), or moderately large, for example in silicon (Si). However, at most, we expect that the stresses may change the overpotential by several thermal voltages. 
If $\gamma$ is small, stress effects in the particle become negligible at leading order (in $\gamma$), and the transport problem collapses to that of simple Fickian diffusion (provided that the chemical potential is monotone in the concentration), decoupling from the mechanics. Therefore, to capture coupling between stresses and electrochemistry in a manner that is consistent with modern cell design, we look at the distinguished limit in which $\Lambda \to 0$, $\gamma=O(1)$. Furthermore, we assume that $\DD^{k}$ is $O(1)$, thereby retaining intraparticle transport limitations. Finally, we remind the reader that the timescale of conduction and diffusion in the electrolyte are usually small compared to the timescale of discharge. This is equivalent to having $\mathcal K^{\mathrm{e}}$ and  $\DD^{\mathrm{e}}$ large and can be used to simplify the DFN model to a single particle model (SPM) and single particle model with electrolyte (SPMe), see \cite{guo,mou,prada,spme}.

\subsection{Discussion of the model objectives}

The model in Section~\ref{sec:2} extends the standard Doyle-Fuller-Newman (DFN) framework by coupling electrochemistry with mechanics across particle (micro), electrode (meso), and cell (macro) scales. This coupling breaks the usual pseudo-two-dimensional (P2D) separation exploited by DFN, where electrolyte and electronic transport are solved on electrode-scale domains while intercalation is treated as a spherically symmetric, one-dimensional diffusion problem within an isolated particle. With mechanics included, this local P2D reduction of the particle problem is no longer valid in its usual form, for three main reasons:
\begin{enumerate}
\item[(i)] \emph{Bidirectional mechano-electrochemical coupling.} Lithiation drives particle swelling, while the resulting stress field modifies the lithium chemical potential and interfacial overpotential. The particle concentration, stress, and electrochemical reaction kinetics are therefore coupled within a single local problem, rather than through diffusion alone. 
\item[(ii)] \emph{Mechanical nonlocality.} The stress state of a particle depends on the deformation transmitted through the non-active matrix and current collectors, and therefore on the response of neighbouring particles and the entire cell stack. Each particle’s surface stress is thus determined by the global mechanical field, rather than by local electrochemical conditions alone.
\item[(iii)] \emph{Loss of spherical symmetry.} Even in simple periodic electrode architectures, the stresses acting on a particle are generally anisotropic. As a result, the particle problem is not radially symmetric, and higher-order deformation modes arise even under uniform lithiation.
\end{enumerate}
Without neglecting the complications (i)--(iii), our objective is to systematically reduce the model to a level of complexity comparable to the original DFN. In doing so, we will prevent the relevant mechanical effects from becoming the computational bottleneck in a battery simulation.

First, we exploit the fact that the non-active matrix is much softer than the active material and expand the solution in the small stiffness ratio by taking $\Lambda \to 0$. This asymptotic approach simplifies the two-way coupling discussed in point~(ii). In particular, we find that at leading order the particle expands freely, and the effect of the non-active matrix appears as a first-order correction, albeit one that has a measurable impact on cell voltages in many relevant scenarios \citep{Cann,Rieger}. This is the focus of Section~\ref{sec:3}. 

Second, to find the non-active matrix stress and determine how it
affects the first-order corrections for the particle states we examine
the intermediate mesoscopic length scale associated with the thickness
of the electrode. Here, we use multiple-scale homogenisation to link
the microscale mechanics of individual particles to the
electrode-scale mechanics. This is the mechanical counterpart of the
argument that can be used to formulate the familiar DFN model, in
which  the electrolyte dynamics is  homogenised along the electrode thickness. The details are presented in Sections~\ref{sec:4}. 

Owing to the extreme aspect ratio of modern electrodes (they are much
slimmer in the dimension perpendicular to the current collectors than
in the direction in-plane with the current collectors) we then show
that the macroscopic electrode deformation can be determined only
using knowledge of the degree of swelling observed on the macroscopic
scale. In particular, in Section~\ref{sec:5}, we show that the
electrodes are strained only in the through-cell direction, since
current collectors prevent in-plane deformations, with the strain
depending on the mechanical properties of each layer in the stack and
the boundary conditions applied on the whole cell, e.g. clamped
boundaries or compressive loading. Section~\ref{sec:3}, \ref{sec:4} and \ref{sec:5} thus allow us to address point~(ii) and enable a clean separation between particle-scale mechanics and the macroscopic mechanical field. 

To address point (iii), in Section~\ref{sec:6} we demonstrate that, although the stress field within a particle is not strictly spherically symmetric due to spatial variations in the local non-active matrix stress, only its radial component influences the first-order correction to the electrochemical problem. Thus, by averaging the coupled mechanical and transport equations over the particle surface, we reduce the description to an effective radial problem, which quantifies how non-active matrix-induced stress modifies the exchange current and thereby impacts electrolyte transport.

In Section~\ref{sec:7}, we collect the most pertinent insights from the preceding sections and show that our
model can be written in a form similar to the DFN but with the macroscale mechanical effects entering as a correction to the chemical potential in the active particle. The resulting problem is a DFN counterpart which includes mechanical effects across the length scales thereby achieving our aim of systematically including multiscale mechanical effects without significantly increasing computational costs.

\section{Soft non-active matrix expansion}
\label{sec:3}

We now consider the solution to the model presented in Section~\ref{sec:2} in the distinguished limit $\Lambda \ll 1$, with all other dimensionless parameters listed in \eqref{dlessparam1}-\eqref{dlessparam2} taken to be $O(1)$. Our aim is to simplify our two-way coupling between electrochemistry and mechanics. To do so, we expand all our variables in powers of $\Lambda$, so that a generic field $\Phi$ becomes
\be
\label{eq:expansion}
    \Phi_i=\Phi_i^{(0)}+\Lambda \Phi_i^{(1)}+ \Lambda^2 \Phi_i^{(2)} +  \cdots.
\ee

\subsection{Leading-order problem: unconstrained particle expansion}

Substituting the expansion \eqref{eq:expansion} in \eqref{eq:mech_bek}-\eqref{eq:totI} allows us to obtain the leading-order problem. We avoid writing the problem again, noting that the equations are the same, except all fields (i.e. strains, displacements, stresses, concentrations, fluxes, currents and potentials) pick up a superscript $(0)$, referring to the leading-order term. The only difference lies in \eqref{eq:mech_bc}, the mechanical boundary conditions at the interface of the particle and non-active matrix, where the stiffness difference asserts that 
\begin{align}
\label{eq:mech_bc_0_surface}
    \skz_{ij} n_j= 0,\qquad \Ukz_i=\Ubekz_i. 
\end{align}
This form of the boundary condition means that a significant simplification occurs: the surface of the particle is, to a good approximation, stress-free, so that the particle and non-active matrix become decoupled at leading order. While, before, the non-active matrix stress was forcing us to consider non-radial deformations, the stress-free condition means we recover a radial problem for concentrations and displacements (just like in DFN). 

In what follows, we write down the leading-order equations exploiting the radial symmetry. To keep our notation uncluttered, we drop explicit reference to electrode index $k$ and will focus our attention on the equations that pertain to the positive electrode. The solution in the negative electrode is identical in structure, but with the elastic moduli, partial molar volume, reference equilibrium potential and concentration corresponding to the negative electrode. We will discuss these later, in Section~\ref{sec:6}.

We consider the radial coordinate $r = (X_1^2+X_2^2+X_3^2)^{1/2}$ for a particle centred at the origin with radius $r=R$, and let $\Urz$ be the radial displacement (the only non-zero leading-order component), so in the particle our stress/strain equations \eqref{eq:mech_bek}-\eqref{eq:mech_bc}, become
\begin{align}
\notag
    \ekz_{rr} &= \frac{\dd \Urz}{\dd r}, \qquad
    \ekz_{\theta\theta} = \ekz_{\phi\phi} = \frac{\Urz}{r},\\
\notag
    \skz_{rr} &= (2G+\lambda)\,\frac{\dd \Urz}{\dd r} + 2\lambda\,\frac{\Urz}{r}
           - (2G+3\lambda)\,g,\\
\label{eq:mech_0}
    \skz_{\theta\theta} &= \lambda \,\frac{\dd \Urz}{\dd r} + 2(G+\lambda)\,\frac{\Urz}{r} - (2G+3\lambda)\,g, \qquad \skz_{\phi\phi} = \skz_{\theta\theta}.
\end{align}
where 
\be
\label{eq:mech_g}
g=\third \V \big(\ckz-\Cref \big).
\ee
Mechanical equilibrium (no body forces) reduces to
\begin{equation}
\label{eq:mech_eq_0}
    \frac{\dd \skz_{rr}}{\dd r} + \frac{2}{r}\big(\skz_{rr}-\skz_{\theta\theta}\big) = 0,
\end{equation}
with regularity at the centre and a stress-free particle surface at leading order implying
\begin{equation}
\label{eq:mech_bc_0}
    \Urz(0)=0, \qquad \skz_{rr}(R)=0.
\end{equation}
In the solid active particle the lithium transport equations become
\begin{equation} 
\label{eq:diff_k_0}
    \frac{\partial \ckz}{\partial t}+\frac{1}{r^2} \p{ (r^2 \Nkz)}{r} =0,\qquad
    \Nkz=-\DD^{k} \ckz \left(1-\ckz \right) \p{\mukz}{r} \quad \textrm{in} \quad \Om.
\end{equation}
with 
\begin{equation}
\label{eq:mu_0}
    \mukz = - \Ueqk (\ckz) - \third\gamma \V \skz_{ll}.  
\end{equation}

The equations in the active particles, namely \eqref{eq:mech_0}--\eqref{eq:mu_0}, show a clear two-way coupling between local mechanical stresses and electrochemistry. This model is analogous to the one studied numerically by \cite{zhang2007numerical}. It is also the basis for similar models concerned with mechanical effects on electrochemistry and fracture, where surface stresses are ignored and internal stresses are generated by concentration gradients during charging and discharging \cite{Wu1,Wu2}.

\subsubsection{Leading-order equations in the non-active matrix}

We now write down the leading-order transport equations in the non-active matrix and crucially, the mechanical equilibrium condition. The former is important to determine the flux on the leading-order particle problem, the latter determines the surface stress on the particle at the next order, as we shall see. 

For the electron conduction through the homogenised non-active mixture we have 
\begin{equation}
    \frac{\partial \ikz_i}{\partial X_i} = 0, \qquad \ikz_i = - \K \hat{\cal B} \frac{\partial \phi^{(0)}}{\partial X_i} \quad \text{in} \quad  \Obe,
  \end{equation}
supplemented by the boundary conditions
\begin{equation}
    \ikz \cdot \vec{n}= -J^{(0)} \chi \sinh\left(\frac{\etakz}{2} \right) \quad \textrm{on} \quad \dO.
\end{equation}
\begin{equation} 
\label{eq:diff_bc_0}
   \Nkz \cdot \hat{\vec{r}}= J^{(0)} \sinh\left(\frac{\etakz}{2} \right) \quad \textrm{on} \quad \dO,
\end{equation}
where
\begin{equation}
\label{eq:J0}
    J^{(0)} =K^{(0)}
    \sqrt{\cez \ckz \left(1-
    \ckz \right)},
\end{equation}
and the overpotential $\etakz$ is given by
\begin{equation} 
\label{eq:eta_0}
    \etakz = \phi^{(0)} - \phiez - \mathrm{U}^{\mathrm{eq}}(\ckz) - \third \gamma \V \skz_{ll}  \quad \textrm{on} \quad \dO.
\end{equation}

The mechanical equations in $\Obe$ are
\be 
\label{eq:mech_be_0}
    \ebekz_{ij} =\frac{1}{2}\left( \frac{\dd \Ubekz_j}{\dd X_i} + \frac{\dd \Ubekz_i}{\dd X_j} \right), \qquad \sbekz_{ij}= 2 G^{\mathrm{na}} \ebekz_{ij} +  \delta_{ij} \lambda^{\mathrm{na}} \ebekz_{ll}, \quad \p{\sbekz_{ij}}{X_j}=0,
\ee
with boundary conditions given by
  \begin{equation}
  \Ubekz_i = \Ukz_i \qquad \mbox{ on } r = R.\label{eq:mech_be_bc_0}
  \end{equation}
Note that the solution to the particle problem in \eqref{eq:mech_0}-\eqref{eq:mu_0} fixes the boundary condition in \eqref{eq:mech_be_bc_0} via the displacement $\Ukz_i$, thus determining the leading-order deformation in the non-active region. 

The electrolyte transport equations are
\begin{equation} 
\label{eq:e_diff_0}
    \frac{\partial (\alpha \cez )}{\partial t}+ \frac{\partial \Nkez_i}{\partial X_i}=0, \qquad \Nkez_i=-  \B\DD^{\mathrm{e}} D^{\mathrm{e}} \frac{\partial \cez}{\partial X_i} +{\tplus}{  \ikez_i}  \quad \textrm{in} \quad \Obe,
\end{equation}
\begin{equation} 
\label{eq:e_current_0}
   \frac{\partial \ikez_i}{\partial X_i} = 0, \qquad
    \ikez_i= - \K^{\mathrm{e}} \B \left( \frac{\partial \phiez}{\partial X_i} -2(1-\tplus)\frac{1}{\cez}\frac{\partial \cez}{\partial X_i}\right)  \quad \textrm{in} \quad \Obe,
\end{equation}
where $D^{\mathrm{e}}$ depends on $\ce$. The boundary conditions are
\begin{equation} 
\label{eq:e_diff_bc_0}
    \Nkez \cdot \hat{\vec{r}}= - J^{(0)} (\chi/\che)  \sinh\left(\frac{\etakz}{2} \right) \quad \textrm{on} \quad \dO,
\end{equation}
\begin{equation} 
\label{eq:e_current_bc_0}
    \ikez \cdot \hat{\vec{r}}= -   J^{(0)}  (\chi/\che) \sinh\left(\frac{\etakz}{2}\right) \quad \textrm{on} \quad \dO.
\end{equation}
At the interface between the electrode and the current collectors we have
\begin{equation}
    \Nkez \cdot {\bf n} = \ikez \cdot {\bf n} = 0,
\end{equation}
with the current satisfying
\begin{equation}
\label{eq:totI0}
    \iint_{{\dOcck}} \ikz \cdot {\bf n} \, dS = I/(m H^k W_2 W_3).
\end{equation}
We also require continuity of ionic current through the separator interface and no electronic current. Since we assume specified current operating conditions, where $I$ is given, this leading order problem determines the leading order voltage $V^{(0)}$ of the cell.


The equations in the non-active matrix can, in principle, be solved numerically. However, it is much more common to homogenise the transport equations \eqref{eq:e_diff_0}--\eqref{eq:totI0} to obtain a Doyle--Fuller--Newman (DFN) model. We do not repeat that homogenisation here, since it has already been carried out by \cite{gilescolin2012}. We will write the full DFN equations, with the inclusion of mechanical corrections, in Section~\ref{sec:7}. Crucially, we note that at this order, the  particle problem is mechanically isolated (due to the free surface boundary condition, \eqref{eq:mech_bc_0_surface}), and the particle behaviour is independent of the non-active matrix around it. Thus, at leading order, we do not have a model that is capable of capturing the multiscale mechanical/electrochemical coupling that we seek.  
We show that the multiscale mechanical coupling is recovered at $O(\Lambda)$.

Before proceeding we observe that although the concentration and stress gradients within the active material must, in general, be determined numerically, we can obtain a closed form solution for the surface displacement and thus for the total expansion of the particle. Due to the linearity of \eqref{eq:mech_0}, the average strain in the particle is equal to the average swelling strain, and is given by:
\be
\label{eq:av-g}
\gav \equiv \frac{\Urz(R)}{R}=\frac{3}{R^3}\int_0^R  g \,s^2 \d s.
\ee
This result is particularly useful when solving the first-order problem in the non-active matrix.

\subsection{First-order correction: the non-active matrix pushes back on the particle}
\label{firstOrderParticle}
Proceeding to the next order, we now uncover how the non-active matrix affects the particle. The stress--strain relation in equation  \eqref{eq:srtress-strain} becomes
\be  
\label{eq:stress-strain_1}
    \ekf_{ij} = \frac{1}{2} \left( \frac{\dd \Ukf_i}{\dd X_j} + \frac{\dd \Ukf_j}{\dd X_i} \right), \qquad {\sigma_{ij}^{(1)} = 2   G \left( \ekf_{ij} - \delta_{ij} \third\V \ckf \right) + \delta_{ij} \lambda \left( \ekf_{ll} -  \delta_{ll}\third\V \ckf \right)} \quad \textrm{in} \quad \Om.
\ee
Equilibrium in \eqref{eq:mech_k} requires
\be
\label{eq:mech_k_1}
    \frac{\dd \sbekf_{ij}}{\dd X_j}=0\quad \textrm{in} \quad \Obek,\qquad \frac{\dd \sigma_{ij}^{(1)} }{\dd X_j} = 0 \quad \textrm{in} \quad \Om,
\ee
while the boundary condition is 
\be 
\label{eq:mech_bc_1}
\sigma_{ij}^{(1)} n_j = \sbekz_{ij} n_j,
\quad \text{on} \quad \dO.
\ee

The first-order mechanical equilibrium equation \eqref{eq:mech_k_1} shows that the leading-order non-active matrix solution to equations \eqref{eq:mech_be_0} and
\eqref{eq:mech_be_bc_0} determines the first-order correction, \eqref{eq:mech_bc_1}, to the
stress on the particle surface. This surface stress will then alter the stress within the particle, affecting the electrochemistry. The coupling between particle
and non-active matrix mechanics therefore enters through this
boundary term. Determining the resulting stresses and strains in the
electrode is non-trivial: the non-active matrix stress depends on the
collective deformation and spatial arrangement of neighbouring
particles. In order to account for these interactions while making a model that is sufficiently simple, we rely on the fact that (i) electrodes are composed of many particles and (ii) that the typical size of a particle is small compared to the thickness of the electrode. These observations allow us to carry out a multiple-scale homogenisation that links the electrode-scale stress to the local non-active matrix stress surrounding the particle, effectively allowing us to bridge the gap between particle mechanics and electrode mechanics. Once this is done, in Section~\ref{sec:6}, we return to the problem in \eqref{eq:mech_k_1}--\eqref{eq:mech_bc_1} and show that the non-active matrix push on the particle affects the intercalation process.

\section{Leading-order mechanical problem in the non-active matrix}
\label{sec:4}

\begin{figure}
\centering
\includegraphics[width=0.8\textwidth]{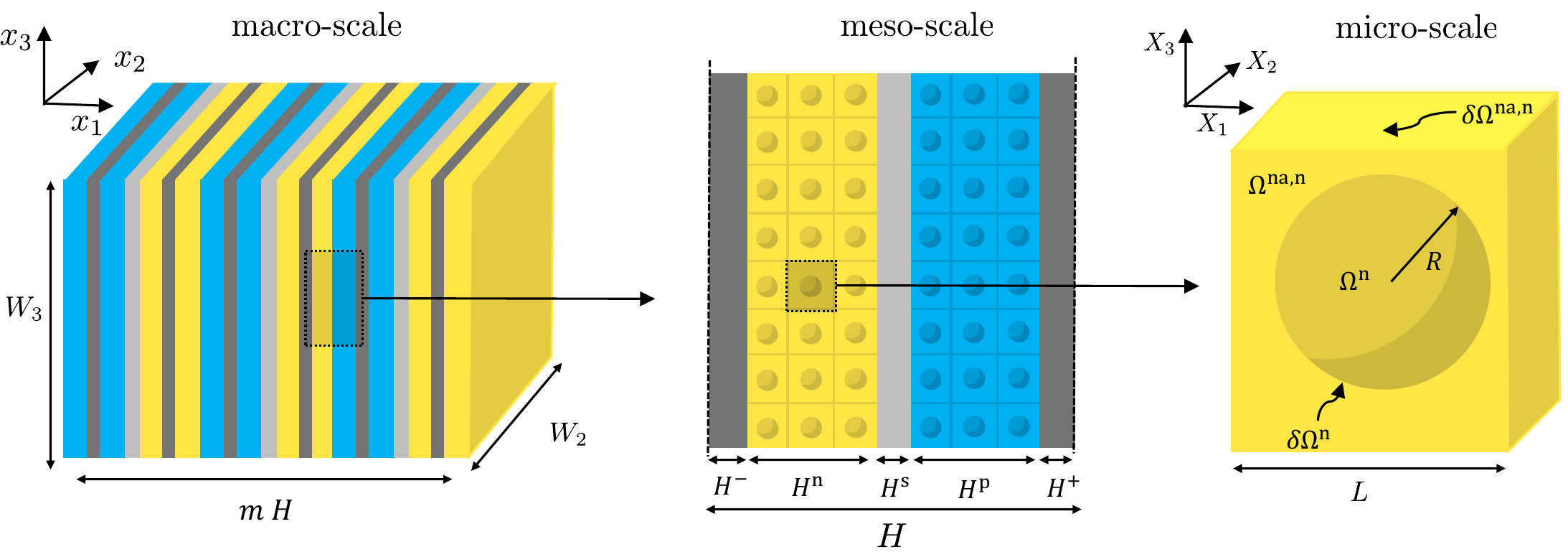}
\caption{Sketch of the multiscale geometry used in the mechanical reduction. The left panel shows the full pouch-cell stack at the macroscale, with through-cell coordinate $x_1$, in-plane coordinates $x_2$ and $x_3$. The dimensions are $W_2$, $W_3$, and total stack thickness $mH$, where $m$ is the number of electrode stacks. The middle panel shows the mesoscale representative stack of thickness $H$, consisting of negative electrode, separator, positive electrode, and current-collector layers with thicknesses $H^{\mathrm{n}}$, $H^{\mathrm{s}}$, $H^{\mathrm{p}}$, $H^{-}$ and $H^{+}$. The right panel shows the microscale unit cell of side length $L$, containing an active particle of radius $R$ in the domain $\Omega^{\mathrm{n}}$ embedded in the non-active matrix $\Omega^{\mathrm{be,n}}$, with particle boundary $\partial\Omega^{\mathrm{n}}$.} 
\label{fig:cell}
\end{figure}

\subsection{Multiple scales expansion}

We now introduce a multiple scale framework to link the particle-scale mechanics to the electrode-scale deformations. The hierarchy of scales used in this reduction is sketched in Figure~\ref{fig:cell}: the right panel shows the microscale particle cell, while the middle panel shows how such cells are repeated across an electrode-scale layer. We focus our attention on homogenisation of the mechanical equations because treatment of the electrochemical relationships has been presented elsewhere, see \cite{gilescolin2012}. Since we have made the modelling choice to restrict ourselves to the small strain setting, the electrochemical homogenisation is largely unchanged from that presented in \cite{gilescolin2012}. The only alteration is via the stress coupling in the Butler-Volmer reaction rate. However, if one were to consider the more challenging nonlinear strain setting then the homogenisation procedure would be substantially different. Our objective is to determine the effective stress in the non-active matrix, which governs the traction on the particle surface and contributes to the electrode-scale force balance.

The description of the effective behaviour of an electrode (porous, wet, viscoelastic with rigid particles) was given in \cite{Foster2025}, building on \cite{Cha15}. The homogenisation we present here is a simplified version of these previous calculations, and makes use of our assumption that cell-cycling timescales are sufficiently long that the changes in stresses associated with (i) the flow of the electrolyte and (ii) viscoelastic relaxation can be neglected.

We idealise the electrode as a periodic composite made of a lattice of geometrically identical representative elements of characteristic size $L^*$, repeated across a mesoscopic electrode thickness $H^*$, as indicated by the middle and right panels of Figure~\ref{fig:cell}. This separation of length scales is quantified by the small parameter
\be
\Delta = \frac{L^*}{H^*} \ll 1,
\ee
and the homogenised limit corresponds to $\Delta \to 0$.

We introduce a mesoscale coordinate $x_i$ in addition to the microscale coordinate $X_i$, related by
\be
\label{eq:two_scales_relation}
X_i = \frac{x_i}{\Delta}.
\ee
We make the ubiquitous assumption of the method of multiple scales and treat $x_i$ and $X_i$ as independent variables, so that spatial derivatives split as
\be
\label{eq:grad_split}
\p{}{X_j} \mapsto \p{}{X_j} + \Delta\p{}{x_j}.
\ee
Note that, while the microscale coordinates and displacements $X_i$ and $U_i$ are scaled by the typical unit-cell dimension $L^*$, the mesoscale coordinates and displacements $x_i$ and $u_i$ are scaled by the typical electrode thickness $H^*$, giving rise to separation of scales.

The $O(\Lambda^0)$ solution, $\Urz$, obtained by solving \eqref{eq:mech_0}--\eqref{eq:eta_0}, describes a single sphere that expands/shrinks radially, up to rigid-body translation and rotation. These motions differ from one unit cell to another in our periodic array and vary slowly across the electrode, reflecting the mesoscale deformation of the material. Thus, the leading order displacement in a particle can be written as
\be
\label{radialUsurf}
\Ukz_i = X_i\,\frac{\Urz}{r} + u_i + \omega_l \,\epsilon_{ijl}\,X_j,
\ee
where $X_i/r$ is the unit radial vector, $\epsilon_{ijl}$ is the Levi--Civita tensor, $u_i$ is the mesoscale rigid-body displacement, and $\omega_l$ is the rotation vector. The quantities $\Urz$, $u_i$ and $\omega_l$ are constant for a given particle but vary on the mesoscale. They are determined by enforcing an overall force and torque balance,
\be
\label{fandtalk}
\int_{\dO} \sbekz_{ij}\,n_j\,\d S = 0,
\qquad
\int_{\dO} \epsilon_{ilm}\,X_m\,\sbekz_{ij}\,\nu_j\,\d S = 0,
\ee
which depend on the non-active matrix stress field.

The solution in the particle, \eqref{radialUsurf}, provides the boundary condition on the non-active matrix displacement \eqref{eq:mech_be_bc_0}. Hence the boundary condition becomes
\be
\label{eq:be_bc}
\Ubekz_i = X_i \gav(t) + u_i + \omega_l \,\epsilon_{ijl}\,X_j,
\ee
where $\gav$ is defined in \eqref{eq:av-g}. To solve the mechanical problem in the non-active matrix, we therefore solve the leading-order problem \eqref{eq:mech_be_0} with boundary condition \eqref{eq:be_bc}.
Since we are now going to perform a new expansion in powers of $\Delta$, we  
remove the superscript $(0)$ (associated with the leading-order problem in $\Lambda$).

\medskip

The extra freedom introduced by treating $x_i$ and $X_i$ as independent is removed by requiring that all fields are periodic with respect to $X_i$, so that slow variations across the cell are captured through dependence on the slow variable $x_i$.

Some care is needed to write the rigid-body translations and rotations of the unit cells, $u_i$ and $\Theta$, in multiple-scales form. These are spatially constant on each unit cell, but may vary from cell to cell, and hence vary on the mesoscale after homogenisation. We therefore impose that $u_i$ and $\Theta$ have no microscale variation, i.e.
\be
\label{UTA}
\p{u_i}{X_j} = 0,
\qquad
\p{\omega_i}{X_j} = 0
\qquad \mbox{in } \Op,
\ee
so that, in multiple-scales form,
\be
\label{UT}
\p{u_i}{X_j} + \Delta \p{u_i}{x_j} = 0,
\qquad
\p{\omega_i}{X_j} + \Delta \p{\omega_i}{x_j} = 0
\qquad \mbox{in } \Op.
\ee
Using \eqref{eq:grad_split}, equation \eqref{eq:mech_be_0} in $\Obe$ becomes
\begin{subequations}
  \label{beqnlo}
\begin{align}
	\ds \frac{\dd \sbe_{ij}}{\dd X_j} + \Delta \frac{\dd \sbe_{ij}}{\dd x_j}& =0,  \label{beqnlo.a} \\ 
	\ds \sbe_{ij} &= 2 G^{\mathrm{na}} \epsb + \delta_{ij}\lambda^{\mathrm{na}} \epsbvol, \\ 
	\ds \epsb &=\frac{1}{2} \left( \frac{\dd U^{\mathrm{na}}_i}{\dd X_j} + \frac{\dd U^{\mathrm{na}}_j}{\dd X_i} \right) + \frac{\Delta}{2} \left( \frac{\dd  U^{\mathrm{na}}_i}{\dd x_j} + \frac{\dd  U^{\mathrm{na}}_j}{\dd x_i} \right),
\end{align} 
\end{subequations}
with boundary condition \eqref{eq:be_bc} on $\dO$ and periodic boundary conditions on the boundary of the unit cell $\dObe$. We now expand all dependent variables in powers of $\Delta$. We will allow displacements of order $O(\Delta^{-1})$ to account for accumulation of displacements over the mesoscale---we originally scaled displacements with length of the unit  cell $L^*$ (comparable to the particle diameter), so at leading order $u \sim L^*/\Delta \sim H^*$ scales like the mesoscale length.  The displacements and strains are expanded as
\be 
\label{binderexp}
    \ub = \frac{1}{\Delta} U^{\mathrm{na},(-1)}_i + \Ubekz_i + \Delta \Ubekf_i +\cdots, \quad \epsb = \ebekz_{ij}+\Delta \ebekf_{ij} + \cdots,
\ee
with all other fields expanded similarly; note that we are reusing the superscript notation here---superscripts now refer to the $\Delta$ expansion and not to the $\Lambda$ expansion.


Inserting the expansion \eqref{binderexp} into \eqref{UT},
\eqref{beqnlo} and \eqref{eq:be_bc} gives the leading-order problem in $\Delta$ of
\begin{align}
\label{beqnloB} 
    \p{u_i^{(-1)}}{X_j} &=0, \qquad \mbox{ in }\Op,\\
    \frac{\dd U^{\mathrm{na},(-1)}_i}{\dd X_j} + \frac{\dd U^{\mathrm{na},(-1)}_j}{\dd X_i} & = 0 \qquad \mbox{ in }\Obe,\\
    U^{(-1)}_i & = u^{(-1)}_i \qquad \mbox{ on }\partial \Om,
\end{align}
with $U^{\mathrm{na},(-1)}_i$ periodic in $X_i$ with period 1. The solution is a rigid body motion, which periodicity constrains to be a translation, i.e.
\be
    U^{(-1)}_i(x_i,X_i) = u_i^{(-1)}(x_i).
\ee
At next order in $\Delta$ we find
\begin{align}
\label{lads}
    \epsilon^{\mathrm{na},(0)}_{ij} & = \frac{1}{2} \left( \frac{\partial
    U^{\mathrm{na},(0)}_i }{\partial X_j} + \frac{\dd U_j^{\mathrm{na},(0)}}{\dd X_i} \right) +
{\frac{1}{2} \left( \frac{\partial u^{(-1)}_i}{\partial x_j} +
    \frac{\dd u^{(-1)}_j}{\dd x_i} \right)}\qquad \mbox{ in } \Obe,\\
\label{equilibrium0}
    \sigma^{\mathrm{na},(0)}_{ij} &= 2 G^{\mathrm{na}} \epsilon^{\mathrm{na},(0)}_{ij} + \delta_{ij} \lambda^{\mathrm{na}} 
\epsilon^{\mathrm{na},(0)}_{ll}, \qquad \frac{\dd \sigma^{\mathrm{na},(0)}_{ij}}{\dd X_j} = 0\qquad \mbox{ in }\Obe,\\
\p{\omega_l^{(0)}}{X_j}& =0, \qquad 
 \frac{\partial u^{(-1)}_i}{\partial x _j} + {\frac{\partial u^{(0)}_i}{\partial X_j}=0}\qquad \mbox{ in }\Op,
\label{inpart}
\end{align}
with boundary conditions
\be
\label{swellforcing}
    U_i^{\mathrm{na},(0)} = X_i \gav + u^{(0)}_i +\omega_k^{(0)}\epsilon_{ijk}
    X_j, \label{bcforce} \qquad
    \int_{\dO} \sbekz_{ij} n_j\, d\Gamma = 0,\qquad \int_{\dO} \epsilon_{mjk}X_m \sbekz_{ij} \nu_j\, d\Gamma = 0,
\ee
 where all variables must be periodic in $X_i$ with period 1.

\subsection{The effective stress in the non-active matrix}

\begin{figure}
\centering
\includegraphics[width=0.975\textwidth]{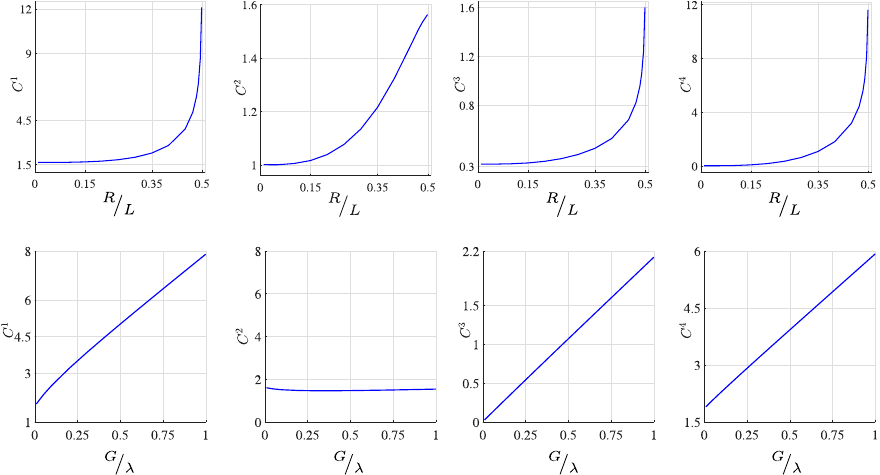}
\caption{Effective elastic properties $C^i$: in the upper panels as a function of the particle size for $G/\lambda = 0.315$, and in the bottom panels as a function of the elastic material for $R/L = 0.45$.}
\label{fig:meso_stress}
\end{figure}

Since (\ref{lads})-(\ref{bcforce}) is a {\it linear} system for $U_i^{\mathrm{na},(0)}$, $u_i^{(0)}$ and $\Theta^{(0)}$, forced by the inhomogeneous terms
\[
    {\frac{1}{2} \left( \frac{\partial u^{(-1)}_i}{\partial x_j} +
      \frac{\dd u^{(-1)}_j}{\dd x_i} \right)}, \qquad
    \frac{\partial u^{(-1)}_i}{\partial x _j} ,\qquad \mbox{ and } \qquad X_i \gav ,
\]
in \eqref{lads}, (\ref{inpart}b) and \eqref{swellforcing} respectively, the solution can be written as
\be 
\label{thisone}
    \sbekz_{ij} = \sigma_{ij}^{kl} \frac{\dd u_k^{(-1)}}{\dd x_l} - \gav(t) \sigma_{ij}^{g},
\ee
where $\sigma_{ij}^{kl}$ and $\sigma_{ij}^{g}$ satisfy the problems stated in Appendix~\ref{app1} and characterise the system's response to the two different forms of excitation, namely macroscopic strain and particle swelling, respectively. We shall refer to these two objects as the cell functions.  A plot of a typical stress field on the surface of the particle is shown in Figure~\ref{fig:meso_stress}. To get an equation for $u_i^{(-1)}$ and thereby close the problem, we need to proceed to one more order in the expansion.

\begin{figure}
\centering
\includegraphics[width=0.6\textwidth]{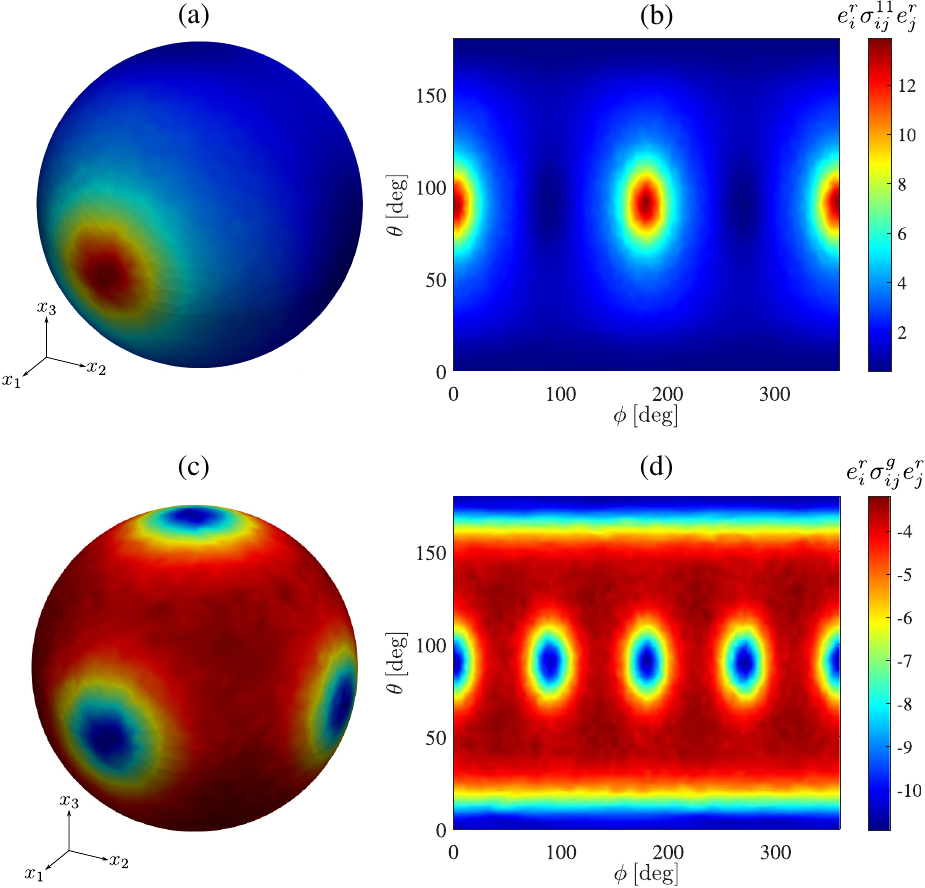}
\caption{Normal traction on the particle surface for the cell functions, with  $G/\lambda = 0.315$ and $R/L = 0.45$. The top row shows the contribution associated with the macroscopic strain: (a) particle-surface view and (b) spherical-coordinate map of $e_i^r \sigma_{ij}^{11} e_j^r$. The bottom row shows the contribution associated with particle swelling: (c) particle-surface view and (d) spherical-coordinate map of $e_i^r \sigma_{ij}^{g} e_j^r$.}
\label{fig:meso_stress00}
\end{figure}

At next order in $\Delta$, equation \eqref{beqnlo.a} yields
\be
\label{lads2}
    \frac{\dd \sbekf_{ij}}{\dd X_j} + \frac{\dd
  \sbekz_{ij}}{\dd x_j} = 0 \qquad \mbox{ in }\Obe.
\ee
Integrating  over $\Obe$ and applying the divergence theorem gives
\be
    -\int_{\dO} \sbekf_{ij} n_j \, \d S + \int_{\dObe} \sbekf_{ij} n_j \, \d S + \frac{\dd}{\dd x_j}\int_{\Obe} \sbekz_{ij}\, \d V = 0.
\ee
The integral over $\dObe$ vanishes due to the periodicity of
$\sbekf_{ij}$. It is tempting to conclude that the integral over
$\dO$ should vanish also, by equating higher-order terms in
(\ref{fandtalk}a). However, care must be taken when handling  
integral constraints in multiple-scale problems. It is shown in
\cite{Cha15} that in fact
\be
    -\int_{\dO} \sbekf_{ij} n_j \, \d S  = \int_{\dO} X_j \frac{\dd\sbekz_{ik}}{\dd x_j} n_k \, \d S.
\ee
Thus the macroscale force balance is
\be 
\label{homstart}
    \frac{\dd \sigmaeff_{ij}}{\dd x_j} = 0,
\ee
where we have introduced the effective stress $\sigmaeff_{ij}$, by defining the averaging operator ${\mathcal A}$, as
\be 
\label{frank}
    \sigmaeff_{ij}  =  \frac{1}{|\Op+\Obe|}\int_{\Obe} \sbekz_{ij} \, \d V + \frac{1}{|\Op+\Obe|}\int_{\dO} X_j \sbekz_{ik}  n_k \, \d S = {\mathcal A} \left( \sbekz_{ij} \right).
\ee
Applying the averaging operator ${\mathcal A}$ to (\ref{thisone}) gives 
\be
\label{sigmadef1}
    \sigmaeff_{ij} = K_{ijkl} \frac{\dd \ueff_k}{\dd x_l} - M_{ij} \gav(t),
\ee
where $K_{ijkl} = {\mathcal A}(\sigma_{ij}^{kl})$, $M_{ij} = {\mathcal A}(\sigma_{ij}^g)$, and we have written $ u^{(-1)}_i=\ueff_i$. Given the symmetry of the unit cell, if we define
\be 
    \epsiloneff_{ij} = \frac{1}{2} \left( \frac{\dd \ueff_i}{\dd x_j} + \frac{\dd \ueff_j}{\dd x_i} \right),
\ee
equation \eqref{sigmadef1} can be concisely written in Voigt notation as
\be 
\label{homend}
    \begin{pmatrix} \sigmaeff_{11} \\ \sigmaeff_{22} \\ \sigmaeff_{33} \\ \sigmaeff_{23} \\ \sigmaeff_{13} \\ \sigmaeff_{12} \end{pmatrix} = 
    \begin{pmatrix} C^{1} & C^{2} & C^{2} & 0 & 0 & 0 \\ C^{2} & C^{1} & C^{2} & 0 & 0 & 0 \\ C^{2} & C^{2} & C^{1} & 0 & 0 & 0 \\ 0 & 0 & 0 & C^{3} & 0 & 0 \\ 0 & 0 & 0 & 0 & C^{3} & 0 \\ 0 & 0 & 0 & 0 & 0 & C^{3} \end{pmatrix} \begin{pmatrix} \epsiloneff_{11} \\ \epsiloneff_{22} \\ \epsiloneff_{33} \\ \epsiloneff_{23} \\ \epsiloneff_{13} \\ \epsiloneff_{12} \end{pmatrix} - \begin{pmatrix} C^{4} \\ C^{4} \\ C^{4} \\ 0 \\ 0 \\ 0 \end{pmatrix} \gav(t),
\ee
where $C^{1},\ldots,C^{4}$ are given by
\be
\label{cells}
    \begin{pmatrix} C^{1} & 0 & 0 \\ 0 & C^{2} & 0 \\ 0 & 0 & C^{2} \end{pmatrix} = {\mathcal A} \left( \sigma^{11}_{ij} \right),\quad
    \begin{pmatrix} 0 & C^{3} & 0 \\ C^{3} & 0 & 0 \\ 0 & 0 & 0 \end{pmatrix} = {\mathcal A} \left( \sigma^{12}_{ij} \right),\quad
    \begin{pmatrix} C^{4} & 0 & 0 \\ 0 & C^{4} & 0 \\ 0 & 0 & C^{4} \end{pmatrix} = {\mathcal A} \left( \sigma^{g}_{ij} \right).
\ee
The effective mechanical parameters $C^{i}$ depend on a geometric parameter, the size of the active particle relative to the unit cell, and a material parameter, the ratio of elastic moduli of the non-active matrix, $\Gbek/\lbek$. The variation of the effective mechanical parameters in terms of the geometric and material parameters is shown in Figure~\ref{fig:meso_stress}.

Equations (\ref{homstart}), (\ref{sigmadef1}) describe the mechanics of the electrode on the mesoscopic (electrode) scale. Note that the effective mechanical properties are cubic-symmetric and, in general, anisotropic rather than isotropic. To understand why, it is sufficient to think about the symmetry of the unit cell composing the material. Compression along the $X_1$, $X_2$ or $X_3$ directions is equivalent due to cubic symmetry. However, diagonal compression for example along the $(1,1,0)$ vector direction is different (and softer, as there is a larger proportion of soft non-active matrix).

\section{Macroscale mechanical problem}
\label{sec:5}

The multiple-scale homogenisation in the previous section allows us to connect the stress in the non-active matrix material surrounding a particle at the microscale with the mesoscale (electrode) stress. 
The mesoscale stress depends on the active particle swelling through the term $M_{ij} g(t)$ in \eqref{sigmadef1}, and also on the boundary conditions imposed on the  equilibrium equation \eqref{homstart}, which depend on how the  entire cell is constrained. Thus, one more step is required to link the mesoscale (electrode) mechanics to the macroscale (whole-cell) mechanics. Fortunately, this work has already been done by \cite{giudici2024mechanical}. They exploited the thin aspect ratio of the electrodes (i.e.\ the typical thickness \( H^* \) is much smaller than the height and width \( W^* \)) and assumed that the current collectors are sufficiently stiff that they do not bend or stretch, so that they move only  in the through-cell direction. Of course, more complex macroscale deformations are possible. Under such assumptions there are only through-cell displacements. The  magnitude of the strains is determined by the electrode swelling and the boundary conditions, with the stresses distributed among electrodes and separator according to their mechanical properties.

In this section, we revisit this result and show how it applies to the material described by (\ref{homstart}) and (\ref{sigmadef1}). The macroscale stack and the representative layer stack used in the reduction are shown in the left and middle panels of Figure~\ref{fig:cell}, respectively; the boundary conditions below are applied on the outer faces normal to the through-cell direction $x_1$. In the first case the cell is rigidly clamped, with no through-cell displacement so the total thickness $NH$ remains unchanged, while in the second case the cell has an applied load, with a specified through-cell pressure $p$ where the current collectors, which remain flat, are displaced in the through-thickness direction, creating a total displacement $U$. Both conditions that we will examine have been studied experimentally; see, for example, \cite{sch25} for fixed-displacement measurements and \cite{moh21} for applied-load or constant-pressure measurements. In practical settings, such as automotive modules, the mechanical constraint is typically neither perfectly rigid nor perfectly load-controlled. Instead, the surrounding bracing has a finite stiffness: the cell is able to expand, thereby deforming the housing or compression system, which in turn exerts a restoring load on the cell \citep{wun19}.

\subsection{Macroscale geometry and assumptions}

To avoid superfluous notation we elect not to introduce a distinct macroscale coordinate system and instead retain the mesoscale coordinates, \( x_i \). As shown in the left panel of Figure~\ref{fig:cell}, the total thickness of the stack is $NH$: $N$ being the number of electrodes ($N/2$ the number of anode-cathode pairs) and $H$ being the average reference thickness of an electrode. \cite{giudici2024mechanical} considered a cell composed of several layers stacked together perpendicular to the \( x_1 \) direction: positive electrodes (\( \mathrm{p} \)), positive current collectors (\(+\)), negative electrodes (\( \mathrm{n} \)), negative current collectors (\(-\)), and separators (\( \mathrm{s} \)), with the representative repeat stack shown in the middle panel of Figure~\ref{fig:cell}. Since separators are typically not adhered to the electrodes, they assumed separators can slide over the electrode, meaning that no shear stresses are transmitted from one electrode to the other. This assumption means only through-cell stresses are transmitted by the separator, and the in-plane stresses in different electrodes pairs decouple. Therefore, it is sensible to divide the cell into two sublayers: a negative electrode pair composed of two negative electrode layers sandwiching their negative current collector (\( \mathrm{n}|-|\mathrm{n} \)), and a positive pair composed of two positive electrodes sandwiching a positive current collector (\( \mathrm{p}|+|\mathrm{p} \)), with half of the separator attached on either side, as represented in the middle panel of Figure~\ref{fig:cell}.

The battery is formed by \( N/2 \) repeating negative and positive pairs separated by separators (\( \mathrm{n}|-|\mathrm{n}|\mathrm{s}|\mathrm{p}|+|\mathrm{p}|\mathrm{s}  \)). Assuming these repeating layers all behave the same, and noting the symmetry, we need only consider one sub-cell, containing one negative electrode, one separator, one positive electrode, and half of each current collector. Due to the soft porous nature of the separator, and since it is not attached to either electrode but allowed to slide, we assume it behaves as a Winkler foundation that does not transmit shear stresses. The change in thickness of the separator is 
\[
\Delta u_1^\mathrm{s} = u_1^\mathrm{s} \big|_{x_1=\frac{1}{2} H^{\mathrm{s}}} - u_1^\mathrm{s} \big|_{x_1=-\frac{1}{2} H^{\mathrm{s}}},
\]
where we recall that $u_1$ is through cell displacement, with the superscript s denoting the separator.
We assume the separator has a general stress-displacement relation given by
\begin{equation}
\label{eq:sep}
    \sigma_{11} = \mathcal{S}(\Delta u_1^\mathrm{s})
\end{equation}
for some function \(\mathcal{S}\) which is monotonically increasing and typically has a steep-shallow-steep progression with increasing strain. The shallow region, almost flat, corresponds to the porous matrix collapsing. The typical shape is discussed in detail in \cite{giudici2024mechanical}.

To distinguish between layers we label the displacement in each layer \(\ueff_i\) by \(u^k_i\), the stress \(\sigmaeff_{ij}\) by \(\sigma_{ij}^k\) and the swelling \(g\) by \(g^{k}\) where \(k\in\{\mathrm{n},\mathrm{p},\mathrm{s}\}\) (with \(g^{\mathrm{s}}=0\)). We also define the width in the $x_2$ and $x_3$ direction as $W_2$ and $W_3$ respectively. We now solve \eqref{homstart} and \eqref{sigmadef1} in each layer subject to (a) stress \(\sigma_{1j}\) and displacements \(u_i\) are continuous across the electrode/current collector interface; (b) normal stress and normal displacements \(\sigma_{11}, u_1\) are continuous across the electrode/separator interface but \(\sigma^{\mathrm{s}}_{1j}=0\) when \(j \neq 1\) since separators do not transmit shear stresses; (c) due to symmetry about the mid-plane of each current collector we assume the current collector remains flat and perpendicular to \(x_1\); (d) zero stress on the lateral boundaries \(x_2 = 0, W_2\), \(x_3 = 0, W_3\); (e) the macroscale boundary conditions along \(x_1\) are either total fixed displacement, \(U\) (typically zero), or fixed applied pressure \(p\). If we centre \(x_1\) in the middle of the separator and choose
\begin{equation}
\label{bc:u}
    u_1(-H^{\mathrm{n}}-\tfrac{1}{2} H^{-} - \tfrac{1}{2} H^{\mathrm{s}}, x_2, x_3) = 0, 
    \quad
    u_1(H^{\mathrm{p}}+\tfrac{1}{2} H^{+} + \tfrac{1}{2} H^{\mathrm{s}}, x_2, x_3) = \Delta u,
\end{equation}
then the two possible conditions in (e) can be respectively written as
\begin{eqnarray}
\label{macrobc}
    \Delta u = \frac{U}{m} \quad \text{for prescribed total thickness change, $U$}; \\
    \iint \sigma_{11}\, \mathrm{d}x_2 \, \mathrm{d}x_3 = F = p W_2 W_3 \quad \text{for prescribed applied pressure, $p$},
\end{eqnarray}
where \(F\) is the externally applied force in the $x_1$ direction.

\subsection{The thin electrode limit}
We can simplify the model of the macroscale stresses by exploiting the fact that the electrodes (the mesoscale) are thin, meaning \(\delta = H/\min(W_2,W_3) \ll 1\). The detailed analysis of the problem in the limit $\delta \to 0$ is given in \cite{giudici2024mechanical}, and shows that
\begin{equation}
\label{eq:cond_sigma11}
    \sigma^k_{12} = \sigma^k_{13} = 0, \qquad \sigma_{11}^k =
    \sigma_{11}(x_2,x_3)\qquad \text{ for }k  \in \{\mathrm{n},-,\mathrm{s},\mathrm{p},+\},
\end{equation}
so that \ \(\sigma_{11}^k\) is independent of \(x_1\) and \(k\). For the in-plane stresses (the $x_2$ and $x_3$ directions), we introduce the tension in each layer,
\begin{align}
\label{def:tension}
    T^k_{ i j }=\int_{H^k} \sigma^k_{ i j }\; \mathrm{d}x_1, \quad k \in \{\mathrm{n},-,\mathrm{p},+\},\quad i,j \in \{2,3\}. 
\end{align}

Using the average stress, at order $\delta^0$, the mechanical equilibrium conditions for each electrode pair are
\begin{equation}
\label{eq:equilibriumF}
    \p{}{x_j} \left(T^{\mathrm{p}}_{ i j }+\tfrac{1}{2} T^{+}_{ i j } \right)=0,\qquad
    \p{}{x_j} \left(T^{\mathrm{n}}_{ i j }+\tfrac{1}{2} T^{-}_{ i j } \right)=0, \quad i,j \in \{2,3\}.
  \end{equation}
The boundary conditions reflecting no shear stresses at the outer edges of the cell require
\begin{equation}
\label{def:Tbcs}
    \left(T^{\mathrm{p}}_{ i j }+\tfrac{1}{2} T^{+}_{ i j } \right) n_j =0,\qquad
    \left(T^{\mathrm{n}}_{ i j }+\tfrac{1}{2} T^{-}_{ i j } \right) n_j =0,
\end{equation}
where $n_j$ are the components of the normals to the battery sides, in the $x_2$ and $x_3$ direction. At leading order the in-plane displacements satisfy
\begin{equation}
\label{condition_u}
    u^{+}_i=u^{\mathrm{p}}_i=u^{\mathrm{p}}_i(x_2,x_3), \qquad 
    u^{-}_i=u^{\mathrm{n}}_i=u^{\mathrm{n}}_i(x_2,x_3),
    \quad i,\, j \in \{2,3\}.
\end{equation}
Note the displacement in the two electrodes need not necessarily be equal since the displacement need not be continuous across the separator. Combining (\ref{eq:cond_sigma11}) and \eqref{eq:equilibriumF} with \eqref{sigmadef1}, \eqref{eq:sep} and one of \eqref{macrobc} gives a set of 10 equations for the 10 unknowns
\(u^{\mathrm{p}}_1, u^{\mathrm{p}}_2, u^{\mathrm{p}}_3, u^{+}_1, u^{\mathrm{n}}_1, u^{\mathrm{n}}_2,
 u^{\mathrm{n}}_3, u^{-}_1, \Delta u_1^\mathrm{s} , \sigma_{11}(x_2,x_3)\).
This is an elliptic problem in $x_1,x_2$ with Neumann boundary conditions and an algebraic constraint coming from the through-cell stress. Further analytical progress can be made by considering the limit in which the current collectors are stiff compared to the electrodes, meaning they remain planar.

\subsection{The stiff current collector limit}

When the current collectors are stiff compared to the electrodes they do not deform. Thus, not only do they remain planar, but the in-plane displacements vanish, so that
\[
u^k_2=u^k_3=0 \quad \text{in all layers.}
\]
Then, since only through-cell displacement is allowed, the stresses in the electrodes are
\begin{equation}
\label{sig11}
    \sigma_{11}=C^{1}\frac{\partial u_1^{k}}{\partial x_1} - C^{4} \bar{g}^{k} , \qquad
    \sigma^{k}_{ i j }=\left(C^{2}\frac{\partial u_{1}^{k}}{\partial x_{1}}
  - C^{4} \bar{g}^{k} \right) \delta_{ i j }, \qquad \text{ for }k \in\{\mathrm{n},\mathrm{p}\},
\end{equation}
where \(\delta_{i j }\) is the Kronecker \(\delta\),  and the particle induced expansion \(\bar{g}^{k}=\bar{g}^{k}(x_1,x_2,x_3,t)\) may in general be a function of all coordinates as well as time. Inverting equation (\ref{sig11}a) we find 
\begin{equation}
\label{pu}
    \frac{\partial u_1^{k}}{\partial x_1} = \frac{\sigma_{11} + C^{4} \bar{g}^{k}}{C^{1}}.
\end{equation}
It is worth pausing and noting that, since all other displacements are zero and the shear strains vanish at leading order in \(\delta\), once we determine \(\sigma_{11}\) we have all the information we need to determine the microscale stress in the non-active matrix.

Integrating \eqref{pu} over the thickness of each layer, we obtain the total through-cell displacement in each layer as
\begin{equation}
\label{u1_sol}
    \Delta u^{\mathrm{n}}_1=\frac{H^{\mathrm{n}}}{C^{1}}  \left(\sigma_{11}+C^{4} \langle g^{\mathrm{n}}\rangle  \right),\qquad
    \Delta u^{\mathrm{p}}_1=\frac{H^{\mathrm{p}}}{C^{1}}  \left(\sigma_{11}+C^{4} \langle g^{\mathrm{p}}\rangle \right),
\end{equation}
where we define 
\begin{equation}
    \langle g^{k} \rangle=\frac{1}{H^k}\int_{H^k} \bar{g}^{k} \, \mathrm{d}x_1.
\end{equation} 
This is the average over the mesoscale thickness of the microscale-averaged expansion $\gk$ and introducing it simplifies the later notation. 

The total change in thickness of the negative electrode, positive electrode and separator is
\[
\Delta u= \Delta u^{\mathrm{n}}_1+\Delta u^{\mathrm{p}}_1+\Delta u^{\mathrm{s}}_1,
\]
(note that \(\Delta u\) is  independent of \(x_2\) and \(x_3\), since the two current collectors remain parallel and flat). Using equations \eqref{u1_sol} and \eqref{eq:sep} we have
\begin{align}
\label{eqn:deltau}
    \Delta u
    &= \frac{ H^{\mathrm{n}}}{C^{1}} \left(\sigma_{11}+C^{4} \langle g^{\mathrm{n}}\rangle\right)
     + \frac{ H^{\mathrm{p}}}{C^{1}} \left(\sigma_{11}+C^{4} \langle g^{\mathrm{p}} \rangle \right)
     + \mathcal{S}^{-1}(\sigma_{11}). 
\end{align}

For a clamped boundary condition, corresponding to fixed \(\Delta u\),
we can solve \eqref{eqn:deltau} for \(\sigma_{11}\) and then substitute
into \eqref{pu} to determine \(\partial u_1^k/\partial x_1\). In the particular case in which the separator is linear elastic, with \( \mathcal{S}(\Delta u_1^\mathrm{s}) = K
\Delta u_1^\mathrm{s} \), we can solve explicitly to find
\begin{align}
\label{eq:c}
    \sigma_{11} &= \frac{C^{1} \Delta u  - C^{4} \left( H^{\mathrm{n}} \langle g^{\mathrm{n}} \rangle
    + H^{\mathrm{p}} \langle g^{\mathrm{p}} \rangle \right)}{H^{\mathrm{n}} + H^{\mathrm{p}} + \frac{C^{1}}{K}},\\
    \frac{\partial u_1^{k}}{\partial x_1} &= \frac{\Delta u}{H^{\mathrm{n}} + H^{\mathrm{p}} + \frac{C^{1}}{K}} + \frac{C^{4} }{C^{1}}\left(g^{k} - \frac{  H^{\mathrm{n}} \langle g^{\mathrm{n}} \rangle
    + H^{\mathrm{p}} \langle g^{\mathrm{p}} \rangle }{H^{\mathrm{n}} + H^{\mathrm{p}} + \frac{C^{1}}{K}}\right),
\end{align}
for \(k \in \{\mathrm{n},\mathrm{p}\}\). If, instead of clamping the system, we apply a load with average pressure \(p\) we find (for the linear elastic separator case)
\begin{align}
\label{eqn:deltauB}
    \Delta u &= \frac{H^{\mathrm{n}}}{C^{1}} \left( p+ C^{4} \overline{\langle g^{\mathrm{n}}
        \rangle} \right) + \frac{H^{\mathrm{p}}}{C^{1}} \left(p+ C^{4}
      \overline{\langle g^{\mathrm{p}} \rangle} \right) + \frac{p}{K} \nonumber \\
    &= \left(\frac{H^{\mathrm{n}}+H^{\mathrm{p}}}{C^{1}} +\frac{1}{K}\right)p +  \frac{C^{4}}{C^{1}}\left(H^{\mathrm{n}} \overline{\langle g^{\mathrm{n}}
        \rangle}  + H^{\mathrm{p}} 
      \overline{\langle g^{\mathrm{p}} \rangle}\right),
\end{align}
where
\[
  \overline{\langle g^{k}
  \rangle}
\equiv
  \frac{1}{H^kW_2W_3}\int_{H^k}\int_{W_2}\int_{W_3} \bar{g}^{k} \, \mathrm{d}x_1\,\mathrm{d}x_2\,\mathrm{d}x_3
\equiv
  \frac{1}{W_2W_3}\int_{W_2}\int_{W_3} \langle g^{k} \rangle \,\mathrm{d}x_2\,\mathrm{d}x_3,
\]
is the average swelling in layer \(k\) over the whole electrode, obtained by taking the mesoscale average of the microscale average expansion $g$. We can then substitute \eqref{eqn:deltauB} into \eqref{eq:c} to give
\begin{align}
    \sigma_{11}& = -p +  \frac{C^{4} \left( H^{\mathrm{n}}( \overline{\langle g^{\mathrm{n}}
  \rangle}-\langle g^{\mathrm{n}} \rangle)
    + H^{\mathrm{p}} ( \overline{\langle g^{\mathrm{p}}
  \rangle}-\langle g^{\mathrm{p}} \rangle) \right)}{H^{\mathrm{n}} + H^{\mathrm{p}} + \frac{C^{1}}{K}},\\
    \frac{\partial u_1^{k}}{\partial x_1} &= -\frac{p}{C^{1}} + \frac{C^{4} }{C^{1}}\left(g^{k} + \frac{  H^{\mathrm{n}}( \overline{\langle g^{\mathrm{n}}
  \rangle}-\langle g^{\mathrm{n}} \rangle)
    + H^{\mathrm{p}} ( \overline{\langle g^{\mathrm{p}}
  \rangle}-\langle g^{\mathrm{p}} \rangle) }{H^{\mathrm{n}} + H^{\mathrm{p}} + \frac{C^{1}}{K}}\right),
\end{align}
for \(k \in \{\mathrm{n},\mathrm{p}\}\).
It is worth observing that in the cases of very slow charging, equipotential current collectors and small resistances, where the current is uniform in the \(x_2\) and \(x_3\) directions, the swelling and the macroscale stress will also be independent of \(x_2\) and \(x_3\), because \(\overline{\langle g^{k}  \rangle} = \langle g^{k} \rangle\).

We note that the analysis above can readily be extended to consider a cell housing that deforms and in turn applies a load back on the electrode stack, as may well be the case in practice, but we do not consider this here.

\section{Feedback of the macroscale strain to the microscale mechanics}
\label{sec:6}

In Section~\ref{sec:4} we have linked the microscale non-active matrix stress surrounding an active particle with the mesoscale electrode stresses. We have also shown that the stresses in each electrode depend on the interaction among the different battery components (separator, current collectors, electrodes) and the macroscale boundary conditions imposed on the whole battery. In Section~\ref{sec:5} we have connected the electrode-level (mesoscale) structure with that of the battery (macroscale) structure and boundary conditions, closing the mechanical problem. In particular, we have shown that the cell undergoes only through-cell strains and that, with knowledge of the active particle swelling, mechanical and geometric properties of electrodes and separator, and of the imposed cell boundary conditions, we can determine the average electrode stress, given by
\be
\label{sigmadef1.1}
    \sigmaeff_{ij} = K_{ij11} \frac{\dd \ueff_1}{\dd x_1} - M_{ij} g.
\ee
and, using (\ref{thisone}), we also have access to the microscale resolved stress
\be
\label{sigmadef2}
    \sbekz_{ij} = \sigma^{11}_{ij} \frac{\dd \ueff_1}{\dd x_1} - \sigma^g_{ij} g,
\ee
where $\sigma^{11}_{ij}$ and $\sigma^g_{ij}$ are cell functions that can be obtained numerically, both positive so that in the common scenario of both expansive swelling and compressive electrode strains we obtain compression of the non-active matrix as expected.  This non-active matrix stress is exactly what is needed in (\ref{eq:mech_bc_1}) for us to solve for the first-order correction to the electrochemo-mechanical problem in the particle, \eqref{eq:mech_k_1}-\eqref{eq:mech_bc_1}.

Thus, in this section we return to the $\Lambda$ expansion introduced in Section~\ref{sec:3}  (arising from the extreme contrast in mechanical properties between the active particles and binder), and examine how the non-active matrix stresses at leading order in $\Lambda$ affect the particle mechanical and electrochemical behaviour at $O(\Lambda)$. Our objective is to write a DFN model that includes the effect of non-active matrix mechanics on the electrochemistry.

To achieve a DFN model it is important to note that we have homogenised the mechanics of the electrode stack, but we have not homogenised the electrochemical transport in the same way. This is deliberate. For the electrolyte concentration and electronic conduction we adopt the standard volume-averaged porous-electrode description (as in \cite{gilescolin2012}): the electrolyte and solid phases are treated as continua on the electrode scale, and the electrochemical reactions do not have to be resolved around each individual particle. Instead, the interfacial reaction appears as a distributed source term in the electrode-scale transport equations, with an intensity equal to the local (DFN) interfacial current density multiplied by the specific surface area of active material, (see equations~(3.4)--(3.5) in \cite{gilescolin2012}).

The overall picture is therefore as follows:
\begin{enumerate}
    \item \textbf{Mechanics} is homogenised up to the cell scale because the deformation is strongly constrained by the battery architecture (current collectors, separator, clamping or applied load). This information is needed to determine the actual stress transmitted by the non-active matrix back onto the particles.
    \item \textbf{Electrochemistry and transport} are left in the standard porous-electrode (DFN-style) form because, once homogenised in the classical sense, the reactions already enter as volumetric source terms. There is no practical benefit in resolving the electrolyte at the same geometric level as the mechanics, unless one wishes to capture local electrolyte starvation or short-scale electrochemical gradients around individual particles.
    \item The two descriptions remain \textbf{fully consistent}: the mechanical model provides the local surface stress at the particle-non-active matrix interface, which feeds into the particle electrochemo-mechanical problem (via stress-dependent chemical potential and overpotential), while the electrochemical model provides the local interfacial current density, which enters as a source term in the electrode-scale DFN transport equations.
\end{enumerate}

Our first task is to demonstrate that, even if the surface stress introduced by the binder is not spatially homogeneous, its effects on the electrochemical problem can be understood by averaging particle transport and mechanical equations over the surface of the sphere, obtaining an effective radial problem. This problem is analogous to the 1D particle problem in DFN, thus achieving an analogous level of complexity.  

\subsection{Surface-integrated particle equations}

Although the microscale problem \eqref{eq:mech_k_1}-\eqref{eq:mech_bc_1} is posed on a  spherical geometry, because of the macroscopic geometry of the battery the boundary condition (\ref{eq:mech_bc_1}) is not spherically symmetric (i.e. $\sigma_{ij}^{\mathrm{na},(0)}$ depends on $\theta$ and $\phi$). Thus,
unlike the leading-order problem \eqref{eq:mech_0}-\eqref{eq:totI0}, in general the solution is not simply a function of $r$ and $t$. However, we recall that in the homogenised electrolyte equations, the source term inherited from particle transport and exchange current is a surface integral of the interfacial exchange current over the particle surface \cite{gilescolin2012} weighted by the surface area density. This means that, to write the homogenised equations, we are only interested in this surface average. In Appendix~\ref{app:radial} we show that if we integrate all the equations of $O(\Lambda)$ over $\theta$ and $\phi$, exploiting the symmetries in the system, then we obtain a radial mechanical and transport problem in the particle that is sufficient to determine the exchange current. This surprising fact means we can find the correction to the chemical potential analytically --- a remarkable outcome given the complexity of the original problem.

The spherically averaged equations are
\begin{align}
\label{eq:radial-sigrr}
    \sigma_{rr}^{k,(1)} &= (2\Gk + \lk)\,\frac{\partial U_{r}^{k,(1)}}{\partial r}
                     + 2 \lk \frac{U_{r}^{k,(1)}}{r}
                     - (2\Gk + 3\lk)\,\linVVk\,\ckf,
\\[4pt]
\label{eq:radial-sigtt}
    \sigma_{\theta\theta}^{k,(1)} &= \lk\,\frac{\partial U_{r}^{k,(1)}}{\partial r}
                     + 2(\Gk+\lk)\,\frac{U_{r}^{k,(1)}}{r}
                     - (2\Gk + 3\lk)\,\linVVk\,\ckf ,
\end{align}
where
\[
U_{r}^{k,(1)}(r,t), \qquad \sigma_{rr}^{k,(1)}(r,t), \qquad \sigma_{\theta\theta}^{k,(1)}(r,t)=\sigma_{\phi\phi}^{k,(1)}(r,t), \qquad \ckf(r,t)
\]
denote, respectively, the first-order corrections to the  radial displacement, radial and hoop stresses, and  lithium concentration, all averaged over the angles as in Appendix~\ref{app:radial}. Since the problem is spherically symmetric after averaging, mechanical equilibrium reduces to
\begin{equation}
\label{eq:radial-mecheq}
    \frac{\partial \sigma_{rr}^{k,(1)}}{\partial r}
    + \frac{2}{r}\bigl(\sigma_{rr}^{k,(1)} - \sigma_{\theta\theta}^{k,(1)}\bigr) = 0.
\end{equation}
The corresponding first-order diffusion equation in the particle becomes
\begin{equation}
\label{eq:radial-diff}
    \frac{\partial \ckf}{\partial t}
    + \frac{1}{r^{2}}\frac{\partial}{\partial r}\!\left(r^{2} N^{k,(1)}_{r}\right) = 0,
    \qquad 0<r<R,
\end{equation}
with radial flux
\begin{equation}
\label{eq:radial-flux}
    N^{k,(1)}_{r} = -\,\Dk \Bigl[
  \ckz\bigl(1 - \ckz\bigr)\,\frac{\partial \mukf}{\partial r}
  - 2 \ckz \ckf \,\frac{\partial \mukz}{\partial r}
    \Bigr].
\end{equation}
The first-order electro-chemo-mechanical potential is
\begin{equation}
\label{eq:radial-mu}
    \mukf = -\,\p{\Ueqk}{c}\bigg|_{c=\ckz}\,\ckf
             - \gamma \linVVk\, \sigma_{ll}^{k,(1)},
    \qquad
    \sigma_{ll}^{k,(1)} = \sigma_{rr}^{k,(1)} + 2\sigma_{\theta\theta}^{k,(1)}.
\end{equation}
The mechanical boundary conditions on the particle are
\begin{equation}
\label{eq:radial-bc-mech}
    U_{r}^{k,(1)}(0,t) = 0,
    \qquad
    \sigma_{rr}^{k,(1)}(R,t)
    = \Sigma_{11}\,\p{u_1^{\mathrm{eff}}}{x_1}
    - \Sigma^{g} \,\frac{U_{r}^{k,(0)}(R,t)}{R},
\end{equation}
where the surface-averaged coefficients
\begin{equation}
\label{eq:Sigma-def}
    \Sigma_{11}
    = \frac{1}{4\pi R^{2}}
    \iint_{\Gamma^{\mathrm{a}}}
    e^{r}_i\sigma^{11}_{ij}e^{r}_j \, \mathrm{d}S,
    \qquad
    \Sigma^{g}
    =  \frac{1}{4\pi R^{2}}
    \iint_{\Gamma^{\mathrm{a}}}
    e^{r}_i\sigma^{g}_{ij}e^{r}_j \, \mathrm{d}S,
\end{equation}
are those obtained from the unit-cell (non-active matrix) problems in Appendix~\ref{app1}, and $\p{u_1^{\mathrm{eff}}}{x_1}$ is the through-cell strain computed in Section~\ref{sec:5}. 

The two parameters in \eqref{eq:Sigma-def} represent the average radial stress generated by the non-active matrix on the particle as a result of electrode strains and particle expansion, respectively.
\begin{figure}
\centering
\includegraphics[width=0.5\textwidth]{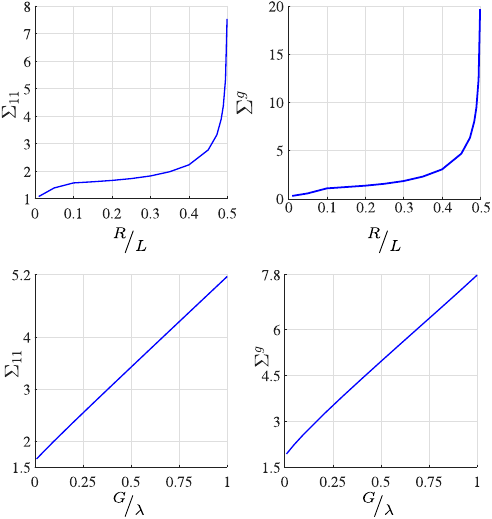}
\caption{Surface-averaged coefficients $\Sigma_{11}$, and $\Sigma^g$: in the upper panels as a function of the particle size for for $G/\lambda = 0.315$, and in the bottom panels as a function of the elastic material for $R/L = 0.45$.}
\label{fig:binder_stress}
\end{figure}

For the electrochemical boundary condition at $r=R$, the flux is exactly the surface average over the sphere, and  reads
\begin{equation}
\label{eq:radial-bc-diff}
N^{k,(1)}_r(R,t)
=
J^{k,(1)}
\sinh\left(\frac{\eta^{k,(0)}}{2}\right)
+
\frac{J^{k,(0)}\eta^{k,(1)}}{2}
\cosh\left(\frac{\eta^{k,(0)}}{2}\right).
\end{equation}
  with the first-order overpotential
\begin{equation}
\label{eq:radial-eta}
    \etakf = \phikf - \phief
               -\,\p{\Ueqk}{c}\bigg|_{c=\ckz}\,\ckf
               - \gamma \linVVk\, \sigma_{ll}^{k,(1)}
               \quad \text{on } r=R.
\end{equation}
 Finally, we have the initial condition 
\be 
\label{iiccofirstorder}
    c^{(1)}|_{t=0} = 0.
\ee

Substituting \eqref{eq:radial-sigrr}-\eqref{eq:radial-sigtt} into \eqref{eq:radial-mecheq} gives the radial ODE
\begin{equation}
\label{eq:radial-ode}
  \,\frac{\d^2 U^{k,(1)}_r}{\d r^2}
  + \frac{2}{r}\,\frac{\d U^{k,(1)}_r}{\d r}
  - \frac{2}{r^2} \, U^{k,(1)}_r
  = \frac{(2\Gk+3\lk)}{(2\Gk+\lk)}\,\linVVk\,\frac{\d \ckf}{\d r}.
\end{equation}
The homogeneous solution of \eqref{eq:radial-ode}, which we identify with a hat, is given by
\[
  \hat{U}^{k,(1)}_r = \Psi \, r,
\]
for some constant $\Psi$, which produces the (isotropic) stress field
\begin{equation}
\label{eq:radial-hom-stress}
  \hat{\sigma}^{k,(1)}_{rr} = \hat{\sigma}^{k,(1)}_{\theta\theta} = \hat{\sigma}^{k,(1)}_{\phi\phi}
  = (2\Gk+3\lk)\,\Psi.
\end{equation}
Choosing
\begin{equation}
\label{eq:alpha-choice}
  \Psi
  = \frac{1}{(2\Gk+3\lk)}
    \left(
      \Sigma_{11}\,\p{\ueff_1}{x_1}
      - \Sigma^{g}\,\frac{U^{k,(0)}_r(R,t)}{R}
    \right)
\end{equation}
ensures that the boundary condition \eqref{eq:radial-bc-mech} is satisfied and the first-order stress trace
\[
  \hat{\sigma}^{k,(1)}_{ll}
   = \hat{\sigma}^{k,(1)}_{rr} + 2\hat{\sigma}^{k,(1)}_{\theta\theta}
   = 3(2\Gk+3\lk)\,\Psi
\]
is then spatially uniform. This means that the homogeneous solution has no radial gradients in $\ckf$. This is not surprising: nothing in the homogeneous problem drives the formation of concentration gradients. Hence, we set the homogeneous solution $\hat{c}^{(1),k} \equiv 0$ to satisfy also the initial condition. 

We now focus on the particular solution. We identify the quantities pertaining to the particular problem with a tilde. 
The mechanical equations for the particular solutions are
\begin{equation}
\label{eq:radial-ode-tilde}
  \,\frac{\d^2 \tilde{U}^{k,(1)}_r}{\d r^2}
  + \frac{2}{r}\,\frac{\d \tilde{U}^{k,(1)}_r}{\d r}
  - \frac{2}{r^2} \, \tilde{U}^{k,(1)}_r
  = \frac{(2\Gk+3\lk)}{(2\Gk+\lk)}\,\linVVk\,\frac{\d \tilde{\ckf}}{\d r}.
\end{equation}
With stress free boundary condition, 
\be
\tilde{\sigma}^{k,(1)}_{rr}=0 \quad \text{on} \quad \dOk.
\ee
The transport equations require
\begin{equation}
\label{eq:radial-diff_a}
    \frac{\partial \tilde{\ckf}}{\partial t}
    + \frac{1}{r^{2}}\frac{\partial}{\partial r}\!\left(r^{2} \tilde{N}^{k,(1)}_{r}\right) = 0,
    \qquad 0<r<R,
\end{equation}
with radial flux
\begin{equation}
\label{eq:radial-flux_a}
    \tilde{N}^{k,(1)}_{r} = -\,\Dk \Bigl[
    \ckz\bigl(1 - \ckz\bigr)\,\frac{\partial \tilde{\mukf}}{\partial r}
    - 2 \ckz \tilde{\ckf} \,\frac{\partial \mukz}{\partial r}
    \Bigr].
\end{equation}
where
\begin{equation}
\label{eq:radial-mu-tilde}
    \tilde{\mu}^{k,(1)} =-\p{\Ueqk}{c}\bigg|_{c=\ckz}\,\tilde{\ckf} - \gamma \linVVk\, \tilde{\sigma}_{ll}^{k,(1)}-\gamma \VVk\, \left(
      \Sigma_{11}\,\p{\ueff_1}{x_1}
      - \Sigma^{g}\,\frac{U^{k,(0)}_r(R,t)}{R}
    \right).
\end{equation}
Note that the factor of $\third$ in front of the last term is gone since $\sigma^{(1),k}_{ll}=3 \sigma^{(1),k}_{rr}$. The homogeneous mechanical stress does not affect the radial flux in the particle,
since it is spatially uniform. However, it contributes to the interfacial
overpotential. The first-order Butler--Volmer condition is therefore
\begin{equation}
\label{eq:radial-bc}
\widetilde N_r^{k,(1)}(R,t)
=
J^{k,(1)}
\sinh\left(\frac{\eta^{k,(0)}}{2}\right)
+
\frac{J^{k,(0)}}{2}
\widetilde\eta^{k,(1)}
\cosh\left(\frac{\eta^{k,(0)}}{2}\right).
\end{equation}
where the first-order correction to the overpotential now reads
\begin{equation}
\label{eq:radial-eta-tilde}
    \tilde{\etakf} = \phikf - \phief
               -\,\p{\Ueqk}{c}\bigg|_{c=\ckz}\,\tilde{\ckf}
               - \gamma \linVVk\, \sigma_{ll}^{k,(1)}-\gamma \VVk\, \left(
      \Sigma^k_{11}\,\p{u^k_1}{x_1}
      - \Sigma^{k,g}\,\frac{U^{k,(0)}_r(R,t)}{R}
    \right)
               \quad \text{on } r=R.
\end{equation}

Crucially, the problem written in \eqref{eq:radial-ode-tilde} - \eqref{eq:radial-eta-tilde} is analogous to the leading order problem in the particle, with stress-assisted diffusion, but no surface stress. In fact, it is exactly the next order in the $\Lambda$ expansion for a free lithiating spherical particle, but with one added difference: a mechanical correction to the potential and overpotential which depends on the mesoscale coordinate $x_1$. This spatial dependence comes from the particle swelling and the non-active matrix reaction force. 

This observation is crucial; it suggests that our model can be thought of as a free expanding particle with a correction in the chemical potential and the overpotential inherited from non-active matrix stresses. Thus, up to errors of $O(\Lambda^2)$, we may write our model directly as a free expanding particle with a spatially varying mechanical correction to the overpotential.

\section{DFN model with multiscale mechanics}
\label{sec:7}

We now write a dimensional DFN model augmented by the leading-order feedback from multiscale mechanics derived in Section~\ref{sec:6}. Readers who have arrived here directly from Section~\ref{intro}, skipping the derivation in the interim sections, can readily make use of our results without worrying about the asymptotic problem. The outcome of the previous sections was to show that, for small swelling strains and a small stiffness ratio of the non-active matrix relative to the active particles, the mechanics enters only as a correction to the chemical potential. Our final problem is therefore of similar computational complexity to the standard DFN model with stress-assisted diffusion, but includes electro-chemo-mechanical coupling from the microscale, namely the particle scale, up to the macroscale, namely the battery scale. In the language of the asymptotic derivation presented above: if one were to expand the model below in $\Lambda$, the $O(1)$ and $O(\Lambda)$ equations reduce exactly to those obtained in Section~\ref{sec:6}.

The through-cell coordinate is $x_1\in[0,H]$, with negative electrode $\Omega^{\mathrm n}=[0,H^{\mathrm n}]$, separator $\Ose=[H^{\mathrm n},H^{\mathrm n}+H^{\mathrm{se}}]$, and positive electrode $\Omega^{\mathrm p}=[H^{\mathrm n}+H^{\mathrm{se}},H]$. In an electrode $k\in\{\mathrm n,\mathrm p\}$ we use the effective transport coefficients from Section~\ref{sec:2}:
\[
\Deffk=\Bk D^{\mathrm e},\qquad
\keffek=\Bk \ke,\qquad
\keffk=\Bkhat\,\kappa^{k}.
\]
We denote by $b^{k}$ the specific interfacial area, that is, the electrochemically active particle surface per electrode volume. Throughout this section, as in the rest of the paper, the electrode label $k\in\{\mathrm n,\mathrm p\}$ is written as a superscript, while subscripts are reserved for spatial directions and tensor indices. This differs only notationally from the common DFN convention in which the electrode label is often written as a subscript. We also retain our convention that $\Jk$ denotes the molar interfacial reaction flux, with units $\mathrm{mol\,m^{-2}\,s^{-1}}$. Thus the reaction current density denoted by $j^k$ in \cite{brosa2022continuum} is $F\Jk$ in the notation used here.

\subsection{Mechanics and transport in the active particles}

In a spherical particle of electrode $k$, with radius $r\in(0,R^{k})$,
\begin{align}
\label{eq:dfn-solid-diff-mech}
    \p{\ck}{t}
    +\frac{1}{r^2}\frac{\partial}{\partial r}\!\big(r^2 N^{k}_{r}\big) &= 0,\\[-0.2em]
    N^{k}_{r} &= -\,\frac{\Ddiffk(\ck)\,\ck}{RT}\!\left(1-\frac{\ck}{\ckmax}\right)
    \frac{\partial \mu^{\mathrm{DFN},k}}{\partial r}.
\end{align}
The DFN chemical potential is
\begin{equation}
\label{eq:mu-dfn-mech-dim-expanded}
    \mu^{\mathrm{DFN},k}
    =
    \mu^0
    - F\,\Ueqk(\ck)
    - \linVk\,\sk_{ll}
    -
    \Vk\,\lbek
    \left(
    \Sigkoneone\,\p{u^{k}_1}{x_1}
    -\Sigkg\,\frac{\Ukr(R^{k})}{R^{k}}
    \right).
\end{equation}
Here $\Sigkoneone$ and $\Sigkg$ are dimensionless quantities defined in \eqref{eq:Sigma-def}. They depend only on two quantities: a material parameter, $\Gbek/\lk$, equivalent to a Poisson-ratio dependence, and a geometric parameter, $R^{k}/L$.

The mechanical state inside the particles obeys
\begin{align}
    \ek_{rr}&= \frac{\d \Ukr}{\d r},\qquad
    \ek_{\theta\theta}=\ek_{\phi\phi}=\frac{\Ukr}{r},\qquad
    \mathrm{tr}\,\ek=\frac{\d \Ukr}{\d r}+\frac{2\,\Ukr}{r},
\nonumber\\[0.2em]
\label{eq:mech_dfn}
    \sk_{rr} &= (2\Gk+\lk)\,\frac{\d \Ukr}{\d r} + 2\lk\,\frac{\Ukr}{r}
    -(2\Gk+3\lk)\,\linVk\,(\ck-\ckref),
\\
    \sk_{\theta\theta}&=\lk\,\frac{\d \Ukr}{\d r}+2(\Gk+\lk)\,\frac{\Ukr}{r}
    -(2\Gk+3\lk)\,\linVk\,(\ck-\ckref),\qquad \sk_{\phi\phi}=\sk_{\theta\theta}.
\nonumber
\end{align}
Thus
\[
    \sk_{ll}=\sk_{rr}+2\sk_{\theta\theta}.
\]
The particle mechanics is supplemented by equilibrium,
\[
    \frac{\d\sk_{rr}}{\d r}+\frac{2}{r}(\sk_{rr}-\sk_{\theta\theta})=0,
\]
regularity at $r=0$, and a stress-free surface at leading order:
\[
    \Ukr(0)=0,\qquad \sk_{rr}(R^{k})=0.
\]

The surface flux and Butler--Volmer kinetics are
\begin{align}
\label{eq:dfn-surface-flux-current}
    N^{k}_{r}(R^{k},t) &= \Jk(x_1,t),\\
\label{eq:dfn-bv-current}
    \Jk &= \Kk
    \sqrt{\frac{\ce}{\cezero}\frac{\ckR}{\ckmax}\!\left(1-\frac{\ckR}{\ckmax}\right)}
    \;\sinh\!\left(\frac{F}{2RT}\,\eta^{\mathrm{DFN},k}\right),
\end{align}
where $\ckR=\ck(R^{k},t)$ and $\Jk$ is positive for deintercalation from the active particle into the electrolyte. The mechanically corrected overpotential is
\begin{equation}
\label{eq:eta-dfn-mech-correct}
    \eta^{\mathrm{DFN},k}
    =
    \phik- \phie
    -\Ueqk(\ckR)
    -\frac{\linVk}{F}\,\sk_{ll}(R^{k},t)
    -
    \frac{\lbek\,\Vk}{F}\,
    \left(
      \Sigkoneone\,\p{u^{k}_1}{x_1}
    - \Sigkg\,\frac{\Ukr(R^{k})}{R^{k}}
    \right).
\end{equation}

\subsection{Charge transport in the electrolyte and non-active matrix}

In each electrode $k$,
\begin{equation}
\label{eq:dfn-ce}
    \p{(\alk\,\ce)}{t}
    =
    \frac{\partial}{\partial x_1}\!\left(\Deffk\,\p{\ce}{x_1}\right)
    +(1-\tplus)\,b^{k}\,\Jk,
    \qquad x_1\in\Omega^k.
\end{equation}
For the phase currents in the different regions we define
\begin{equation}
\label{eq:currents-def}
    \ike 
    = -\,\keffek
    \left(
    \p{\phie}{x_1}
    -\frac{2RT}{F}(1-\tplus)\,\p{\ln\ce}{x_1}
    \right),
    \qquad    
    \ik
    = -\,\keffk\,\p{\phik}{x_1}.
\end{equation}
Current conservation can be written in divergence form as
\begin{align}
\label{eq:dfn-phie}
    \frac{\partial}{\partial x_1}\!\left[
    \keffek
    \left(
    \p{\phie}{x_1}
    -\frac{2RT}{F}(1-\tplus)\,\p{\ln\ce}{x_1}
    \right)
    \right]
    &=-\,F b^{k}\,\Jk,
\\
\label{eq:dfn-phik}
    \frac{\partial}{\partial x_1}\!\left(\keffk\,\p{\phik}{x_1}\right)
    &= F b^{k}\,\Jk.
\end{align}
Equivalently, these equations state that
\[
    \p{\ike}{x_1}=F b^k \Jk,
    \qquad
    \p{\ik}{x_1}=-F b^k \Jk,
\]
so that the total current is conserved. This is precisely where the factor of $F$ enters: $\Jk$ is a molar flux, whereas the phase currents $\ike$ and $\ik$ are electrical current densities.

In the separator, $\Ose$,
\begin{align}
\label{eq:dfn-separator}
    \p{(\alpha^{\mathrm{se}}\,\ce)}{t}
    &=
    \frac{\partial}{\partial x_1}\!\left(D^{\mathrm{e,eff},\mathrm{se}}\,\p{\ce}{x_1}\right),
\\
    \frac{\partial}{\partial x_1}\!\left[
    \kappa^{\mathrm{e,eff},\mathrm{se}}
    \left(
    \p{\phie}{x_1}
    -\frac{2RT}{F}(1-\tplus)\,\p{\ln\ce}{x_1}
    \right)
    \right]
    &=0,
\end{align}
with
\[
    D^{\mathrm{e,eff},\mathrm{se}}=\mathcal{B}^{\mathrm{se}}D^{\mathrm e},
    \qquad
    \kappa^{\mathrm{e,eff},\mathrm{se}}=\mathcal{B}^{\mathrm{se}}\ke.
\]

\subsection{Interfaces, boundaries, and initial data}

At electrode--separator interfaces we impose continuity of $\ce$ and of the electrolyte concentration flux $D^{\mathrm{e,eff}}\p{\ce}{x_1}$, continuity of $\phie$ and $\ike$, and zero solid current in the separator. At current collectors, we set
\[
    \phin=0,\qquad \phip=V,
\]
with galvanostatic current density $\bar I(t)=I/(W_2W_3)$ imposed by
\[
    i^{\mathrm e}=0\quad \text{at}\ x_1=0,H,\qquad
    i^{\mathrm{n}}(0,t)=\bar I(t),\quad i^{\mathrm{p}}(H,t)=\bar I(t).
\]
The initial data are
\[
    \ck(r,0)=c^{k,0}(r),
    \qquad
    \ce(x_1,0)=\cezero.
\]

\subsection{Mechanical closure at the battery scale}

The mechanical inputs in \eqref{eq:mu-dfn-mech-dim-expanded}--\eqref{eq:eta-dfn-mech-correct}, namely the through-cell strains $\p{u^{k}_1}{x_1}$, are supplied by the reduced mechanics of Section~\ref{sec:5}. In particular, with layerwise moduli $C^{k,1}$, $C^{k,2}$ and chemo-elastic couplings $C^{k,4}$, the through-cell stress satisfies
\[
    \p{\sigma^k_{11}}{x_1}=0,
    \qquad
    \sigma^k_{11}=C^{k,1}\,\p{u^k_1}{x_1}-C^{k,4}\,\gav^{k}(x_1,t).
\]
The displacements $u^k_1$ and stresses $\sigma^k_{11}$ are continuous across all interfaces. The macroscale boundary condition is either a prescribed total displacement,
\begin{equation}
\label{eq:clamped-total-displacement}
    \sum_{k\in\{\mathrm n,\mathrm{se},\mathrm p\}}
    \int_{\Omega^k} \p{u^k_1}{x_1}\,\d x_1=0,
\end{equation}
or an imposed applied normal load $p(t)$, giving
\begin{equation}
\label{eq:applied-load-stress}
    \sigma^k_{11}=-p(t).
\end{equation}
where we take the convention used in experiments that a positive applied pressure is compressive. 
Here
\[
    \gav^{k}(x_1,t)=\frac{\Ukr(R^{k},t)}{R^{k}}
    \quad \text{in the electrodes},
    \qquad
    \gav^{\mathrm{se}}\equiv0
    \quad \text{in the separator}.
\]

Since $\sigma_{11}^k$ is independent of $x_1$ in each layer and, by stress continuity, is the same in all layers, we denote this common stress by $\sigma_{11}(t)$. The local through-cell strain is therefore
\begin{equation}
\label{eq:local-strain-from-stress}
    \varepsilon_{11}^k(x_1,t)
    =
    \p{u_1^k}{x_1}
    =
    \frac{\sigma_{11}(t)+C^{k,4}\gav^k(x_1,t)}{C^{k,1}} .
\end{equation}

For a clamped cell, substituting \eqref{eq:local-strain-from-stress} into \eqref{eq:clamped-total-displacement} gives
\begin{equation}
\label{eq:clamped-common-stress}
    \sigma_{11}(t)
    =
    -
    \frac{
    \displaystyle
    \sum_{k\in\{\mathrm n,\mathrm p\}}
    \frac{C^{k,4}}{C^{k,1}}
    \int_{\Omega^k}\gav^k(x_1,t)\,\d x_1
    }{
    \displaystyle
    \sum_{k\in\{\mathrm n,\mathrm{se},\mathrm p\}}
    \frac{H^k}{C^{k,1}}
    } .
\end{equation}
The corresponding strain field is
\begin{equation}
\label{eq:clamped-strain-field}
    \varepsilon_{11}^k(x_1,t)
    =
    \frac{C^{k,4}}{C^{k,1}}\gav^k(x_1,t)
    -
    \frac{1}{C^{k,1}}
    \frac{
    \displaystyle
    \sum_{\ell\in\{\mathrm n,\mathrm p\}}
    \frac{C^{\ell,4}}{C^{\ell,1}}
    \int_{\Omega^\ell}\gav^\ell(x_1,t)\,\d x_1
    }{
    \displaystyle
    \sum_{\ell\in\{\mathrm n,\mathrm{se},\mathrm p\}}
    \frac{H^\ell}{C^{\ell,1}}
    } .
\end{equation}

For a cell subject to an imposed normal load $p(t)$, the common stress is instead prescribed,
\[
    \sigma_{11}(t)=-p(t),
\]
and hence
\begin{equation}
\label{eq:loaded-strain-field}
    \varepsilon_{11}^k(x_1,t)
    =
    \frac{-p(t)+C^{k,4}\gav^k(x_1,t)}{C^{k,1}} .
\end{equation}
These formulae provide the strain $\p{u_1^k}{x_1}$ that enters the mechanically corrected chemical potential and overpotential in \eqref{eq:mu-dfn-mech-dim-expanded} and \eqref{eq:eta-dfn-mech-correct}.

This completes the statement of the full DFN-style electro-chemo-mechanically coupled multiscale model. Apart from the mechanical parameters characterising the mechanical response of electrodes and separator, and the partial molar volume of the active particles, two new parameters arise in the model: $\Sigkoneone$ and $\Sigkg$. They identify the magnitude of the stress on the surface of the particle generated by the non-active matrix as a consequence of electrode strain and particle expansion, respectively. They are obtained by solving the cell-function problems \eqref{lads}--\eqref{bcforce} and integrating their radial component over the surface of the sphere, as in \eqref{eq:Sigma-def}.

\subsection{Slow-charging single-particle limit: SPM with mechanics}
\label{sec:7-spm}

As is the case with the standard DFN model, it is often useful to consider the single-particle-model reduction, appropriate when the cell is operating relatively slowly. In these conditions, the electrolyte and concentration within the active particles are relatively unpolarised. Thus, we assume that $\ce(x_1,t)\approx\cezero$, $\p{\ln\ce}{x_1}\approx0$, and $\ck(r,t)\approx\overline{c}^{k}(t)$ is uniform in each particle. The layer-average interfacial molar flux is uniform in $x_1$ and, with $\ik$ defined positive in the $+x_1$ direction,
\begin{equation}
\label{eq:spm-j-avg}
    J^{\mathrm n}(t)
    =
    \frac{\bar I(t)}{F b^{\mathrm n}H^{\mathrm n}},
    \qquad
    J^{\mathrm p}(t)
    =
    -\,\frac{\bar I(t)}{F b^{\mathrm p}H^{\mathrm p}} .
\end{equation}
Thus, under positive discharge current, $J^{\mathrm n}>0$ in the negative electrode, corresponding to deintercalation, and $J^{\mathrm p}<0$ in the positive electrode, corresponding to intercalation.

Conservation of Li, together with the assumption that the concentration in the particles is uniform, yields
\begin{equation}
\label{eq:spm-mass}
    \frac{\d \overline{c}^{k}}{\d t}
    =
    -\,\frac{3}{R^{k}}\,J^{k}(t),
    \qquad k\in\{\mathrm n,\mathrm p\}.
\end{equation}
Equations \eqref{eq:spm-j-avg} and \eqref{eq:spm-mass} close the concentration ODEs.

The interfacial molar flux relates to the overpotential through Butler--Volmer kinetics, evaluated at the uniform states:
\begin{equation}
\label{eq:spm-bv}
  J^{k}(t)
    =
    \Kk
    \sqrt{\frac{\overline{c}^{k}}{\ckmax}\!\left(1-\frac{\overline{c}^{k}}{\ckmax}\right)}
    \,\sinh\!\left(\frac{F}{2RT}\,\eta^{\mathrm{SPM},k}(t)\right).
\end{equation}
The SPM overpotential is
\begin{align}
\label{eq:spm-eta}
    \eta^{\mathrm{SPM},k}(t)
    &=
    \phi^{\mathrm{SPM},k}(t)-\phie(t)
    -\Ueqk(\overline{c}^{k}),
\\
\label{eq:spm-phik}
    \phi^{\mathrm{SPM},k}(t)
    &=
    \phik(t)
    -\frac{\linVk}{F}\,\sk_{ll}(t)
    -\frac{\Vk\,\lbek}{F}
    \left(
    \Sigkoneone\,\varepsilon^{k}_{11}(t)
    -\Sigkg\,\gav^{k}(t)
    \right).
\end{align}
Here $\varepsilon^{k}_{11}(t)=\p{u^k_1}{x_1}$ is the through-cell strain in layer $k$, and $\gav^{k}(t)=\Ukr(R^{k},t)/R^{k}$. In the stress-free particle limit of Section~\ref{sec:6}, the local particle stress contribution satisfies $\sk_{ll}\equiv0$.

The mechanics reduces to a uniform through-cell stress $\sigma_{11}(t)$ and piecewise-constant strains $\varepsilon^{k}_{11}(t)$:
\[
    \sigma_{11}(t)
    =
    C^{k,1}\,\varepsilon^{k}_{11}(t)-C^{k,4}\,\gav^{k}(t),
    \qquad
    \varepsilon^{k}_{11}(t)
    =
    \frac{\sigma_{11}(t)+C^{k,4}\,\gav^{k}(t)}{C^{k,1}},
\]
with
\[
    \gav^{k}(t)=
    \begin{cases}
        \Ukr(R^{k},t)/R^{k}, & k=\mathrm n,\mathrm p,\\[0.2em]
        0, & k=\mathrm{se}.
    \end{cases}
\]
The common stress is closed either by the clamped condition
\[
    \sum_{k\in\{\mathrm n,\mathrm{se},\mathrm p\}} H^k\,\varepsilon^{k}_{11}(t)=0,
\]
or by an imposed normal load
\[
    \sigma_{11}(t)=-p(t).
\]

\paragraph{Voltage with mechanical correction}

In the slow-charging limit we neglect electrolyte concentration and Ohmic drops, set $\phie$ constant, and take $\phi^{\mathrm{SPM},k}$ from \eqref{eq:spm-phik}. The mechanical contribution to the terminal voltage is
\begin{align}
    \Delta V(t)
    &=
    \big[\phi^{\mathrm{SPM},\mathrm p}(t)-\phi^{\mathrm{SPM},\mathrm n}(t)\big]
    -
    \big[\phip(t)-\phin(t)\big]
\nonumber \\
    &=
    -\,\frac{1}{F}
    \left[
    \Vkp\lambda^{\mathrm{na},\mathrm p}
    \big(
    \Sigma^{\mathrm p}_{11}\varepsilon^{\mathrm p}_{11}
    -\Sigma^{\mathrm p,g}\gav^{\mathrm p}
    \big)
    -
    \Vkn\lambda^{\mathrm{na},\mathrm n}
    \big(
    \Sigma^{\mathrm n}_{11}\varepsilon^{\mathrm n}_{11}
    -\Sigma^{\mathrm n,g}\gav^{\mathrm n}
    \big)
    \right]
\nonumber\\
    &\quad
    -\,\frac{1}{F}
    \left[
    \linVkp\sigma^{\mathrm p}_{ll}
    -
    \linVkn\sigma^{\mathrm n}_{ll}
    \right].\label{eq:spm-dVmech}
\end{align}
In the soft binder limit, the stress-free particle has $\sigma^{\mathrm p}_{ll}=\sigma^{\mathrm n}_{ll}=0$, so the final line in \eqref{eq:spm-dVmech} vanishes.

\begin{figure}
\centering
\includegraphics[width=\textwidth]{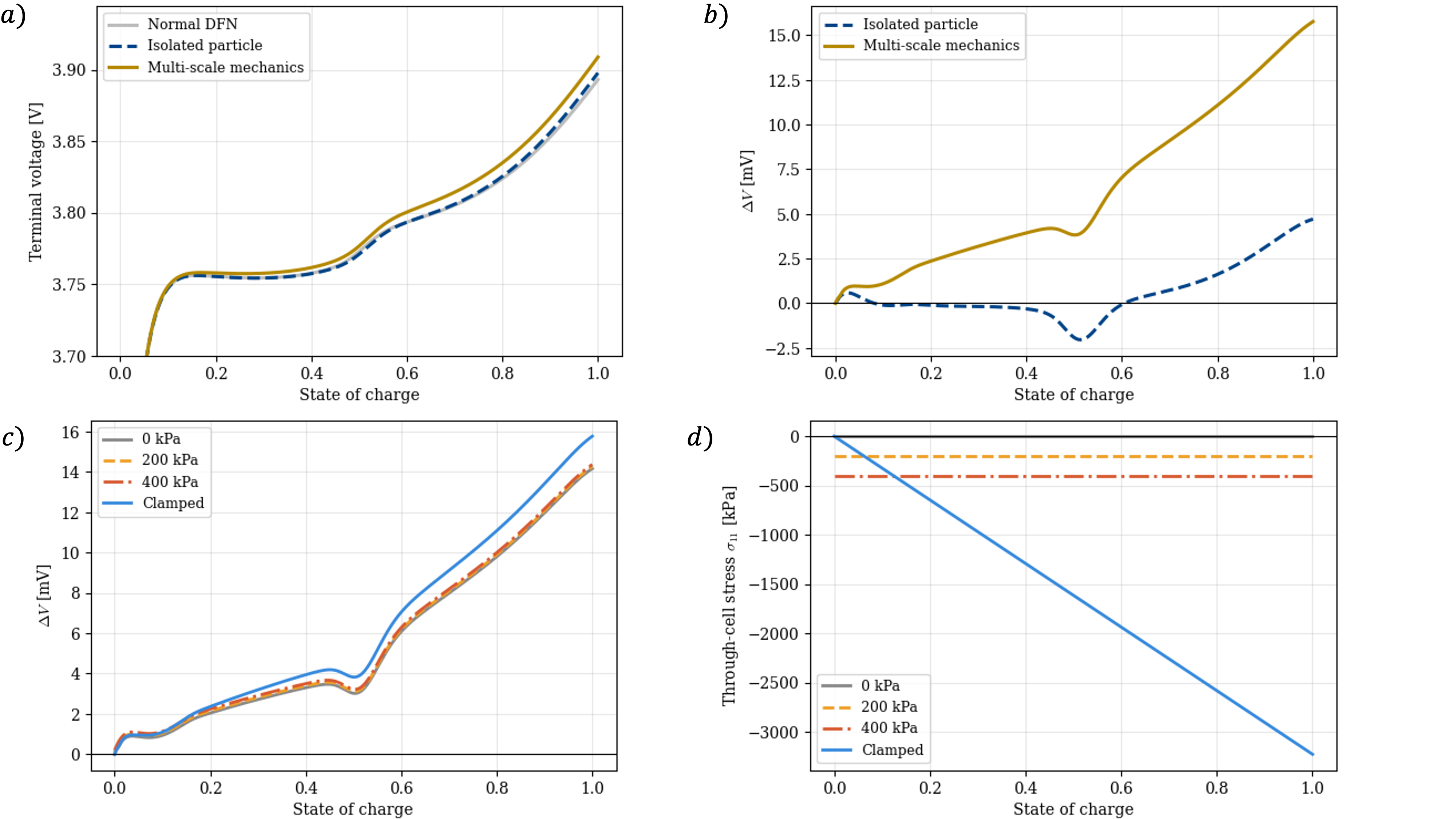}
\caption{
PyBaMM implementation of the mechanically corrected DFN model. 
(a) Terminal voltage during a $0.5$C charge for the standard DFN model (grey), DFN with stress-assisted diffusion in isolated particle (dashed blue) and full multiscale-mechanics stress-assisted diffusion and cell-scale mechanical coupling (yellow). 
(b) Mechanical voltage contribution $\Delta V$, measured relative to the standard DFN voltage, for the isolated-particle and multiscale-mechanics models. 
(c) Mechanical voltage contribution for the multiscale-mechanics model under different macroscale mechanical boundary conditions, with applied pressures of $0\,\mathrm{kPa}$, $200\,\mathrm{kPa}$ and $400\,\mathrm{kPa}$, and a perfectly clamped cell. 
(d) Corresponding through-cell stress $\sigma_{11}$ for the same boundary conditions. Positive applied pressure is plotted as compressive stress, so that $\sigma_{11}<0$. }
\label{fig:pybamm-dfn-mechanics}
\end{figure}

\subsection{PyBaMM simulations}
\label{sec:pybamm-simulations}

We now illustrate how mechanical effects affect battery operation by implementing the reduced model in PyBaMM. 
The comparison is shown in Figure~\ref{fig:pybamm-dfn-mechanics}. 
We consider three DFN-type simulations. 
The first is a standard DFN model with no mechanics. 
The second includes stress-assisted diffusion in isolated active particles \cite{Deshpande2012,Ai}, so that concentration gradients generate internal particle stresses which feed back on the lithium chemical potential, but the particle does not feel the surrounding non-active matrix. 
The third is the full multiscale-mechanics model derived above, in which the same stress-assisted diffusion is retained and the particle chemical potential is also corrected by the stress transmitted through the non-active matrix and by the macroscale through-cell mechanical state.

Figure~\ref{fig:pybamm-dfn-mechanics}(a) shows that the voltage shift caused by mechanics is small on the scale of the absolute terminal voltage, but systematic. 
The isolated-particle correction remains close to the standard DFN voltage for most of the charge, whereas the multiscale correction produces a visibly larger voltage increase, especially at high states of charge. 
This is clearer in Figure~\ref{fig:pybamm-dfn-mechanics}(b), where the voltage difference relative to the standard DFN model is plotted directly. 
The isolated-particle stress-assisted-diffusion contribution is only a few millivolts and changes sign over part of the charge. 
By contrast, the multiscale-mechanics contribution grows almost monotonically and reaches approximately $15\,\mathrm{mV}$ at full charge. 
This confirms that the non-active matrix and the cell-scale mechanical boundary condition can alter the electrochemical response even when the mechanical correction is formally higher order in the stiffness-ratio expansion.

Figures~\ref{fig:pybamm-dfn-mechanics}(c,d) isolate the role of the macroscale mechanical boundary condition. 
When the cell is subject to a prescribed applied pressure, the through-cell stress is approximately constant during the charge, with larger applied pressure producing a larger compressive stress. 
The voltage correction, however, changes only weakly between the $0$, $200$, and $400\,\mathrm{kPa}$ load-controlled cases. 
The clamped cell behaves differently: because the total thickness change is constrained to vanish, swelling generates a compressive through-cell stress whose magnitude grows throughout the charge. 
Consequently, the clamped case gives the largest voltage correction, reaching about $16\,\mathrm{mV}$ at full charge. 
The simulations therefore show that the dominant mechanical effect is not merely the presence of an externally applied pressure, but the way in which swelling is converted into stress by the global mechanical constraint. All parameters used in the simulations are shown in Table~\ref{tab:pybamm-parameters}.

\begin{table}[t]
\centering
\scriptsize
\setlength{\tabcolsep}{4pt}
\renewcommand{\arraystretch}{0.92}
\caption{Parameters used in the PyBaMM simulations shown in Figure~\ref{fig:pybamm-dfn-mechanics}. Electrochemical and transport parameters not listed explicitly are the default PyBaMM DFN parameters.}
\label{tab:pybamm-parameters}
\begin{tabular}{p{0.37\textwidth}p{0.42\textwidth}p{0.13\textwidth}}
\toprule
Quantity & Value & Units \\
\midrule
Model & DFN with Fickian particle diffusion & -- \\
Initial state of charge & $0$ & -- \\
Charge rate & $0.5$C & -- \\
Final time & $7.20\times10^{3}$ & s \\
Number of output times & $301$ & -- \\
Applied current & $-0.340$ & A \\
Nominal capacity & $0.681$ & A h \\
Temperature & $298$ & K \\
Voltage cut-off & $3.11$--$4.10$ & V \\
\midrule
Negative/separator/positive thickness & $1.00\times10^{-4}$ / $2.50\times10^{-5}$ / $1.00\times10^{-4}$ & m \\
Negative/positive particle radius & $1.00\times10^{-5}$ / $1.00\times10^{-5}$ & m \\
Initial negative/positive concentration & $4.58\times10^{3}$ / $4.92\times10^{4}$ & mol m$^{-3}$ \\
Maximum negative/positive concentration & $2.50\times10^{4}$ / $5.12\times10^{4}$ & mol m$^{-3}$ \\
Initial electrolyte concentration & $1.00\times10^{3}$ & mol m$^{-3}$ \\
Negative/separator/positive porosity & $0.300$ / $1.00$ / $0.300$ & -- \\
Negative/positive active volume fraction & $0.600$ / $0.500$ & -- \\
Negative/positive Bruggeman coefficient & $1.50$ / $1.50$ & -- \\
Negative/positive electronic conductivity & $100$ / $10.0$ & S m$^{-1}$ \\
Cation transference number & $0.400$ & -- \\
Thermodynamic factor & $1.00$ & -- \\
\midrule
Negative/positive Young's modulus & $15.0$ / $150$ & GPa \\
Negative/positive Poisson ratio & $0.300$ / $0.300$ & -- \\
Negative/positive partial molar volume & $6.10\times10^{-6}$ / $1.30\times10^{-6}$ & m$^3$ mol$^{-1}$ \\
Non-active matrix Lam\'e coefficient & $0.500$ & GPa \\
Negative/separator/positive through-cell modulus & $2.00$ / $0.100$ / $2.00$ & GPa \\
Negative/positive swelling-stress coupling & $1.00$ / $1.00$ & GPa \\
Cell-function strain coefficient & $-40.0$ / $-40.0$ & -- \\
Cell-function swelling coefficient & $30.0$ / $30.0$ & -- \\
Applied-load cases & $0$, $200$, $400$ & kPa \\
Clamped case & Zero total through-cell displacement & -- \\
\bottomrule
\end{tabular}
\end{table}
\section{Summary and conclusion}
\label{sec:conclusion}

In this work we have derived a reduced electro-chemo-mechanical model for lithium-ion batteries that incorporates mechanical coupling across the particle, electrode, and cell scales while retaining a structure close to standard DFN-type models. The central result is that, in the limit where the non-active matrix is much softer than the active particles, the leading-order particle problem is the familiar stress-assisted diffusion problem for a freely expanding spherical particle, while the mechanical influence of the surrounding non-active material enters as an effective correction to the active-particle chemical potential and interfacial overpotential. In the full DFN formulation this correction appears explicitly in \eqref{eq:mu-dfn-mech-dim-expanded} and \eqref{eq:eta-dfn-mech-correct}; in the slow-charging limit it gives the voltage correction \eqref{eq:spm-dVmech}.

A useful way of interpreting this result is that the lithium chemical potential is not an intrinsically chemical quantity, but a thermodynamic potential that contains chemical, electrical, and mechanical contributions. In other words, the battery also stores mechanical energy. In the usual DFN model the mechanical part is absent, or at most restricted to the local stress generated by concentration gradients within an isolated particle. Here, by contrast, the chemical potential also contains the mechanical work associated with the stress transmitted by the non-active matrix. Thus, the equilibrium potential entering Butler--Volmer kinetics is shifted not only by the local state of lithiation, but also by the mechanical state of the surrounding electrode and cell stack.

The multiscale structure of the model allows these quantities to be computed systematically. At the microscale, cell problems determine how the non-active matrix transmits stress to the particle surface. These cell problems define the coefficients $\Sigma^k_{11}$ and $\Sigma^{k,g}$ in \eqref{eq:Sigma-def}, which quantify the contributions of electrode-scale strain and particle swelling, respectively. At the electrode and cell scales, homogenised mechanics and thin-cell asymptotics determine the through-cell strain field from the swelling of the electrodes and from the mechanical boundary condition applied to the whole cell. This gives the local strain in \eqref{eq:local-strain-from-stress}, with the common stress determined either by the clamped condition \eqref{eq:clamped-common-stress} or by an imposed load through \eqref{eq:loaded-strain-field}.

This is also the point at which the electrodes become mechanically coupled. In standard electrochemical models, the positive and negative electrodes communicate through the electrolyte, the electrical circuit, and the global current constraint. In the present model they also communicate mechanically, albeit weakly, through the common through-cell stress and strain field. Swelling in one electrode contributes to the global mechanical state and therefore changes the strain experienced by the other electrode. This mechanical coupling is small in the stiffness-ratio expansion, but it provides a direct route by which the lithiation state of one electrode can alter the chemical potential, overpotential, and voltage contribution of the other.

The resulting formulation can be written as a mechanically corrected DFN model. The standard electrochemical equations are retained, but the active-particle chemical potential and Butler--Volmer overpotential acquire an additional contribution from the stress transmitted through the non-active matrix. The same correction can also be reduced to an SPM or SPMe setting in the appropriate slow-charging limits, as in \eqref{eq:spm-eta}--\eqref{eq:spm-phik}. This provides a direct route for incorporating cell-scale mechanics into commonly used battery models without solving the full three-dimensional mechanical problem during every electrochemical simulation.

We have also shown how the model can be implemented in \textsc{PyBaMM}. The required modifications are relatively modest: one adds a mechanically corrected active-particle chemical potential, retains the local stress-assisted diffusion contribution inside the particle, and couples the electrodes through the macroscale strain field. The resulting simulations show that the multiscale mechanical correction produces a measurable shift in the cell voltage, particularly at high states of charge where swelling stresses are largest. This confirms that stresses transmitted through the non-active matrix can affect electrochemical performance even when the underlying mechanical strains are small.

The model deliberately relies on a number of simplifying assumptions. We have treated the active particles as spherical, the non-active matrix and separator as linearly elastic homogenised solids, and the particle--matrix interface as perfectly bonded. We have also neglected fluid pressure in the pores, plasticity, fracture, viscoelasticity, evolving porosity, SEI mechanics, and large-deformation effects. These assumptions are appropriate for isolating the leading mechanism by which electrode- and cell-scale mechanics modify the electrochemical potential, but they should not be interpreted as a complete description of all mechanical processes in a working cell.

A particularly important limitation concerns microstructure. In this paper the geometry enters through idealised periodic unit cells, and the mechanical correction is summarised by the effective coefficients $\Sigma^k_{11}$ and $\Sigma^{k,g}$. These coefficients depend on the geometry of the particle arrangement, the particle volume fraction, and the elastic properties of the non-active matrix. Consequently, different lattice arrangements, such as simple-cubic, body-centred-cubic, or face-centred-cubic packings, will generally generate different particle surface stresses for the same average swelling and electrode strain. Similarly, explicit particle--particle contact, contact through binder bridges, or loss of contact during cycling could substantially alter the stress transmitted to each particle and hence the electrochemical correction derived here.

These geometric and contact effects are therefore a natural next step. In a follow-up paper we will examine how different particle arrangements and contact mechanics modify the effective cell functions and the resulting voltage correction. This will allow us to distinguish the universal part of the multiscale coupling identified here from the microstructure-specific contribution associated with realistic electrode architectures.

\section*{Acknowledgments}
This work was generously supported by the EPSRC Faraday Institution Multi-Scale Modelling project (EP/S003053/1, grant number 
FIRG059).

\bibliographystyle{plainnat} 
\bibliography{references.bib}

\section{Cell functions}
\label{app1}

This appendix states the unit-cell problems used in \eqref{thisone}, \eqref{sigmadef1}, \eqref{cells}, and \eqref{eq:Sigma-def}.  The sign convention is the one used in the main text: tensile stress is positive, compression is negative, and the unit normal on the particle surface is
\begin{equation}
\label{eq:app-normal-def}
    n_i=e_i^r=\frac{X_i}{R^k},
\end{equation}
pointing out of the active particle and into the non-active matrix.  The traction exerted by the non-active matrix on the particle is therefore $\sbekz_{ij}n_j$, consistently with \eqref{eq:mech_bc_1}.

Let $Y$ be the periodic unit cell, let $\Omega^k$ be the active particle, let $\Obek$ be the non-active matrix region, and let $\partial\Omega^k$ denote the particle--matrix interface.  The outer boundary of the unit cell is denoted by $\partial\Obek$.  All cell functions below are periodic on $\partial\Obek$, and we fix the arbitrary rigid translation by imposing zero mean displacement over $\Obek$.

The linearity of \eqref{lads}--\eqref{bcforce} allows the leading-order non-active matrix displacement to be written as
\begin{equation}
\label{eq:app-displacement-decomposition}
    \Ubekz_i
    =
    U_i^{kl}\p{u_k^{(-1)}}{x_l}
    -
    \gav^k U_i^g .
\end{equation}
Here $U_i^{kl}$ is the displacement cell function generated by a unit component of the mesoscale displacement gradient $\partial u_k^{(-1)}/\partial x_l$, while $U_i^g$ is the displacement cell function generated by a unit contraction of the particle surface.  The minus sign in the swelling term is deliberate: positive particle swelling corresponds to minus the unit-contraction solution.

The corresponding strain and stress decompositions are
\begin{equation}
\label{eq:app-strain-decomposition}
    \ebekz_{ij}
    =
    \epsilon_{ij}^{kl}\p{u_k^{(-1)}}{x_l}
    -
    \gav^k \epsilon_{ij}^{g},
\end{equation}
and
\begin{equation}
\label{eq:app-cell-decomposition}
    \sbekz_{ij}
    =
    \sigma_{ij}^{kl}\p{u_k^{(-1)}}{x_l}
    -
    \gav^k\sigma_{ij}^{g}.
\end{equation}
This is the sign convention used in \eqref{thisone}.

\subsection{Cell functions forced by electrode-scale strain}

For each ordered pair $(k,l)$ of coordinate directions, the displacement cell function $U_i^{kl}$ is defined by
\begin{align}
\label{eq:app-strain-cell-eps}
    \epsilon_{ij}^{kl}
    &=
    \frac{1}{2}\left(
    \frac{\partial U_i^{kl}}{\partial X_j}
    +
    \frac{\partial U_j^{kl}}{\partial X_i}
    \right)
    +
    \frac{1}{2}\left(
    \delta_{ik}\delta_{jl}
    +
    \delta_{jk}\delta_{il}
    \right),
    &&\text{in }\Obek,\\
\label{eq:app-strain-cell-sig}
    \sigma_{ij}^{kl}
    &=
    2\Gbek\epsilon_{ij}^{kl}
    +
    \delta_{ij}\lbek\epsilon_{mm}^{kl},
    \qquad
    \frac{\partial\sigma_{ij}^{kl}}{\partial X_j}=0,
    &&\text{in }\Obek.
\end{align}
The additional Kronecker-delta term in \eqref{eq:app-strain-cell-eps} is the imposed unit mesoscale displacement gradient.  Thus $U_i^{kl}$ is the local correction displacement, not the total affine displacement.

On the particle surface the active particle moves as a rigid body at this order.  Since the imposed affine displacement gradient has already been included in \eqref{eq:app-strain-cell-eps}, the correction displacement must cancel it inside the particle.  Hence \eqref{inpart} gives
\begin{equation}
\label{eq:app-particle-rigid-strain}
    \frac{\partial u_i^{kl}}{\partial X_j}
    =
    -\delta_{ik}\delta_{jl},
    \qquad \text{in }\Omega^k.
\end{equation}
Therefore the displacement imposed on the matrix at $\partial\Omega^k$ is
\begin{equation}
\label{eq:app-strain-cell-bc}
    U_i^{kl}
    =
    a_i^{kl}
    -
    \delta_{ik}X_l
    +
    \omega_m^{kl}\epsilon_{ijm}X_j,
    \qquad \text{on }\partial\Omega^k.
\end{equation}
Here $a_i^{kl}$ and $\omega_m^{kl}$ are unknown rigid-body translation and rotation constants.  They are determined by requiring zero resultant force and zero resultant torque on the particle,
\begin{align}
\label{eq:app-strain-cell-force}
    \int_{\partial\Omega^k}\sigma_{ij}^{kl}n_j\,\d S&=0,\\
\label{eq:app-strain-cell-torque}
    \int_{\partial\Omega^k}\epsilon_{imn}X_m\sigma_{nj}^{kl}n_j\,\d S&=0.
\end{align}
The displacement $U_i^{kl}$ and the traction $\sigma_{ij}^{kl}n_j$ are periodic on $\partial\Obek$.  Since only the symmetric part of the imposed displacement gradient enters the strain, only the six symmetric strain modes need to be solved.  For a spherical particle in a cubic unit cell, symmetry further reduces the independent computations to one normal mode, for example $(k,l)=(1,1)$, and one shear mode, for example $(k,l)=(1,2)$; the remaining modes are obtained by permuting the coordinate axes.

\subsection{Cell function forced by particle swelling}

The swelling cell function is defined as the response to a unit contraction of the particle surface.  It satisfies
\begin{align}
\label{eq:app-growth-cell-eps}
    \epsilon_{ij}^{g}
    &=
    \frac{1}{2}\left(
    \frac{\partial U_i^{g}}{\partial X_j}
    +
    \frac{\partial U_j^{g}}{\partial X_i}
    \right),
    &&\text{in }\Obek,\\
\label{eq:app-growth-cell-sig}
    \sigma_{ij}^{g}
    &=
    2\Gbek\epsilon_{ij}^{g}
    +
    \delta_{ij}\lbek\epsilon_{mm}^{g},
    \qquad
    \frac{\partial\sigma_{ij}^{g}}{\partial X_j}=0,
    &&\text{in }\Obek.
\end{align}
The particle displacement is rigid apart from the imposed unit contraction, so
\begin{equation}
\label{eq:app-growth-cell-bc}
    U_i^{g}
    =
    -X_i
    +
    a_i^{g}
    +
    \omega_m^{g}\epsilon_{ijm}X_j,
    \qquad \text{on }\partial\Omega^k.
\end{equation}
Here $a_i^{g}$ and $\omega_m^{g}$ are determined by requiring zero resultant force and zero resultant torque on the particle,
\begin{align}
\label{eq:app-growth-cell-force}
    \int_{\partial\Omega^k}\sigma_{ij}^{g}n_j\,\d S&=0,\\
\label{eq:app-growth-cell-torque}
    \int_{\partial\Omega^k}\epsilon_{imn}X_m\sigma_{nj}^{g}n_j\,\d S&=0.
\end{align}
The displacement $U_i^g$ and the traction $\sigma_{ij}^g n_j$ are periodic on $\partial\Obek$.  Positive particle swelling corresponds to the stress $-\gav^k\sigma_{ij}^{g}$ in \eqref{eq:app-cell-decomposition}.

\subsection{Effective coefficients and surface-averaged tractions}

The averaging operator used in the electrode-scale stress is
\begin{equation}
\label{eq:app-A-operator}
    {\mathcal A}(\tau_{ij})
    =
    \frac{1}{|Y|}\int_{\Obek}\tau_{ij}\,\d V
    +
    \frac{1}{|Y|}\int_{\partial\Omega^k}X_j\tau_{ik}n_k\,\d S.
\end{equation}
Hence
\begin{equation}
\label{eq:app-KM-def}
    K_{ij\,lm}^{k}
    =
    {\mathcal A}\!\left(\sigma_{ij}^{lm}\right),
    \qquad
    M_{ij}^{k}
    =
    {\mathcal A}\!\left(\sigma_{ij}^{g}\right),
\end{equation}
and the homogenised stress in electrode $k$ is
\begin{equation}
\label{eq:app-eff-stress}
    \sigma_{ij}^{\mathrm{eff},k}
    =
    K_{ij\,lm}^{k}\p{u_l^{(-1)}}{x_m}
    -
    M_{ij}^{k}\gav^k.
\end{equation}
For the cubic unit cell this reduces to the Voigt form stated in \eqref{homend}, with
\begin{equation}
\label{eq:app-Ci-def}
    {\mathcal A}(\sigma_{ij}^{11})
    =
    \begin{pmatrix}
    C^{k,1}&0&0\\
    0&C^{k,2}&0\\
    0&0&C^{k,2}
    \end{pmatrix},
    \qquad
    {\mathcal A}(\sigma_{ij}^{12})
    =
    \begin{pmatrix}
    0&C^{k,3}&0\\
    C^{k,3}&0&0\\
    0&0&0
    \end{pmatrix},
\end{equation}
and
\begin{equation}
\label{eq:app-Cg-def}
    {\mathcal A}(\sigma_{ij}^{g})
    =
    \begin{pmatrix}
    C^{k,4}&0&0\\
    0&C^{k,4}&0\\
    0&0&C^{k,4}
    \end{pmatrix}.
\end{equation}
The quantities needed in the first-order particle problem are the surface-averaged radial tractions
\begin{equation}
\label{eq:app-Sigma-general}
    \Sigma^{k}_{lm}
    =
    \frac{1}{4\pi (R^k)^2}
    \int_{\partial\Omega^k}
    n_i\sigma_{ij}^{lm}n_j\,\d S,
    \qquad
    \Sigma^{k,g}
    =
    \frac{1}{4\pi (R^k)^2}
    \int_{\partial\Omega^k}
    n_i\sigma_{ij}^{g}n_j\,\d S.
\end{equation}
There is no additional minus sign in \eqref{eq:app-Sigma-general}.  The minus sign associated with particle swelling has already been included in the stress decomposition \eqref{eq:app-cell-decomposition}.  In the thin-cell reduction used in Sections~\ref{sec:5}--\ref{sec:7}, only the through-cell strain is retained, so the radial traction applied to the particle is
\begin{equation}
\label{eq:app-radial-traction-final}
    \sigma_{rr}^{k,(1)}(R^k,t)
    =
    \Sigma^{k}_{11}\p{u_1^{(-1)}}{x_1}
    -
    \Sigma^{k,g}\gav^k.
\end{equation}
This is the form used in \eqref{eq:radial-bc-mech}.
```

\section{First-order correction: reduction to a radial particle problem}
\label{app:radial}

The first-order particle problem is not pointwise spherically symmetric, because the non-active matrix traction on $\partial\Omega^k$ depends on the angular position.  However, the homogenised DFN source term requires only the surface integral of the interfacial flux.  We therefore derive the closed problem satisfied by the spherical averages.

For any scalar $f(r,\theta,\phi)$ define
\begin{equation}
\label{eq:app-spherical-average}
    \langle f\rangle(r)
    =\frac{1}{4\pi}\int_{S^2}f(r,\theta,\phi)\,\d\Omega,
    \qquad \d\Omega=\sin\theta\,\d\theta\,\d\phi.
\end{equation}
For a vector displacement $U_i^{k,(1)}$ and stress $\sigma_{ij}^{k,(1)}$, define the averaged radial displacement and averaged radial and hoop stresses by
\begin{align}
\label{eq:app-avg-Ur}
    U_r^{k,(1)}(r,t)&=\left\langle U_i^{k,(1)}e_i^r\right\rangle,\\
\label{eq:app-avg-srr}
    \sigma_{rr}^{k,(1)}(r,t)&=\left\langle e_i^r\sigma_{ij}^{k,(1)}e_j^r\right\rangle,\\
\label{eq:app-avg-stt}
    \sigma_{\theta\theta}^{k,(1)}(r,t)
    &=\sigma_{\phi\phi}^{k,(1)}(r,t)
      =\frac{1}{2}\left\langle
      (\delta_{ij}-e_i^r e_j^r)\sigma_{ij}^{k,(1)}
      \right\rangle .
\end{align}
We also write $\ckf(r,t)=\langle c^{k,(1)}\rangle$.  Since the constitutive law is linear and the swelling is isotropic, the averaged stresses are
\begin{align}
\label{eq:app-radial-srr}
    \sigma_{rr}^{k,(1)}
    &=(2\Gk+\lk)\frac{\partial U_r^{k,(1)}}{\partial r}
      +2\lk\frac{U_r^{k,(1)}}{r}
      -(2\Gk+3\lk)\linVVk\ckf,\\
\label{eq:app-radial-stt}
    \sigma_{\theta\theta}^{k,(1)}
    &=\lk\frac{\partial U_r^{k,(1)}}{\partial r}
      +2(\Gk+\lk)\frac{U_r^{k,(1)}}{r}
      -(2\Gk+3\lk)\linVVk\ckf .
\end{align}
Averaging $\partial\sigma_{ij}^{k,(1)}/\partial X_j=0$ over the sphere gives the radial equilibrium equation
\begin{equation}
\label{eq:app-radial-eq}
    \frac{\partial\sigma_{rr}^{k,(1)}}{\partial r}
    +\frac{2}{r}\bigl(\sigma_{rr}^{k,(1)}-\sigma_{\theta\theta}^{k,(1)}\bigr)=0.
\end{equation}
The angular-derivative terms vanish because they are surface divergences on the closed sphere.

The averaged mechanical boundary conditions are
\begin{equation}
\label{eq:app-radial-mech-bc}
    U_r^{k,(1)}(0,t)=0,
    \qquad
    \sigma_{rr}^{k,(1)}(R^k,t)=T^k(x_1,t),
\end{equation}
where
\begin{equation}
\label{eq:app-Tk-def}
    T^k(x_1,t)
    =\Sigma^{k}_{11}\frac{\partial u_1^k}{\partial x_1}
    -\Sigma^{k,g}\gav^k(x_1,t),
    \qquad
    \gav^k=\frac{U_r^{k,(0)}(R^k,t)}{R^k}.
\end{equation}
For the sign convention in Appendix~\ref{app1}, both $\Sigma^{k}_{11}$ and $\Sigma^{k,g}$ are positive.  Thus $T^k<0$ corresponds to a compressive radial traction on the particle.

Substitution of \eqref{eq:app-radial-srr}--\eqref{eq:app-radial-stt} into \eqref{eq:app-radial-eq} gives
\begin{equation}
\label{eq:app-radial-U-ode}
    \frac{\partial^2 U_r^{k,(1)}}{\partial r^2}
    +\frac{2}{r}\frac{\partial U_r^{k,(1)}}{\partial r}
    -\frac{2}{r^2}U_r^{k,(1)}
    =
    \frac{2\Gk+3\lk}{2\Gk+\lk}\,\linVVk\frac{\partial \ckf}{\partial r}.
\end{equation}
It is useful to split the solution into a homogeneous part, which carries the non-active matrix traction, and a particular part, which is traction-free:
\begin{equation}
\label{eq:app-split}
    U_r^{k,(1)}=\widehat U_r^{k,(1)}+\widetilde U_r^{k,(1)},
    \qquad
    \ckf=\widetilde c^{k,(1)}.
\end{equation}
The homogeneous solution is
\begin{equation}
\label{eq:app-hom-U}
    \widehat U_r^{k,(1)}=\frac{T^k}{2\Gk+3\lk}\,r,
\end{equation}
which produces the uniform stress
\begin{equation}
\label{eq:app-hom-stress}
    \widehat\sigma_{rr}^{k,(1)}
    =\widehat\sigma_{\theta\theta}^{k,(1)}
    =\widehat\sigma_{\phi\phi}^{k,(1)}=T^k,
    \qquad
    \widehat\sigma_{ll}^{k,(1)}=3T^k.
\end{equation}
Since this stress is independent of $r$, it does not contribute to radial diffusion, but it does shift the particle chemical potential and the interfacial overpotential.

The averaged first-order transport equation is
\begin{equation}
\label{eq:app-radial-diff}
    \frac{\partial \widetilde c^{k,(1)}}{\partial t}
    +\frac{1}{r^2}\frac{\partial}{\partial r}
      \left(r^2\widetilde N_r^{k,(1)}\right)=0,
    \qquad 0<r<R^k,
\end{equation}
with the corrected linearised flux
\begin{equation}
\label{eq:app-radial-flux-correct}
    \widetilde N_r^{k,(1)}
    =-\Ddiffk\left[
      \ckz(1-\ckz)\frac{\partial \widetilde\mu^{k,(1)}}{\partial r}
      +(1-2\ckz)\widetilde c^{k,(1)}\frac{\partial \mukz}{\partial r}
    \right].
\end{equation}
The factor $(1-2\ckz)$ comes from the linearisation of the mobility $c(1-c)$.  The averaged first-order chemical potential is
\begin{equation}
\label{eq:app-mu-tilde}
    \widetilde\mu^{k,(1)}
    =-\left.\frac{\partial\Ueqk}{\partial c}\right|_{c=\ckz}\widetilde c^{k,(1)}
     -\gamma\linVVk\widetilde\sigma_{ll}^{k,(1)}
     -\gamma\VVk T^k,
\end{equation}
where $\widetilde\sigma_{ll}^{k,(1)}=\widetilde\sigma_{rr}^{k,(1)}+2\widetilde\sigma_{\theta\theta}^{k,(1)}$.  The last term is the contribution of the homogeneous stress trace $3T^k$.

At the particle surface, the first-order Butler--Volmer condition is
\begin{equation}
\label{eq:app-radial-BV}
    \widetilde N_r^{k,(1)}(R^k,t)
    =
    J^{k,(1)}\sinh\left(\frac{\eta^{k,(0)}}{2}\right)
    +\frac{J^{k,(0)}\widetilde\eta^{k,(1)}}{2}
    \cosh\left(\frac{\eta^{k,(0)}}{2}\right),
\end{equation}
with
\begin{equation}
\label{eq:app-radial-eta}
    \widetilde\eta^{k,(1)}
    =\phikf-\phief
     -\left.\frac{\partial\Ueqk}{\partial c}\right|_{c=\ckz(R^k,t)}\widetilde c^{k,(1)}(R^k,t)
     -\gamma\linVVk\widetilde\sigma_{ll}^{k,(1)}(R^k,t)
     -\gamma\VVk T^k .
\end{equation}
If the reaction prefactor $\Kk$ is independent of $\Lambda$, then
\begin{equation}
\label{eq:app-J1}
    J^{k,(1)}
    =\frac{J^{k,(0)}}{2}\left[
      \frac{\cef}{\cez}
      +\frac{(1-2\ckz(R^k,t))\widetilde c^{k,(1)}(R^k,t)}
      {\ckz(R^k,t)(1-\ckz(R^k,t))}
    \right],
\end{equation}
with the obvious additional term if $\Kk$ is also perturbed.  The initial condition is
\begin{equation}
\label{eq:app-c1-initial}
    \widetilde c^{k,(1)}(r,0)=0.
\end{equation}
Equations \eqref{eq:app-radial-U-ode}--\eqref{eq:app-c1-initial} are a radial particle problem of the same type as the leading-order stress-assisted diffusion problem, except for the extra spatially uniform correction $-\gamma\VVk T^k$ in \eqref{eq:app-mu-tilde} and \eqref{eq:app-radial-eta}.  This is why the non-active matrix can be incorporated in the DFN model as a correction to the chemical potential and overpotential without solving a non-radial particle problem at every time step.

\section{PyBaMM implementation}
\label{app:pybamm}

This appendix summarises how the mechanically corrected DFN model in Section~\ref{sec:7} can be implemented in PyBaMM.  The implementation does not require solving the full three-dimensional cell problem during the electrochemical simulation.  The cell problems of Appendix~\ref{app1} are solved once, offline, to obtain the dimensionless coefficients
\[
    C^{k,1},\quad C^{k,4},\quad \Sigma^{k}_{11},\quad \Sigma^{k,g},
    \qquad k\in\{\mathrm n,\mathrm p\}.
\]
During the DFN simulation one computes, for each electrode,
\begin{equation}
\label{eq:app-pybamm-gbar}
    \gav^k(x_1,t)
    =\frac{U_r^k(R^k,x_1,t)}{R^k}
    =\frac{3}{(R^k)^3}\int_0^{R^k}
      \linVk\bigl(c^k(r,x_1,t)-\ckref\bigr)r^2\,\d r.
\end{equation}
The one-dimensional mechanical closure is then
\begin{equation}
\label{eq:app-pybamm-stress}
    \sigma_{11}^k
    =C^{k,1}\frac{\partial u_1^k}{\partial x_1}
     -C^{k,4}\gav^k,
\end{equation}
with continuity of $u_1$ and $\sigma_{11}$ across interfaces.  For an imposed compressive pressure $p(t)>0$ we use the sign convention
\begin{equation}
\label{eq:app-pybamm-pressure}
    \sigma_{11}^k=-p(t),
    \qquad
    \frac{\partial u_1^k}{\partial x_1}
    =\frac{C^{k,4}\gav^k-p(t)}{C^{k,1}}.
\end{equation}
For a clamped cell, the common through-cell stress is instead found from the zero-total-displacement condition
\begin{equation}
\label{eq:app-pybamm-clamp}
    \sum_{k\in\{\mathrm n,\mathrm{se},\mathrm p\}}
    \int_{\Omega^k}\frac{\partial u_1^k}{\partial x_1}\,\d x_1=0,
\end{equation}
with $\gav^{\mathrm{se}}=0$ in the separator.  Equivalently, in each layer
\begin{equation}
\label{eq:app-pybamm-strain-common-stress}
    \frac{\partial u_1^k}{\partial x_1}
    =\frac{\sigma_{11}+C^{k,4}\gav^k}{C^{k,1}},
\end{equation}
and \eqref{eq:app-pybamm-clamp} determines the common stress $\sigma_{11}(t)$.

The non-active matrix contribution to the active-particle chemical potential is
\begin{equation}
\label{eq:app-pybamm-dmu}
    \Delta\mu_{\mathrm{na}}^k
    =-\Vk\lbek
    \left(
      \Sigma^{k}_{11}\frac{\partial u_1^k}{\partial x_1}
      -\Sigma^{k,g}\gav^k
    \right).
\end{equation}
Thus the mechanically corrected particle chemical potential is
\begin{equation}
\label{eq:app-pybamm-mu}
    \mu^{\mathrm{DFN},k}
    =\mu^0-F\Ueqk(c^k)
     -\linVk\sigma_{ll}^k
     +\Delta\mu_{\mathrm{na}}^k.
\end{equation}
The same correction enters the Butler--Volmer overpotential as
\begin{equation}
\label{eq:app-pybamm-eta}
    \eta^{\mathrm{DFN},k}
    =\phik-\phie-\Ueqk(c^k_R)
     -\frac{\linVk}{F}\sigma_{ll}^k(R^k,t)
     +\frac{1}{F}\Delta\mu_{\mathrm{na}}^k,
\end{equation}
where $c^k_R=c^k(R^k,x_1,t)$.  Since $\Delta\mu_{\mathrm{na}}^k$ depends on $x_1$ and $t$ but not on the particle radial coordinate $r$, it shifts the interfacial overpotential but does not add a radial gradient to the particle diffusion flux.  The only stress term that contributes directly to radial stress-assisted diffusion is therefore the local particle stress $\sigma_{ll}^k(r,x_1,t)$ generated by non-uniform swelling inside the particle.

Equivalently, one may implement the correction by replacing the open-circuit potential in the Butler--Volmer expression by
\begin{equation}
\label{eq:app-pybamm-Ueff}
    \Ueqk_{\mathrm{eff}}
    =\Ueqk
     +\frac{\linVk}{F}\sigma_{ll}^k(R^k,t)
     -\frac{1}{F}\Delta\mu_{\mathrm{na}}^k.
\end{equation}
Equations \eqref{eq:app-pybamm-gbar}--\eqref{eq:app-pybamm-Ueff} are the minimal additions needed to move from a standard stress-assisted DFN implementation to the multiscale-mechanics DFN model derived in the main text.

\section{Parameters}\label{sec:params}
In this section, we present the parameter sets for three electrode materials: LiMn$_2$O$_4$ (LMO), LiNi$_{0.8}$Co$_{0.1}$Mn$_{0.1}$O$_2$ (NMC811), and graphite (LiC$_6$). The binder (PVDF) has Young's modulus $E=4$\,\si{GPa} and Poisson's ratio $\nu=0.38$~\cite{Guney2005}, giving a shear modulus $G=1.45$\,\si{GPa} and first Lam\'e parameter $\lambda=4.59$\,\si{GPa}. Because the representative element consists of a single spherical particle
inscribed in a cubic cell, the active-material volume fraction cannot exceed the simple-cubic packing limit $\epsilon_s=\pi/6\approx0.524$; larger values
would place the particle outside its own cell ($L^*<2R_p$). Although the measured LG M50 fractions are higher (0.665 and 0.75 for NMC and graphite), we adopt $\epsilon_s=0.52$ for both, the largest value consistent with the
single-particle cubic geometry, as shown in Tables~\ref{tab:params_NMC}
and~\ref{tab:params_graphite}.


\begin{figure}[htbp]
\centering
\includegraphics[width=5cm]{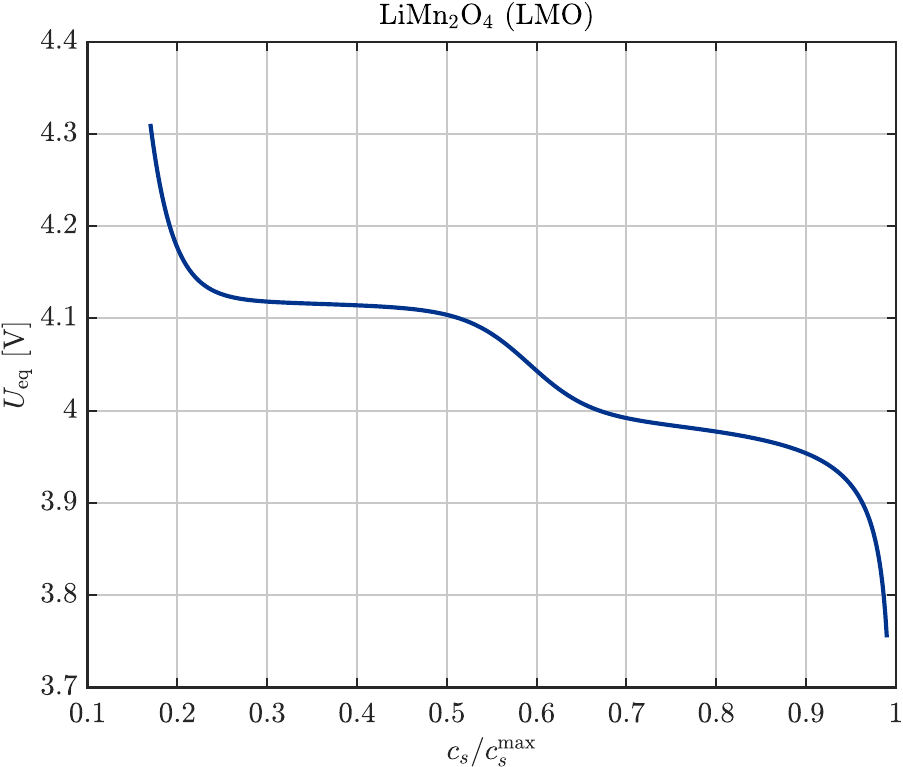}
\includegraphics[width=5cm]{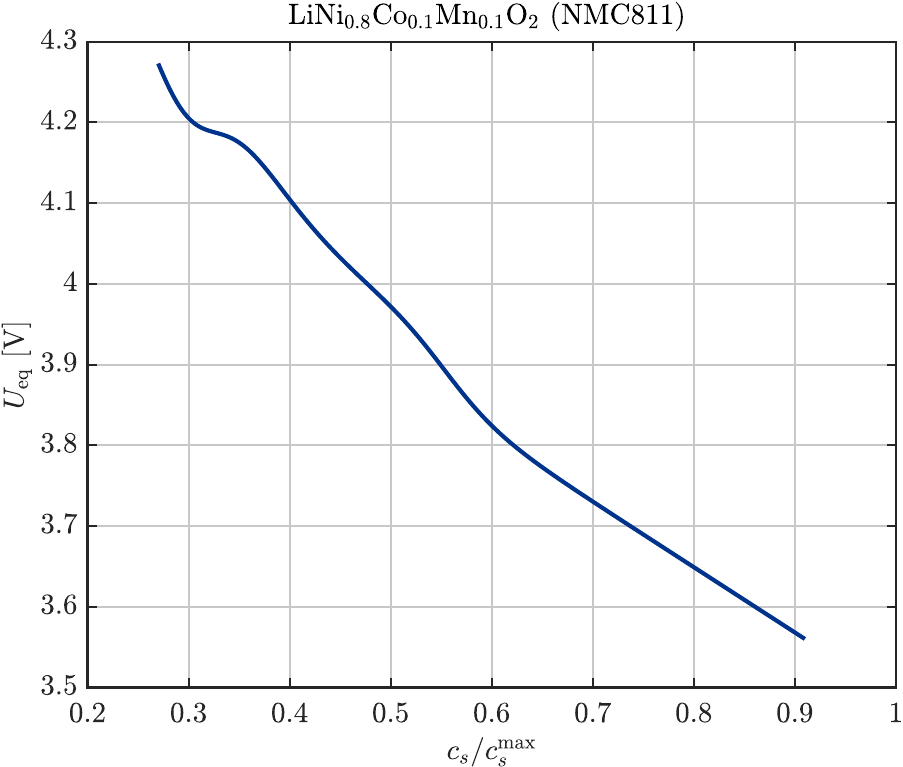}
\includegraphics[width=5cm]{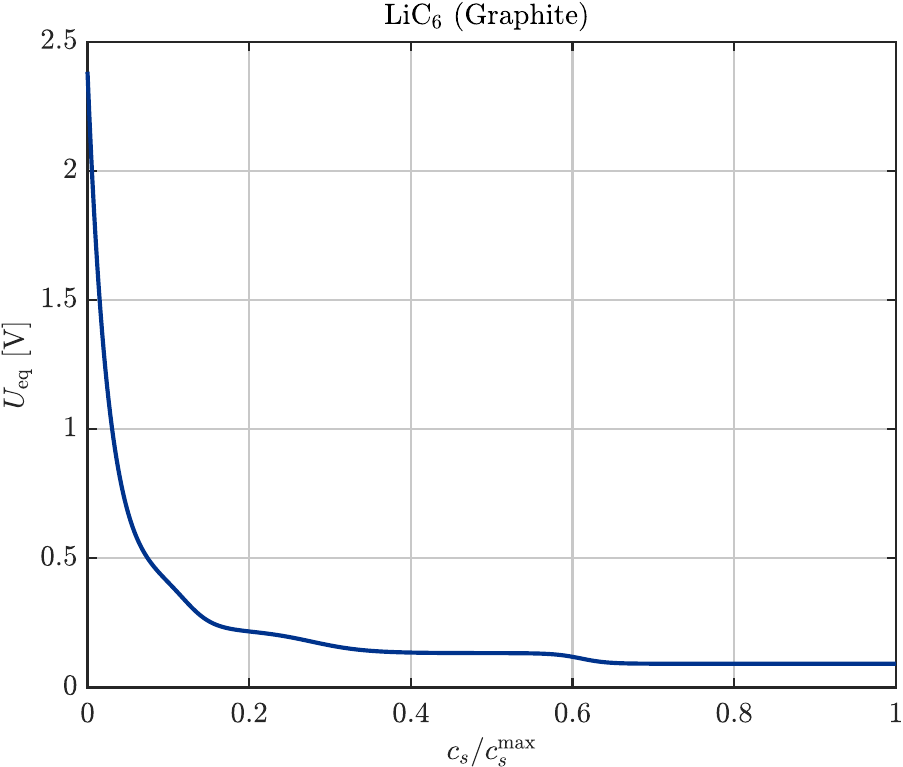}
\\
\includegraphics[width=5cm]{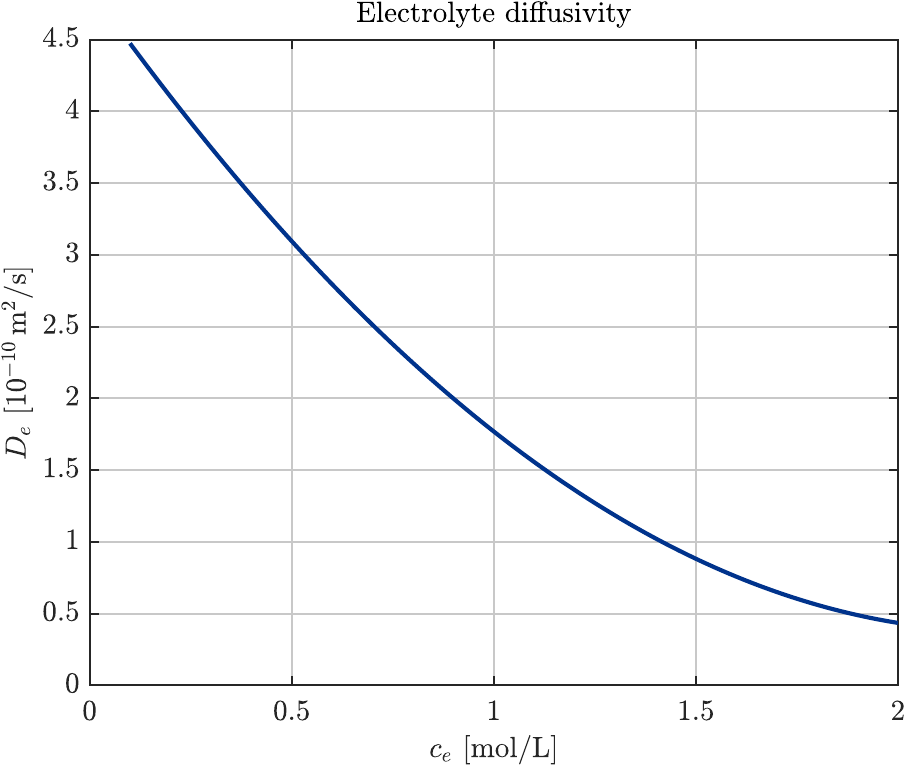}
\includegraphics[width=5cm]{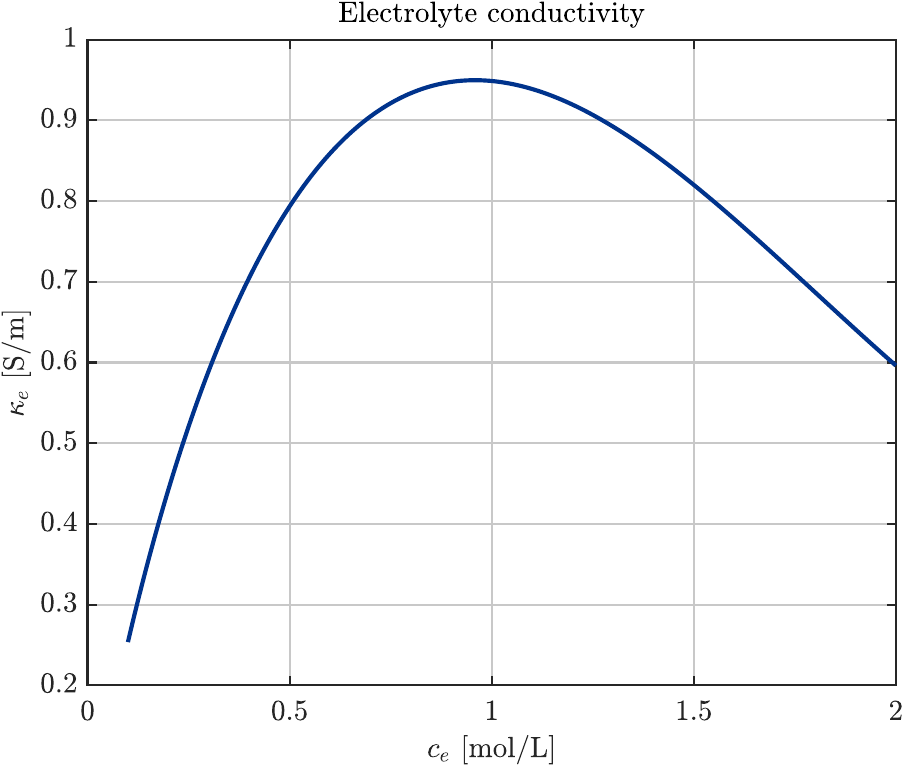}
\caption{On the top row are the plots of open-circuit potential for LMO. NMC811 and graphite and on the bottom row are the plots electrolyte diffusivity and conductivity }\label{fig:params}
\end{figure}
\begin{table}[htbp]
\centering
\scriptsize
\setlength{\tabcolsep}{5pt}
\renewcommand{\arraystretch}{1.18}
\begin{tabular}{l l l l}
\hline
Parameter & Symbol & Value & Unit \\ [0.4ex]
\hline
\multicolumn{4}{l}{Dimensional parameters}\\
\hline
Maximum Li concentration & $c^{\mathrm{k,max}}$ & 22860\,\cite{DoyleNewman1996,Chandra2024} & \si{mol/m^3} \\
Solid diffusivity & $\Dk$ & $1\times10^{-13}$\,\cite{Taleghani_2017,DoyleNewman1996} & \si{m^2/s} \\
Electronic conductivity & $\kappa^k$ & 3.8\,\cite{DoyleNewman1996} & \si{S/m} \\
Open-circuit potential & $\Ueqk$ & Fig.~\ref{fig:params} (spinel)\,\cite{DoyleNewman1996} & \si{V} \\
Particle radius & $R_p$ & 8.5\,\cite{DoyleNewman1996,Klinsmann2016} & \si{\micro m} \\
Reaction-rate constant & $\Kk$ & $2.0\times10^{-11}$\,\cite{DoyleNewman1996} & \si{m^{5/2}.mol^{-1/2}.s^{-1}} \\
Porosity & $\alpha$ & 0.30\,\cite{DoyleNewman1996} & --- \\
Active-material fraction & $\epsilon_s$ & 0.297\,\cite{DoyleNewman1996} & --- \\
Bruggeman exponent & $b$ & 1.5\,\cite{DoyleNewman1996} & --- \\
Electrode thickness & $H$ & 174\,\cite{DoyleNewman1996} & \si{\micro m} \\
Electrode width & $W$ & 9.5\,\cite{Gunter2022} & \si{cm} \\
Cell Area & $A$ & $9.5\times51$\,\cite{Gunter2022} & \si{cm^2} \\
Number of electrode pairs & $m$ & 18\textsuperscript{\dag} & --- \\
Cell capacity & $Q$ & 46\textsuperscript{\dag} & \si{Ah} \\
Current per pair at 1C & $I$ & 2.53\textsuperscript{\dag} & A \\
Initial electrolyte conc. & $c_e^0$ & 1000\,\cite{DoyleNewman1996} & \si{mol/m^3} \\
Typical electrolyte diffusivity & $D^{\mathrm{e}}$ & $1.77\times10^{-10}$\,\cite{Nyman2008} & \si{m^2/s} \\
Typical electrolyte conductivity & $\kappa^{\mathrm{e}}$ & 0.95\,\cite{Nyman2008} & \si{S/m} \\
Transference number & $t^+$ & 0.363\,\cite{DoyleNewman1996} & --- \\
Length of a representative element & $L$ & 20.54\textsuperscript{\dag} & \si{\micro m} \\
Electrolyte diffusivity & $D^{\mathrm{e}}$ & $7.5\times10^{-9}$\,\cite{DoyleNewman1996} ($10^{-9}-10^{-11}$) & \si{m^2/s} \\
Electrolyte conductivity & $\kappa^{\mathrm{e}}$ & 1\,\cite{DoyleNewman1996} & \si{S/m} \\
Young's modulus & $E$ & 93\,\cite{Klinsmann2016} & \si{GPa} \\
Poisson's ratio & $\nu$ & 0.30\,\cite{LMO_elastic2013,Klinsmann2016} & --- \\
Bulk modulus & $K$ & 100--200\,\cite{LMO_elastic2013} & \si{GPa} \\
Shear modulus & $G$ & 35.8 ($E{=}93,\nu{=}0.3$)\textsuperscript{\dag} & \si{GPa} \\
First Lam\'e parameter & $\lambda$ & 53.7 ($E{=}93,\nu{=}0.3$)\textsuperscript{\dag} & \si{GPa} \\
Partial molar volume of Li & $V_c$ & $+3.497\times10^{-6}$\,\cite{Klinsmann2016} & \si{m^3/mol} \\
Universal gas constant & $R$ & 8.314 & \si{J/molK} \\
Temperature & $T$ & 298.15 & \si{K} \\
Faraday constant & $F$ & 96487 & \si{A.s/mol} \\
\hline
\end{tabular}
\caption{Parameters for LiMn$_2$O$_4$ (LMO), using \cite{DoyleNewman1996}
for the electrochemical, transport and geometric parameters,
\cite{Klinsmann2016} for the mechanical parameters, and \cite{Gunter2022}
for the pouch-cell area and capacity. \textsuperscript{\dag}~computed
($L=R_p(4\pi/3\epsilon_s)^{1/3}=20.54$\,\si{\micro m} with $\epsilon_s=0.297$;
$G,\lambda$ from $E,\nu$). The LMO cell uses the same pouch format as the
NMC/graphite cells ($m=18$ pairs); with LMO's lower $c^{\mathrm{k,max}}$
and active fraction the capacity is $\sim46$\,\si{Ah}, giving a per-pair
1C current of $2.53$\,\si{A}. }
\label{tab:params_lmo}
\end{table}
\begin{table}[htbp]
\centering
\scriptsize
\setlength{\tabcolsep}{5pt}
\renewcommand{\arraystretch}{1.18}
\begin{tabular}{l l l l}
\hline
Parameter & Symbol & Value & Unit \\ [0.4ex]
\hline
\multicolumn{4}{l}{Dimensional parameters for NMC811}\\
\hline
Maximum Li concentration & $c^{\mathrm{k,max}}$ & 63104\,\cite{Chen2020} & \si{mol/m^3} \\
Solid diffusivity & $\Dk$ & $1.48\times10^{-15}$\,\cite{Chen2020} & \si{m^2/s} \\
Electronic conductivity & $\kappa^k$ & 0.18\,\cite{Chen2020} & \si{S/m} \\
Open-circuit potential & $\Ueqk$ & Fig.~\ref{fig:params} (Eq.~8)\,\cite{Chen2020} & \si{V} \\
Particle radius & $R_p$ & 5.22\,\cite{Chen2020} & \si{\micro m} \\
Reaction-rate constant & $\Kk$ & $3.42\times10^{-6}$\,\cite{Chen2020} & \si{A.m^{-2}(m^3.mol^{-1})^{1.5}} \\
Porosity & $\epsilon$ & 0.335\,\cite{Chen2020} & --- \\
Active-material fraction & $\epsilon_s$ & 0.5\textsuperscript{\dag} & --- \\
Bruggeman exponent & $b$ & 1.5\,\cite{Chen2020} & --- \\
Typical electrolyte conc. & $c_e^0$ & 1000\,\cite{Chen2020} & \si{mol/m^3} \\
Transference number & $t^+$ & 0.2594\,\cite{Chen2020,Nyman2008} & --- \\
Length of a representative element & $L$ & 10.602\textsuperscript{\dag} & \si{\micro m} \\
Electrode thickness & $H$ & 75.6\,\cite{Chen2020} & \si{\micro m} \\
Electrode width & $W$ & 9.5\,\cite{Gunter2022} & \si{cm} \\
Cell area & $A$ & $9.5\times51$\,\cite{Gunter2022} & \si{cm^2} \\
Number of electrode pairs & $m$ & 18\textsuperscript{\dag} & --- \\
Cell capacity & $Q$ & 78\,\cite{Gunter2022} & \si{Ah} \\
Current per pair at 1C & $I$ & 4.34\textsuperscript{\dag} & A \\
Typical electrolyte diffusivity & $D^{\mathrm{e}}$ & $1.77\times10^{-10}$\,\cite{Nyman2008} & \si{m^2/s} \\
Typical electrolyte conductivity & $\kappa^{\mathrm{e}}$ & 0.95\,\cite{Nyman2008} & \si{S/m} \\
Young's modulus & $E$ & 184\,\cite{Xu2019} & \si{GPa} \\
Poisson's ratio & $\nu$ & 0.26\,\cite{Xu2019} & --- \\
Shear modulus & $G$ & 73.0 ($E{=}184,\nu{=}0.26$)\textsuperscript{\dag} & \si{GPa} \\
First Lam\'e parameter & $\lambda$ & 79.1 ($E{=}184,\nu{=}0.26$)\textsuperscript{\dag} & \si{GPa} \\
Partial molar volume of Li & $V_c$ & $+7.88\times10^{-7}$\,\cite{Ai2022} & \si{m^3/mol} \\
Universal gas constant & $R$ & 8.314 & \si{J/molK} \\
Temperature & $T$ & 298.15 & \si{K} \\
Faraday constant & $F$ & 96487 & \si{A.s/mol} \\
\hline
\end{tabular}
\caption{Parameters for NMC811, using \cite{Chen2020} for the electrochemical,
transport and geometric parameters, \cite{Ai2022} and \cite{Xu2019} for the
mechanical parameters, \cite{Nyman2008} for the electrolyte, and \cite{Gunter2022} for the
pouch-cell area and capacity. \textsuperscript{\dag}~computed
($L=R_p(4\pi/3\epsilon_s)^{1/3}=10.602$\,\si{\micro m} with $\epsilon_s=0.5$;
$G,\lambda$ from $E,\nu$). The NMC811 cell uses the same pouch format as the
LMO/graphite cells ($m=18$ pairs), giving a capacity of $\sim78$\,\si{Ah}
and a per-pair 1C current of $4.34$\,\si{A}.}
\label{tab:params_NMC}
\end{table}

\begin{table}[htbp]
\centering
\scriptsize
\setlength{\tabcolsep}{5pt}
\renewcommand{\arraystretch}{1.18}
\begin{tabular}{l l l l}
\hline
Parameter & Symbol & Value & Unit \\ [0.4ex]
\hline
\multicolumn{4}{l}{Dimensional parameters}\\
\hline
Maximum Li concentration & $c^{\mathrm{k,max}}$ & 33133\,\cite{Chen2020} & \si{mol/m^3} \\
Solid diffusivity & $\Dk$ & $3.3\times10^{-14}$\,\cite{Chen2020} & \si{m^2/s} \\
Electronic conductivity & $\kappa^k$ & 215\,\cite{Chen2020} & \si{S/m} \\
Open-circuit potential & $\Ueqk$ & Fig.~\ref{fig:params} (Eq.~9)\,\cite{Chen2020} & \si{V} \\
Particle radius & $R_p$ & 5.86\,\cite{Chen2020} & \si{\micro m} \\
Reaction-rate constant & $\Kk$ & $6.48\times10^{-7}$\,\cite{Chen2020} & \si{A.m^{-2}(m^3.mol^{-1})^{1.5}} \\
Porosity & $\epsilon$ & 0.25\,\cite{Chen2020} & --- \\
Active-material fraction & $\epsilon_s$ & 0.5\textsuperscript{\dag} & --- \\
Bruggeman exponent & $b$ & 1.5\,\cite{Chen2020} & --- \\
Typical electrolyte conc. & $c_e^0$ & 1000\,\cite{Chen2020} & \si{mol/m^3} \\
Transference number & $t^+$ & 0.2594\,\cite{Chen2020,Nyman2008} & --- \\
Length of a representative element & $L$ & 11.902\textsuperscript{\dag} & \si{\micro m} \\
Electrode thickness & $H$ & 85.2\,\cite{Chen2020} & \si{\micro m} \\
Electrode width & $W$ & 9.5\,\cite{Gunter2022} & \si{m} \\
Cell area & $A$ & $9.5\times51.2$\,\cite{Gunter2022} & \si{cm^2} \\
Number of electrode pairs & $m$ & 18\textsuperscript{\dag} & --- \\
Cell capacity & $Q$ & 78\,\cite{Gunter2022} & \si{Ah} \\
Current per pair at 1C & $I$ & 4.34\textsuperscript{\dag} & A \\
Typical electrolyte diffusivity & $D^{\mathrm{e}}$ & $1.77\times10^{-10}$\,\cite{Nyman2008} & \si{m^2/s} \\
Typical electrolyte conductivity & $\kappa^{\mathrm{e}}$ & 0.95\,\cite{Nyman2008} & \si{S/m} \\
Young's modulus & $E$ & 15\,\cite{Qi2010} & \si{GPa} \\
Poisson's ratio & $\nu$ & 0.30\,\cite{Qi2010} & --- \\
Shear modulus & $G$ & 5.77 ($E{=}15,\nu{=}0.3$)\textsuperscript{\dag} & \si{GPa} \\
First Lam\'e parameter & $\lambda$ & 8.65 ($E{=}15,\nu{=}0.3$)\textsuperscript{\dag} & \si{GPa} \\
Partial molar volume of Li & $V_c$ & $+3.20\times10^{-6}$\citep{Ai2022} & \si{m^3/mol} \\
Universal gas constant & $R$ & 8.314 & \si{J/molK} \\
Temperature & $T$ & 298.15 & \si{K} \\
Faraday constant & $F$ & 96487 & \si{A.s/mol} \\
\hline
\end{tabular}
\caption{Parameters for graphite (LiC$_6$), using \cite{Chen2020} for the
electrochemical, transport and geometric parameters, Qi et al.~(2010) for
the elastic moduli, \cite{Nyman2008} for the electrolyte, and \cite{Gunter2022}
for the pouch-cell area and capacity. \textsuperscript{\dag}~computed
($L=R_p(4\pi/3\epsilon_s)^{1/3}=11.902$\,\si{\micro m} with $\epsilon_s=0.5$;
$G,\lambda$ from $E,\nu$). 
}
\label{tab:params_graphite}
\end{table}

\end{document}